\documentclass{pasa}

 \usepackage[authoryear]{natbib}
 \usepackage{url}
 \usepackage{hyperref}
 \usepackage[usenames]{color}
 \usepackage{colortbl}
 \bibpunct{(}{)}{;}{a}{}{,}
\newcommand{\kms}{\ensuremath{\mbox{km~s}^{-1}}}

\newcommand{\kpc}{\ensuremath{\rm kpc}}
\newcommand{\Msun}{\ensuremath{\rm M_\odot}}
\newcommand{\Lsun}{\ensuremath{\rm L_\odot}}
\newcommand{\Rsun}{\ensuremath{\rm R_\odot}}
   \newcommand{\aj}{AJ}                      
   
   \newcommand{\araa}{ARA\&A}                  
   \newcommand{\apj}{ApJ}                      
   \newcommand{\apjl}{ApJL}                    
   \newcommand{\apjs}{ApJS}                    

   \newcommand{\aap}{A\&A}                      
   
   \newcommand{\mnras}{MNRAS}

   \newcommand{\pasa}{Publications of the Astron. Soc. of Australia}

   \newcommand{\ssr}{Space Science Reviews}

\title[Properties  of dwarfs in Cyg~OB2]{Properties of dwarf stars in Cygnus~OB2}

\author[Maryeva et al.]{Olga Maryeva $^{1}$\thanks{E-mail:
olga.maryeva@gmail.com},  S.~Yu.~Parfenov$^2$,  M.~V.~Yushkin$^1$,  A.~S.~Shapovalova$^2$ and  S.~Yu.~Gorda$^2$\\
\affil{$^{1}$Special Astrophysical Observatory of the Russian Academy of Sciences, Nizhnii Arkhyz, 369167, Russia}
\affil{$^{2}$Ural Federal University, 51 Lenin Str., Ekaterinburg 620000, Russia
}}
\jid{PASA}
\doi{10.1017/pas.\the\year.xxx}
\jyear{\the\year}
 
\begin{document}
 
 \begin{abstract}
We present the results of investigation of five stars, originally classified as dwarfs, belonging to Cyg~OB2 association, their stellar and wind properties. Using both {\sc tlusty} and  {\sc cmfgen} codes we derived effective temperatures, surface gravities, chemical abundances, mass-loss rates and projected rotation velocities. Due to the fact that distance to the stars is well known, we were able to  estimate their luminosities. Using evolutionary models we estimated the ages of these sample stars and find that lower mass ones -- MT282 and MT343 -- belong to older population of the association. Their ages are greater than 10 Myr. The ages of three other stars   -- MT317, MT299, MT259 -- are between 4-6 Myr. 

\end{abstract}
\maketitle%


\section{Introduction}

Cygnus~OB2 (Cyg~OB2), a stellar association discovered by M{\"u}nch and Morgan in 1953 \citep{MunchMorgan}, is now one of the leaders in number of massive stars among Galactic OB-associations. According to the estimates of \citet{Knodlseder} it includes $2600\pm400$ stars and about 100 of them are O-stars located in the one square degree region \citep{Comeron02,Wright,Wright2015}. Due to it the central part of association is the region with the highest concentration of hot stars in the Galaxy. The interest of researchers to individual stars and to the association as a whole is not fading. A lot of articles dedicated to the investigation of Cyg~OB2 stellar population  (like \citet{Drew,Comeron12}), estimation of interstellar extinction (like \citet{Guarcello}), history of star formation (like \citet{Wright13,Wright2015}) have already been published and continue to appear.

Modelling of atmospheres of the stars belonging to an association is an important area of investigation, since it both gives the parameters of individual stars and allows to construct an association' Hertzsprung-Russell (H-R) diagram as a whole. The H-R diagram in turn provides the information about stellar ages, masses and history of star formation. Main works devoted to the numerical modelling of stellar atmospheres in Cyg~OB2 association are \citet{Herrero1999,Herrero02,Negueruela}. Moreover, individual supergiants were studied by \citet{Najarro2011obj7,me2013,me2014}.

In the present work we model the stellar atmospheres of five stars initially assumed to have V luminosity class (dwarfs). 
Star \#6 or MT317\footnote{\#6, \#16 and \#21 are the names of the stars according to the catalogue of \citet{Schulte}, while MT is the catalogue of Cyg~OB2 members by \citet{MT91}} may be called the native resident of association Cyg~OB2. It belongs to the first dozen of blue stars noticed by \citet{MunchMorgan}. Based on spatial closeness of these objects M{\"u}nch and {Morgan} have suggested the existence of OB-association in about $2^\circ$ around $\gamma$~Cygni. \citet{JohnsonMorgan} classified MT317 as O8 (V). The contemporary researchers \citep{KiminkiAv,Chentsov} also consider that MT317 is O8~V. \citet{Morgan16} suggested that star \#16 (MT299) is probably the faint member of the association Cyg~OB2. Then \citet{Schulte} included the star \#21 (MT259) into the association. \citet{Chentsov} classified MT299 as O7.5~V, while  MT259 as B0~V. \citet{MT91} also included MT282 and MT343 stars into the association. The latter was subsequently classified by \citet{KiminkiAv} as B1~V, while the first spectrum of  MT282 was obtained only recently on the Russian 6-m telescope, and MT282  was classified  as B1~IV-V \citep{Chentsov2015}. Among these five stars only the MT259 had previously been studied by \citet{Herrero1999} using stellar atmosphere models.

\begin{table*}
\begin{center}
\caption{Observational log for the spectral data obtained on the Russian 6-m telescope used in the
  work. The Column~1 quotes the names of the objects from the catalogue of \citet{Schulte}, 
  while the Column~2 -- the ones from \citet{MT91}. 
  The penultimate column lists the signal-to-noise ratio (S/N) in the vicinity of He{\scriptsize I} $\lambda$5876 line (for NES data) or near 4900~\AA\  (for SCORPIO).} 
\label{tab:nes}
\begin{tabular}{lccccccccc}
\hline 
 Schulte     &  ~~MT~~  &  Spectral   &$m_V$  & $A_V$   & Date       & Spectro-     &Spectral        & S/N   &     Ref.$^+$        \\ 
             &          &  type$^*$   &[mag]  & [mag]   & dd/mm/year &  graph       & range, [\AA]   &       &             \\ 
\hline 
             &          &             &       &         &            &              &                &       &              \\
   6         & 317      &   O8~V      & 10.68 &  4.6    & 03.06.2010 & NES          & 5300-6690      &  60   &     1          \\ 
             &          &             &       &         & 04.06.2010 & NES          & 4465-5930      &  40   &               \\ 
   16        & 299      &   O7.5~V    & 10.84 &  4.4    & 10.06.2011 & NES          & 4850-6240      &  60   &     1          \\ 
   21        & 259      &   B0~V      & 11.42 &  3.7    & 18.11.2010 & NES          & 5215-6690      &  100  &     1         \\ 
             &          &             &       &         &            &              &                &       &               \\
             &  282     &   B1~IV-V   &15.03  &8.12     & 31.05.2013 & SCORPIO      & 4050-5850      &  50   &     2            \\
             &  343     &   B1~V      &14.5   &6.76     & 03.08.2014 & SCORPIO      & 4050-5850      &  70   &     2            \\ 
\hline
\multicolumn{10}{l}{$^*$ Spestral types  are taken from  literature (see text). } \\
\multicolumn{10}{l}{$^+$ Photometric data $m_V$ and $A_V$ are taken from 1 -- \citet{KiminkiAv}, 2 -- \citet{Chentsov2015}.} \\
\end{tabular}
\end{center}
\end{table*}
 
        The paper is organized as follows. Observational data and main steps of data reduction are described in the next section. Data analysis and determination of different stellar parameters are in Section~\ref{sec:analysis}. The Section~\ref{sec:results} presents the results of modelling, while in Section~\ref{sec:discuss}  we will discuss the locations of the stars on the H-R diagram and the estimated masses, ages and mass loss rates.  In the Section~\ref{sec:conclusions} short conclusions are given. The testing of the method of automatic determination of effective temperature and surface gravity is presented in Appendices.

\section{Observational Data}\label{sec:obs}

         The high resolution spectra of 13 stars in Cyg~OB2 had been thoroughly described by \citet{Chentsov}, who also studied spectral variability of the MT304 (Schulte~\#12) hypergiant. In our work we analyzed all three stars MT317, MT299 and MT25 classified by \citet{Chentsov} as dwarfs and we used the same spectral data. 
         These data were obtained with the high-resolution Nasmyth Echelle Spectrograph (NES) \citep{NES1} on the Russian 6-m telescope. The observations were conducted using the image slicer \\ \citep{NES1}  and a $2048\times2048$ CCD. The extraction of one-dimensional vectors from the two-dimensional echelle spectra was performed with the aid of the  {\sc echelle} context of the {\sc midas} software package, modified by \citet{Yushkin}. The removal of the traces of cosmic ray particles was done by means of median filtration of image sequences using standard {\sc midas} routines. Wavelength calibration was carried out using the spectra of a Th-Ar hollow cathode lamp. Continuum normalization of spectra was performed manually using {\sc dech} software package  \citep{DECH}.          
          
         Spectra of MT282 and MT343 were also obtained on the Russian 6-m telescope, but with Spectral Camera with Optical  Reducer for Photometric and Interferometric Observations (SCORPIO) in the long-slit mode \citep{scorpio}.  VPHG1200G grism was used providing the spectral range of 4050-5850~\AA\AA. All SCORPIO spectra were reduced using the {\tt ScoRe} package\footnote{http://www.sao.ru/hq/ssl/maryeva/score.html}. {\tt ScoRe} was written by Maryeva and Abolmasov in IDL language for SCORPIO long-slit data reduction. Package includes all the standard stages of long-slit data reduction process. The final spectra of MT282 and MT343 have spectral resolution  $\sim 6$~\AA\ (weakly dependent on wavelength) and  signal-to-noise ratio per resolution element in continuum near 4900~\AA\ $\sim 50$ and $\sim 70$, correspondingly. 
         The data used in this work is summarized in Table~\ref{tab:nes}.
               
\section{Methods of data analysis}\label{sec:analysis}

       For estimation of physical parameters of the sample stars we used two atmospheric modelling codes -- {\sc tlusty} \citep{HubenyLanz1995,LanzHubeny2003} and {\sc cmfgen} \citep{Hillier5}. 

       Effective temperature $T_{\rm{eff}}$, surface gravity log$\,\textsl{g}$ and projected rotation velocities ${\it v\sin i}$  were estimated using stellar atmosphere models computed with {\sc tlusty}. {\sc tlusty} is the code for computing plane-parallel, horizontally homogeneous model stellar atmospheres in radiative and hydrostatic equilibrium, developed by Hubeny and Lanz \citep{HubenyLanz1995,LanzHubeny2003}. Departures from local thermodynamic equilibrium (LTE) are allowed for a set of occupation numbers of selected atomic and ionic energy levels. Latest version of the program accounts for a fully consistent, non-LTE metal line blanketing, as well as convection (\url{http://nova.astro.umd.edu/index.html}). For the analysis we used the grids of {\sc tlusty} stellar atmosphere models and synthetic spectra computed by us and two precomputed grids. These precomputed grids are {\sc ostar2002} grid with solar metallicity and microturbulent velocity $v_{\rm{turb}}=10~\kms $ \citep{LanzHubeny2003} and {\sc bstar2006} grid with solar metallicity and $v_{\rm{turb}}=2~\kms $ \citep{LanzHubeny2007}.
 
       For determination of wind parameters, bolometric luminosity and chemical composition we used {\sc cmfgen} code \citep{Hillier5}. This code is modelling spherically symmetric extended outflows using 
       the full comoving-frame solution of the radiative transfer equation. {\sc cmfgen} incorporates line blanketing, the effects of Auger ionization and  clumping. The level populations are calculated through the rate equations. A super-level approach is used to reduce the scale of the problem (and thus the computing time). 
       
\subsection{{Determination of projected rotational velocities}}\label{sec:vsini}

        For estimation of  rotational velocities ${\it v\sin i}$ we used synthetic spectra from the {\sc tlusty} grids mentioned above.
        From these grids we chose the spectrum computed for the effective temperature $T_{\rm{eff}}$ and surface gravity log$\,\textsl{g}$ corresponding to  the spectral type and luminosity class of the considered star. Values of $T_{\rm{eff}}$ and log$\,\textsl{g}$ for a given spectral type were obtained from calibration of \citet{MartinsBC} for O-stars and from \citet{Humphreys} for B-stars. The chosen spectrum was convolved with the instrumental profile and rotational profile computed with different values of ${\it v\sin i}$. The convolution of spectra was performed using the {\sc rotin}3 program (distributed together with {\sc tlusty}). 
        
        The projected rotational velocity and its uncertainty were then estimated  by visual comparison   of shapes  of the observed spectral line profiles of metals and helium with the ones in the convolved  synthetic spectra. These estimates are independent on the microturbulent velocity $v_{\rm{turb}}$ that only affects the equivalent width of line and not line profile shape. 
       
%

%

\subsection{{Determination of effective temperature and surface gravity}}\label{sec:tlusty}

To estimate $T_{\mathrm{eff}}$ and log$\,\textsl{g}$ we used the automatic method. 
The method is based on the comparison of the observed and synthetic spectra. Synthetic spectra for different values of $T_{\rm{eff}}$, log$\,\textsl{g}$, $v_{\rm{turb}}$ and helium abundance He/H had been computed for the models from the {\sc tlusty} grids using {\sc synspec48} code \citep{HubenyLanz1992} and then convolved with rotational profile for the given value of ${\it v\sin i}$  and with the instrumental profile. To characterize how well the observed spectral lines or regions are approximated by model spectra the $\chi^2$ quantity was evaluated as:

\begin{center} 
      \begin{equation} 			
              \chi^2=\frac{1}{N_\mathrm{lines}}\,\sum_{i=1}^{N_\mathrm{lines}} 
              \frac{1}{N_{\nu}^i}\,\sum^{N_{\nu}^i}_{j=1}\left(\frac 
              {F_j^i-F^i_{j\,\mathrm{obs}}}{\sigma^i}\right) ^{2} 
      \label{Eq:criterium} 
		\end{equation} 
\end{center}

    \noindent where $N_{\nu}^i$ is the number of wavelength points in the spectral line $i$, $N_\mathrm{lines}$ is the number of spectral lines used in the analysis, $F_{j\,\mathrm{obs}}^i$ is the observed normalized flux in the $j$-th point of the spectral line $i$, $F_j^i$ is the synthetic normalized flux,  
    and $\sigma^i=\mathrm{(S/N)}^{-1}$ accounts for the signal-to-noise ratio (S/N) for the spectral line $i$. The S/N had been calculated as the mean flux divided by the standard deviation of the flux in the manually selected continuum region close to a given line. 

    Initial estimates of $T_{\rm{eff}}$, log$\,\textsl{g}$,  He/H and $v_{\rm{turb}}$ were obtained as parameters of the model with the minimal $\chi^2$ value. The synthetic spectrum computed with these initial parameters was used to manually refine the continuum  (as the initial continuum estimation is being done manually and is prone to errors due to its ill-defined nature)  and to check the previously estimated ${\it v\sin i}$ value. To refine the continuum we also used the $T_{\rm{eff}}$-log$\,\textsl{g}$ diagnostic diagram (see e.g \citet{Herrero92}) plotted automatically on the basis of $\chi^2$ values of the models with fixed He/H and $v_{\rm{turb}}$ equal to their initial estimates.

   The {\sc ostar2002} and {\sc bstar2006} grids has a relatively rough step on $T_{\rm{eff}}$ and log$\,\textsl{g}$ (2500 K on $T_{\rm{eff}}$ and 0.25 dex on log$\ \textsl{g}$) that could lead to relatively large uncertainty of estimated $T_{\rm{eff}}$ and log$\,\textsl{g}$ values. To refine $T_{\rm{eff}}$ and log$\,\textsl{g}$ estimates obtained at the previous step we computed the grid of stellar atmosphere models with varied values of $T_{\rm{eff}}$ and log$\,\textsl{g}$ and with fixed He/H and $v_{\rm{turb}}$ values. These fixed values were close to those obtained on the previous step. The new grid had steps of 1000~K and 0.1 dex on $T_{\rm{eff}}$ and log$\,\textsl{g}$, respectively. Computations of the new grid were performed with the {\sc tlusty200} code that was slightly modified to allow using of \citet{Kurucz94} line data for nickel (Ni). Atomic models were the same as for stellar atmosphere models from the {\sc ostar2002} grid. To save computer time all stellar atmosphere models for the new grid were computed with the smaller number of lines of iron-peak elements, with smaller wavelength range and smaller maximum number of global iterations compared with those used by \citet{LanzHubeny2003}.

    The new atmosphere models grid was used to compute a number of synthetic spectra for different values of $T_{\rm{eff}}$, log$\,\textsl{g}$, He/H and $v_{\rm{turb}}$. These synthetic spectra were compared with observed ones using the $\chi^2$ quantity. 
    To obtain resulting estimates of the varied stellar parameters we have selected all models with $\chi^2$ smaller than the threshold value $\chi^2_{\rm t}$:
    
    \begin{equation}
          \chi^2_{\mathrm{t}}=\chi^2_{\mathrm{min}} + 0.3\chi^2_{\mathrm{min}},
           \label{eq:chit}
    \end{equation}

    \noindent where $\chi^2_{\mathrm{min}}$ is the minimal $\chi^2$ value. A very similar criterion for the selection of models, with a bit different threshold value, was used by \citet{Castro12} for the analysis of their low resolution spectra. Our choice of $\chi^2_{\rm t}$ threshold is discussed in the Appendix~\ref{sec:chi}. Resulting values of $T_{\rm{eff}}$, log$\,\textsl{g}$, He/H and $v_{\rm{turb}}$ were obtained by averaging parameters of selected models weighted by $\exp\left(-0.5\chi^2\right)$. Application of such weights implies normal distribution of $\chi^2$ that is the good assumption when the number of degrees of freedom is large. Errors of stellar parameters were obtained as weighted standard deviations of parameters of selected models. The spectrum computed with parameters closest to resulting ones was then used again to check the continuum and to verify ${\it v\sin i}$ value. 
    If needed, we repeated the estimation of parameters with the stellar atmospheres grid computed earlier. 
    
    The method of automatic determination of $T_{\rm{eff}}$ and log$\,\textsl{g}$ was common for both high and low resolution observed spectra. In case of low resolution  model spectra were convolved with instrumental profile only, neglecting ${\it v\sin i}$, since instrumental broadening dominates over rotational broadening. Moreover stellar atmosphere models with smaller steps on $T_{\mathrm{eff}}$ and log$\,\textsl{g}$ were not computed. The testing of the method for both high and low resolution spectra is presented in the Appendix~\ref{sec:test}.

\begin{figure*}
\begin{center}
\includegraphics[scale=0.23,viewport=25 0 554 560,clip]{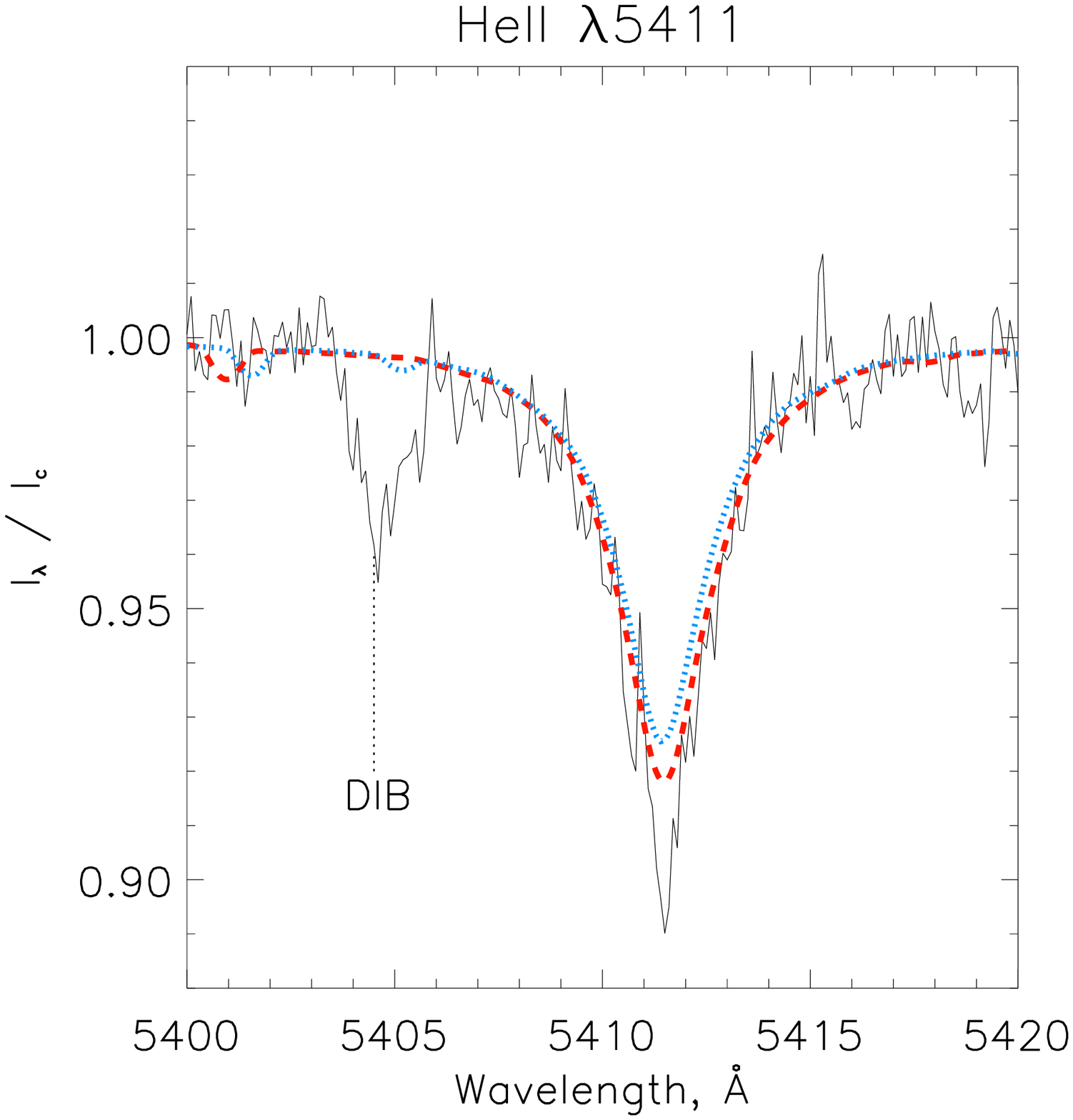}
\includegraphics[scale=0.23,viewport=25 0 554 560,clip]{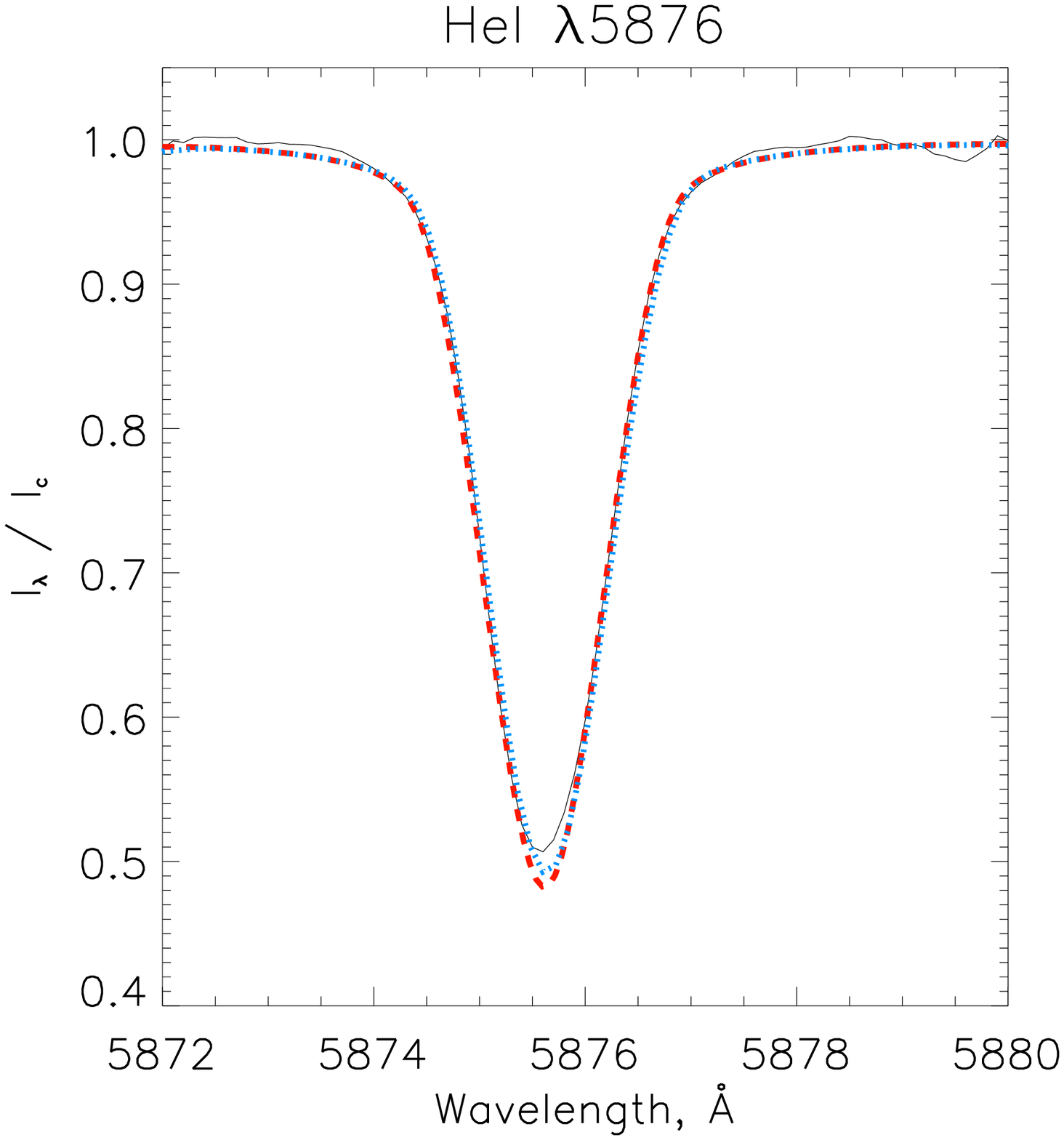}
\includegraphics[scale=0.23,viewport=25 0 554 560,clip]{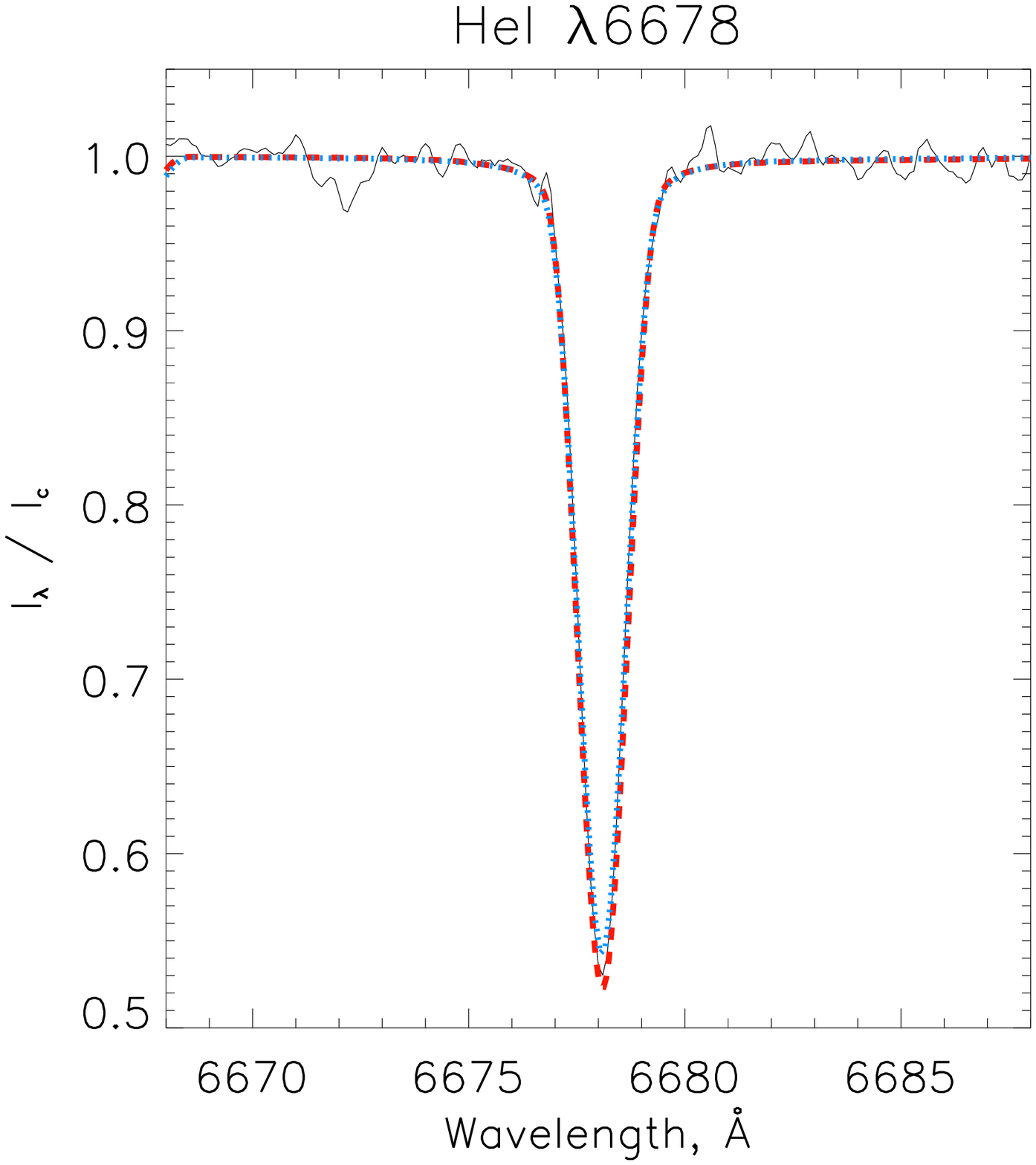}
\includegraphics[scale=0.23,viewport=25 0 554 560,clip]{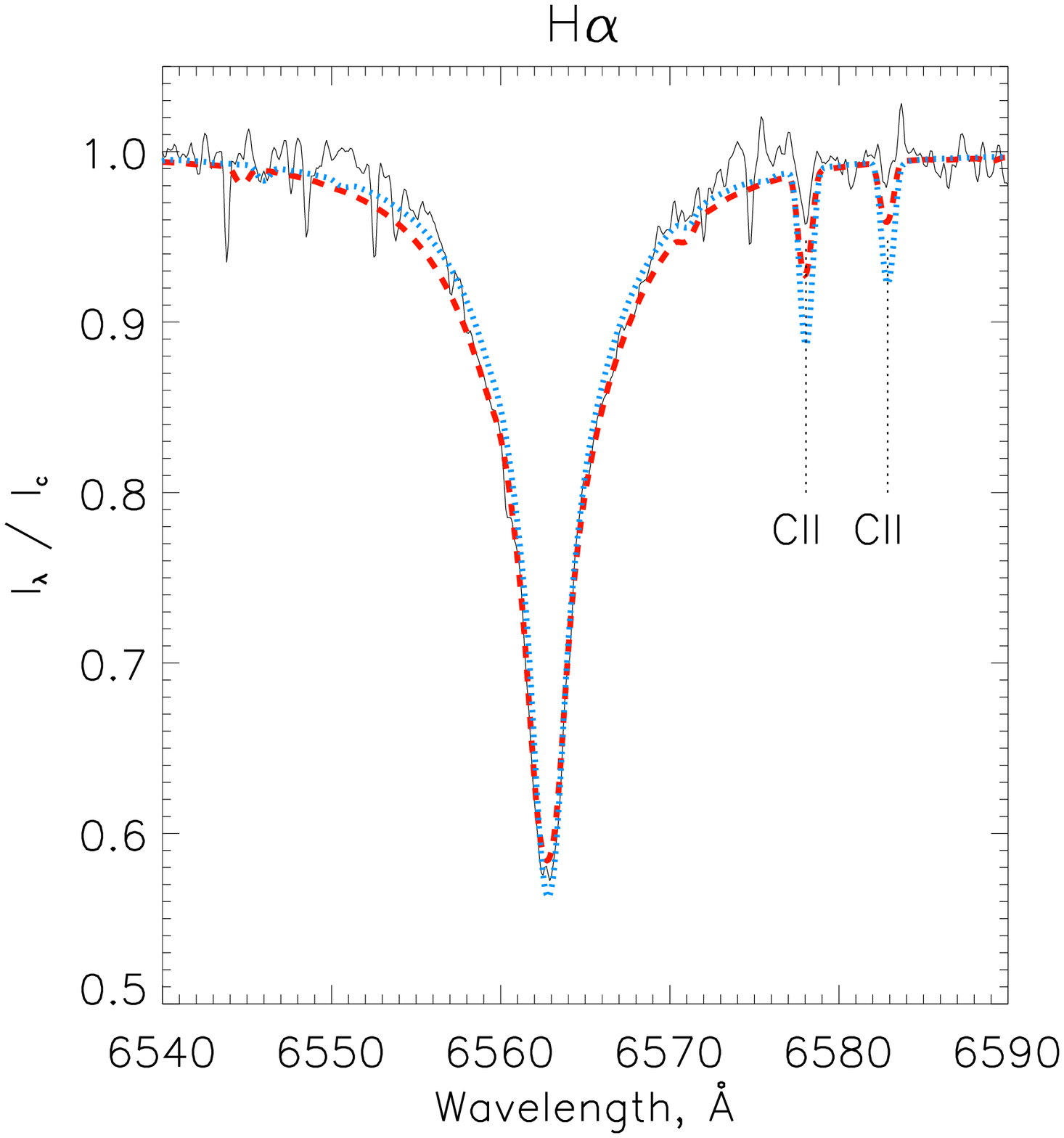}
\includegraphics[scale=0.23,viewport=25 0 554 560,clip]{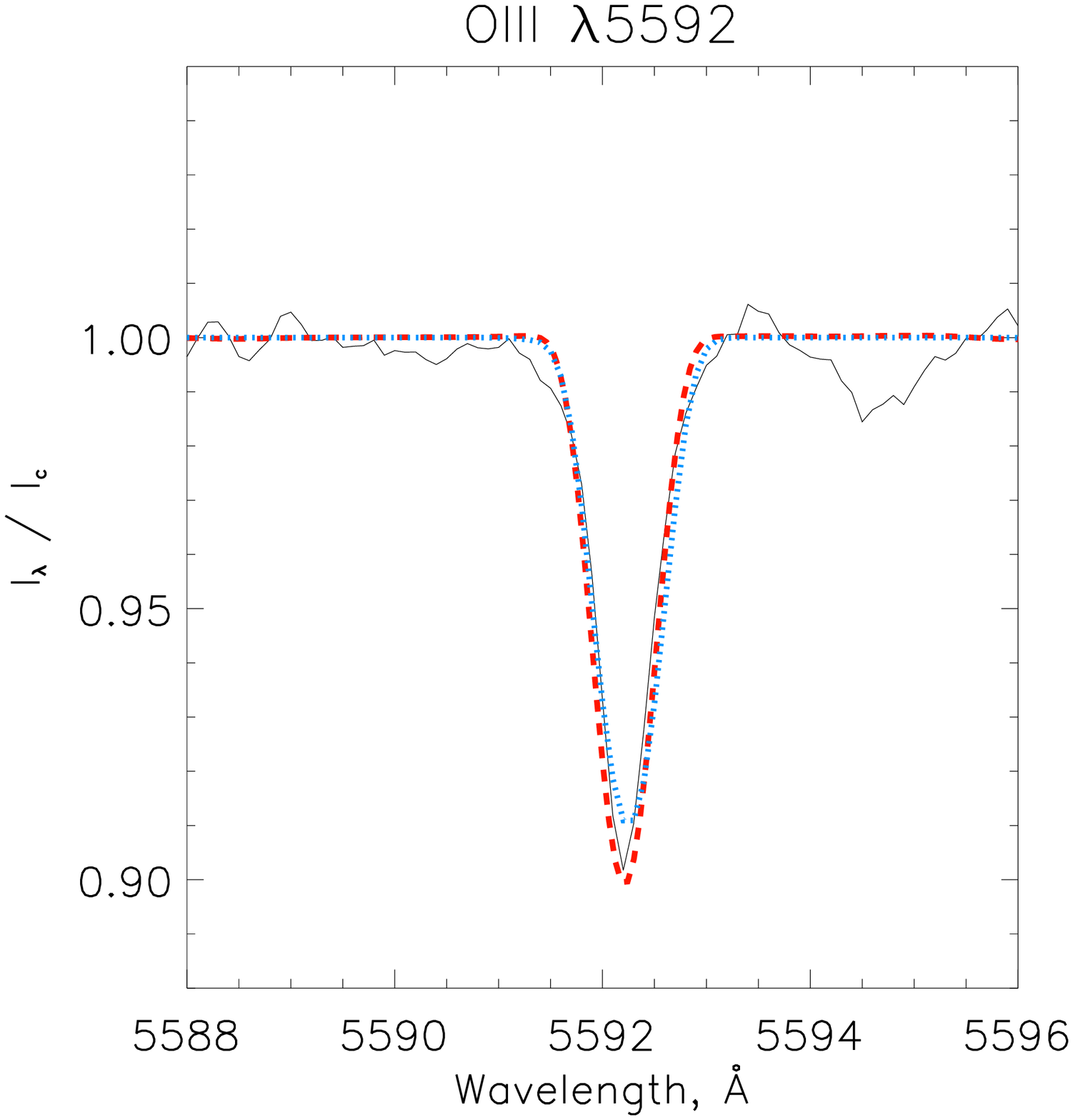}
\includegraphics[scale=0.23,viewport=25 0 554 560,clip]{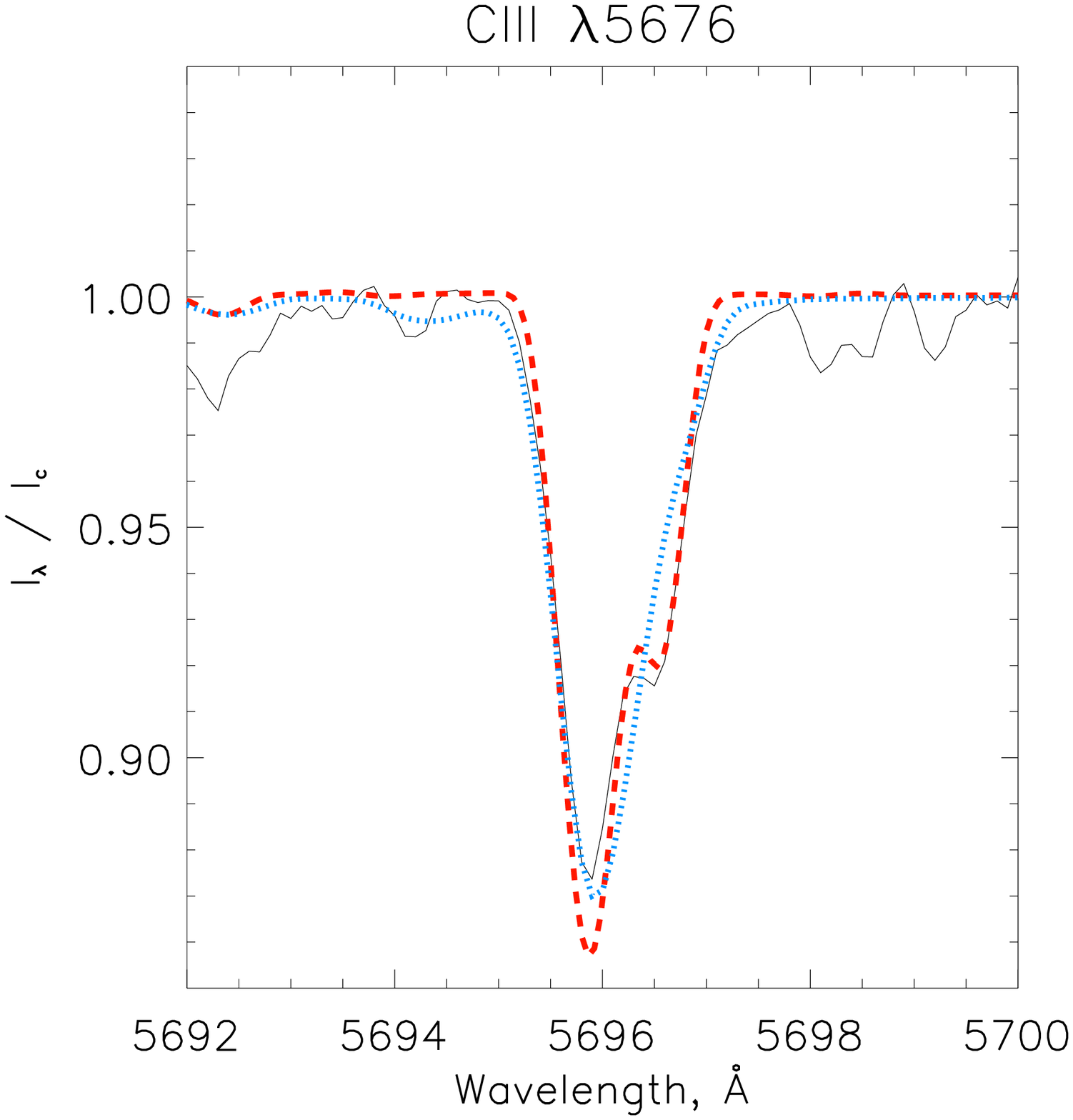}
\includegraphics[scale=0.23,viewport=25 0 554 560,clip]{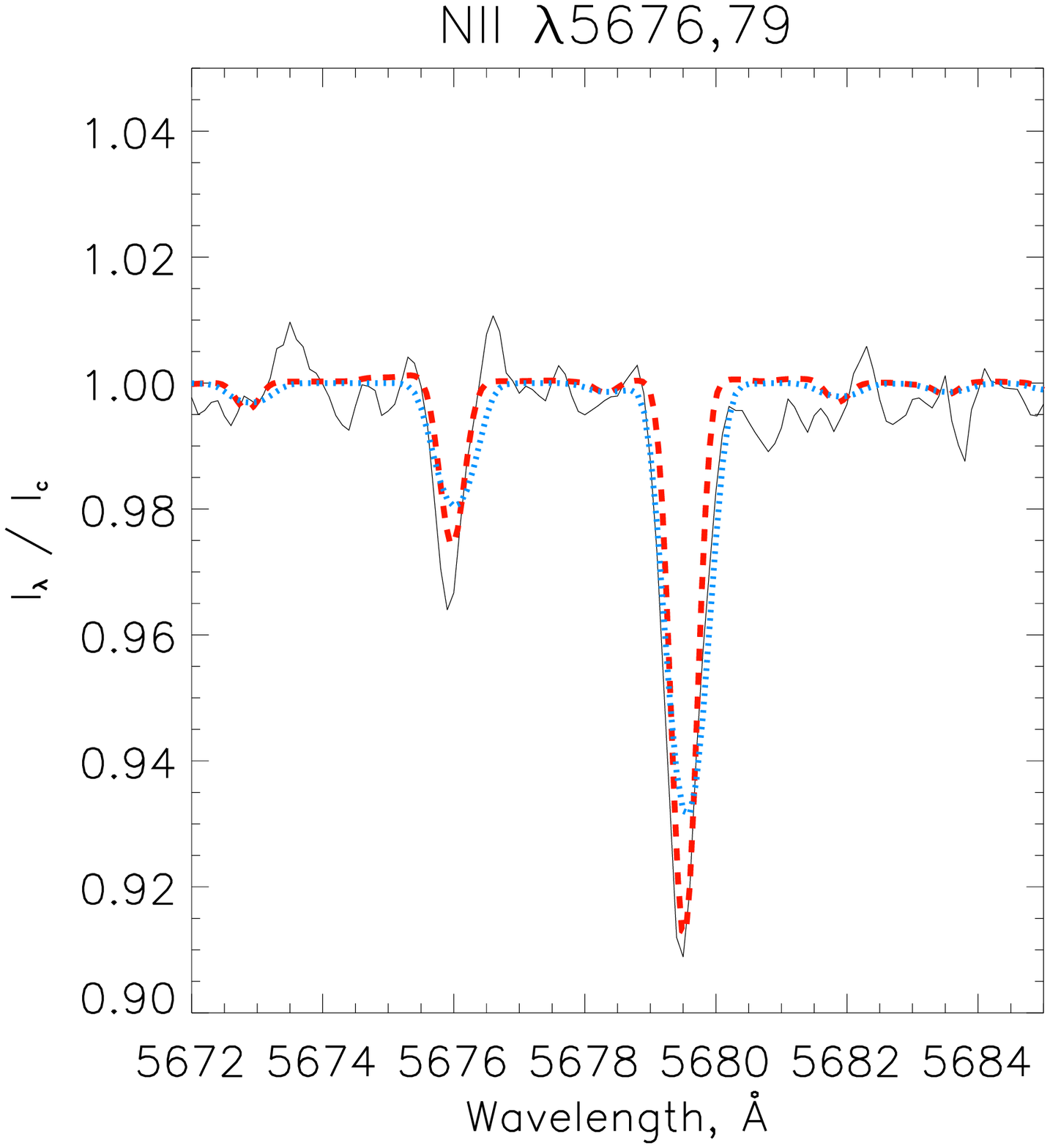}
\includegraphics[scale=0.23,viewport=25 0 554 560,clip]{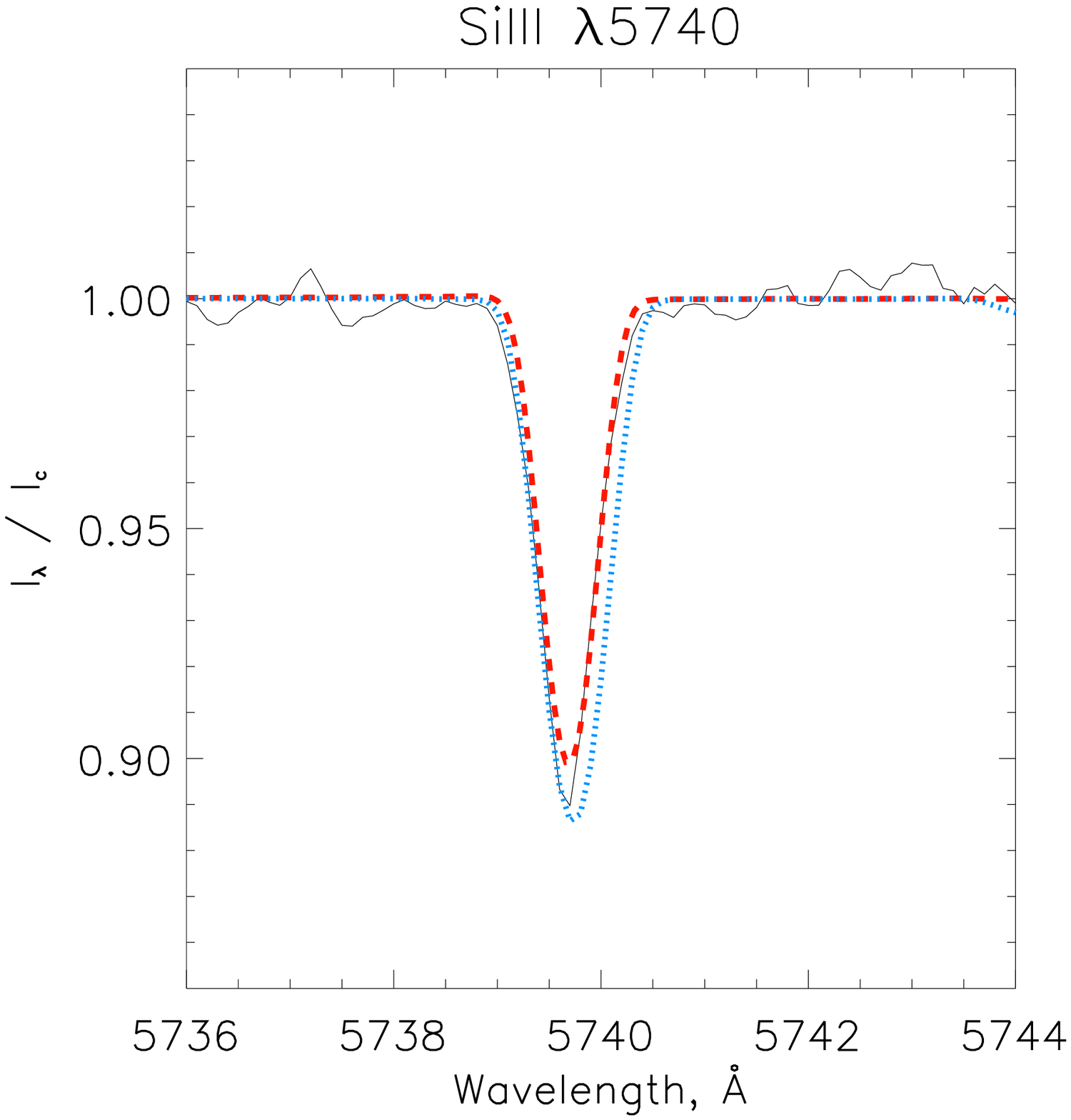}
\caption{Modeling of MT259 (B0 V). The solid line shows the observed profile,  the blue dotted line -- the TLUSTY-model and the red dashed line -- the CMFGEN-model.}
\label{fig:spectrum21}
\end{center}
\end{figure*}
%
\begin{figure*}
\begin{center}
\includegraphics[scale=0.23,viewport=25 0 554 560,clip]{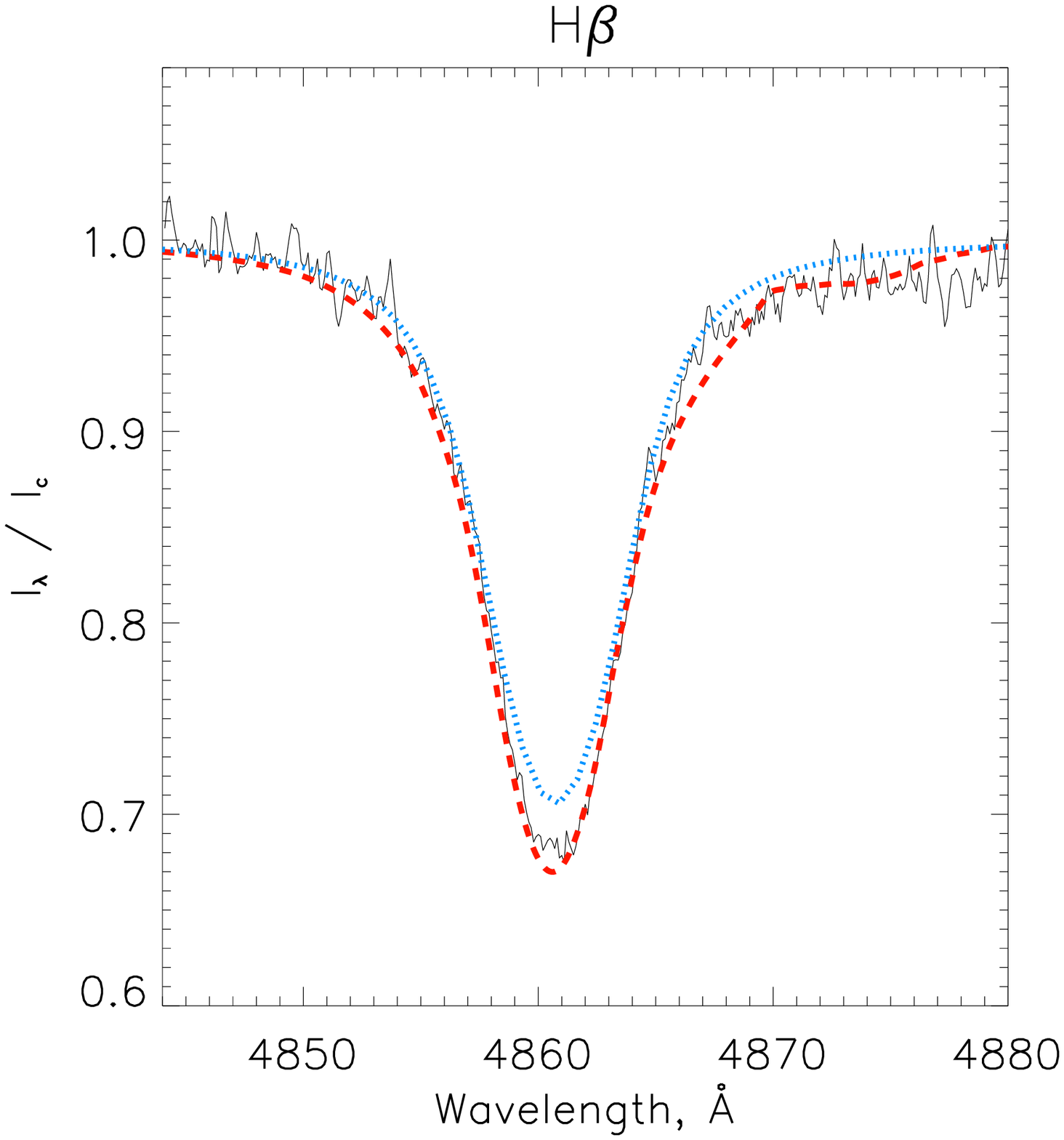}
\includegraphics[scale=0.23,viewport=25 0 554 560,clip]{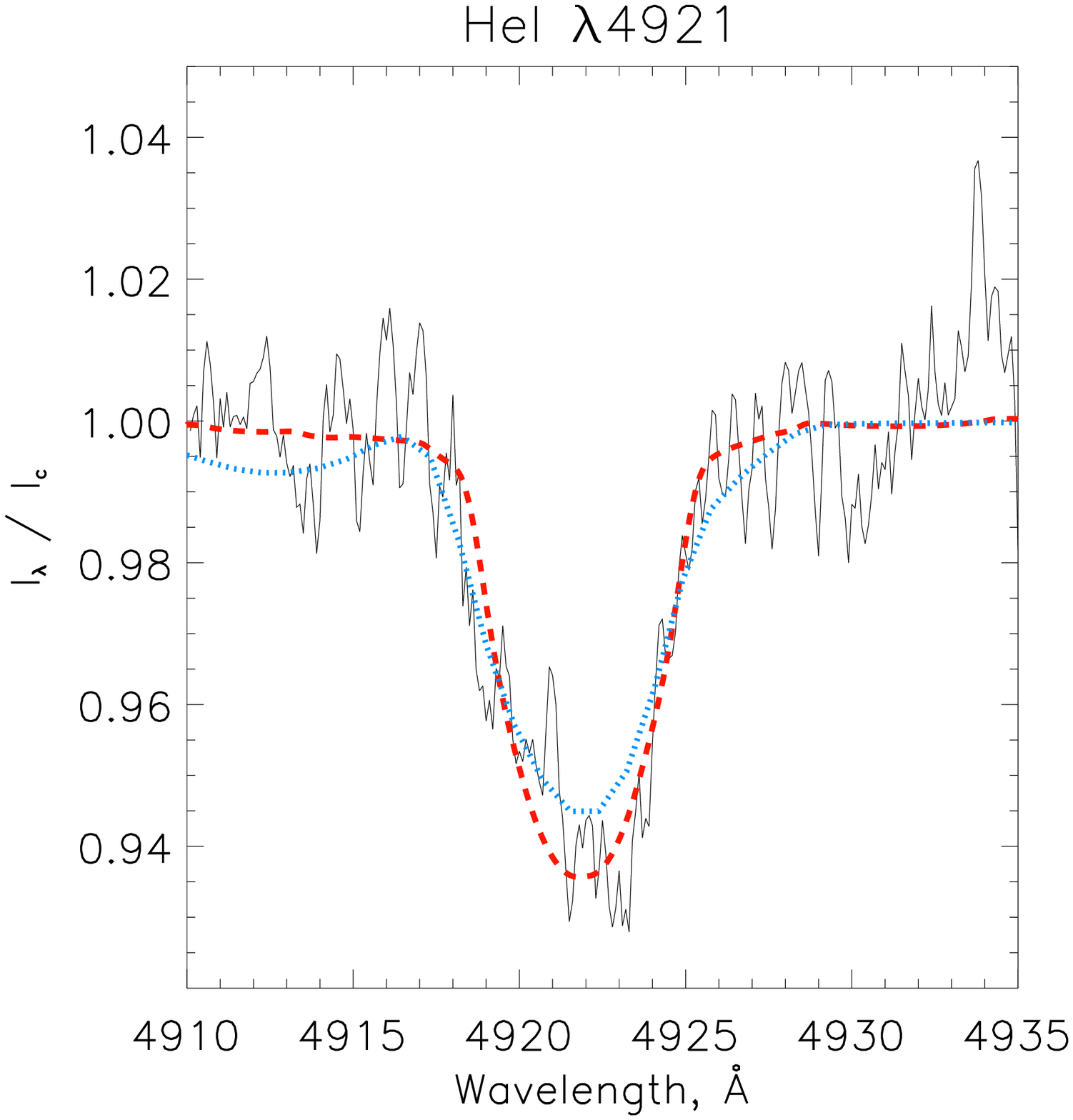}
\includegraphics[scale=0.23,viewport=25 0 554 560,clip]{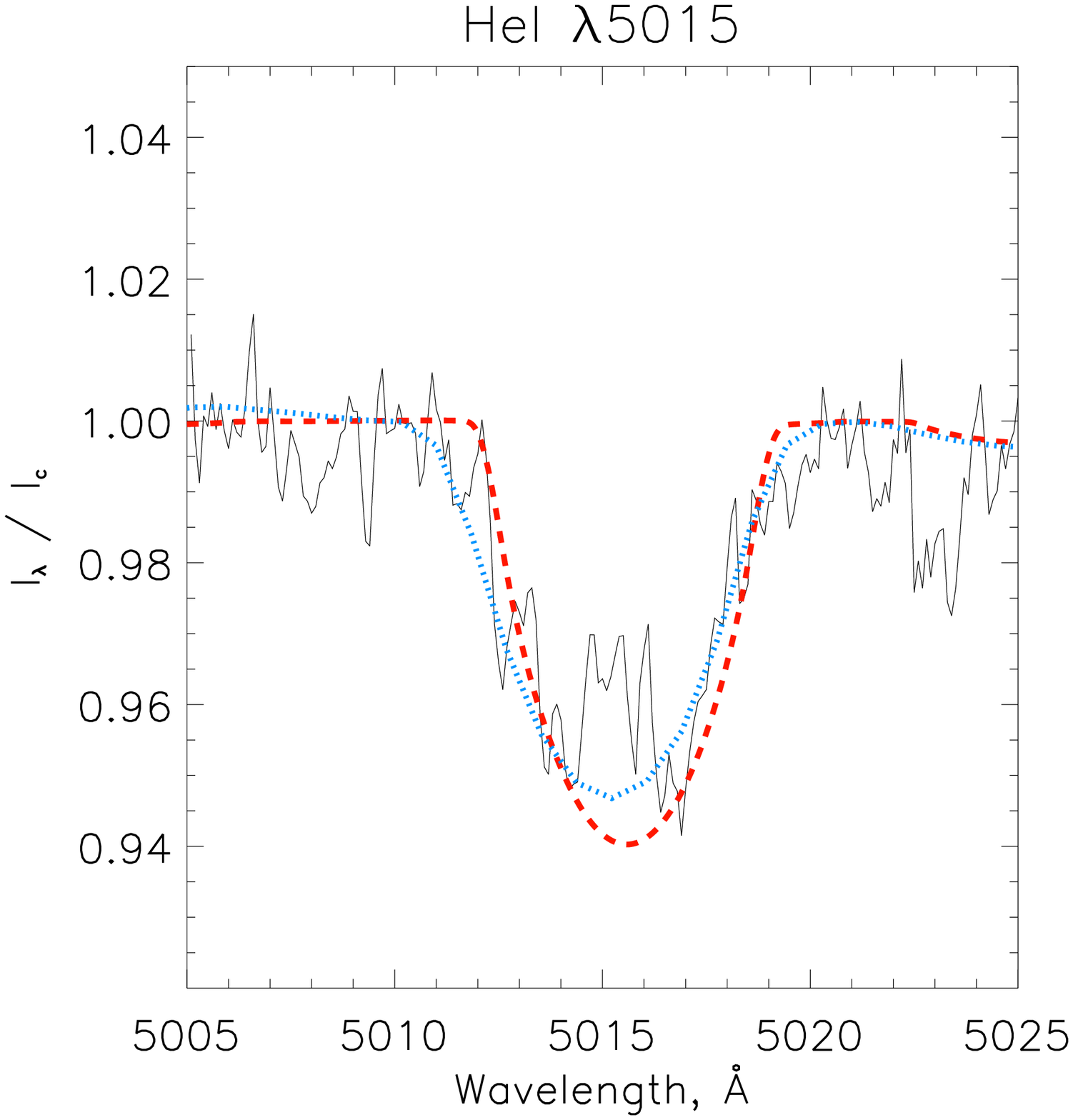}
\includegraphics[scale=0.23,viewport=25 0 554 560,clip]{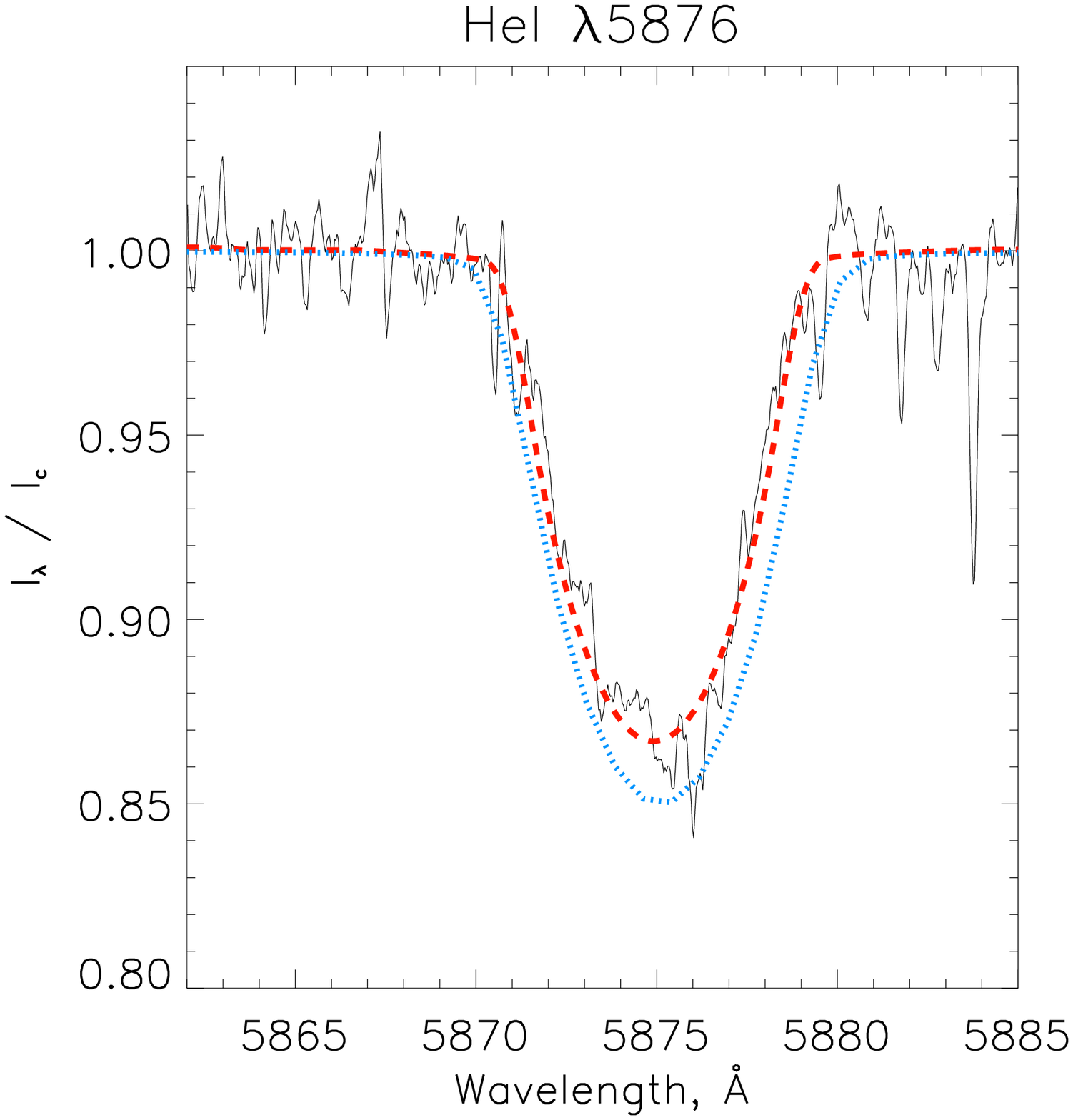}
\caption{Modeling of MT299 (O7.5 V). The solid line shows the observed profile,  the blue dotted line -- the TLUSTY-model and the red dashed  line -- the CMFGEN-model. }       
\label{fig:spectrum16}
\end{center}
\end{figure*}
%
\begin{figure*}
\begin{center}
\includegraphics[scale=0.23,viewport=25 0 554 560,clip]{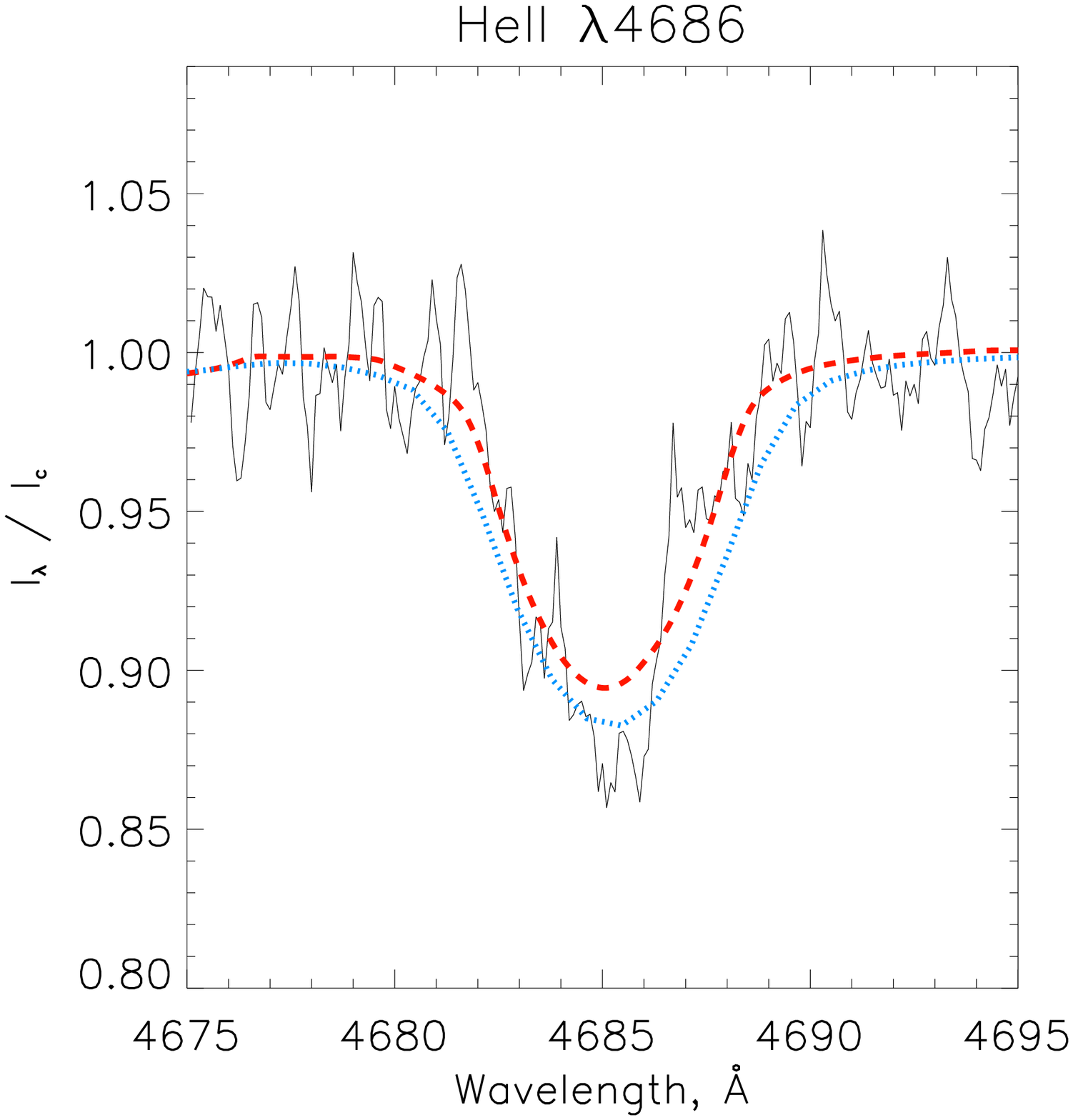}  
\includegraphics[scale=0.23,viewport=25 0 554 560,clip]{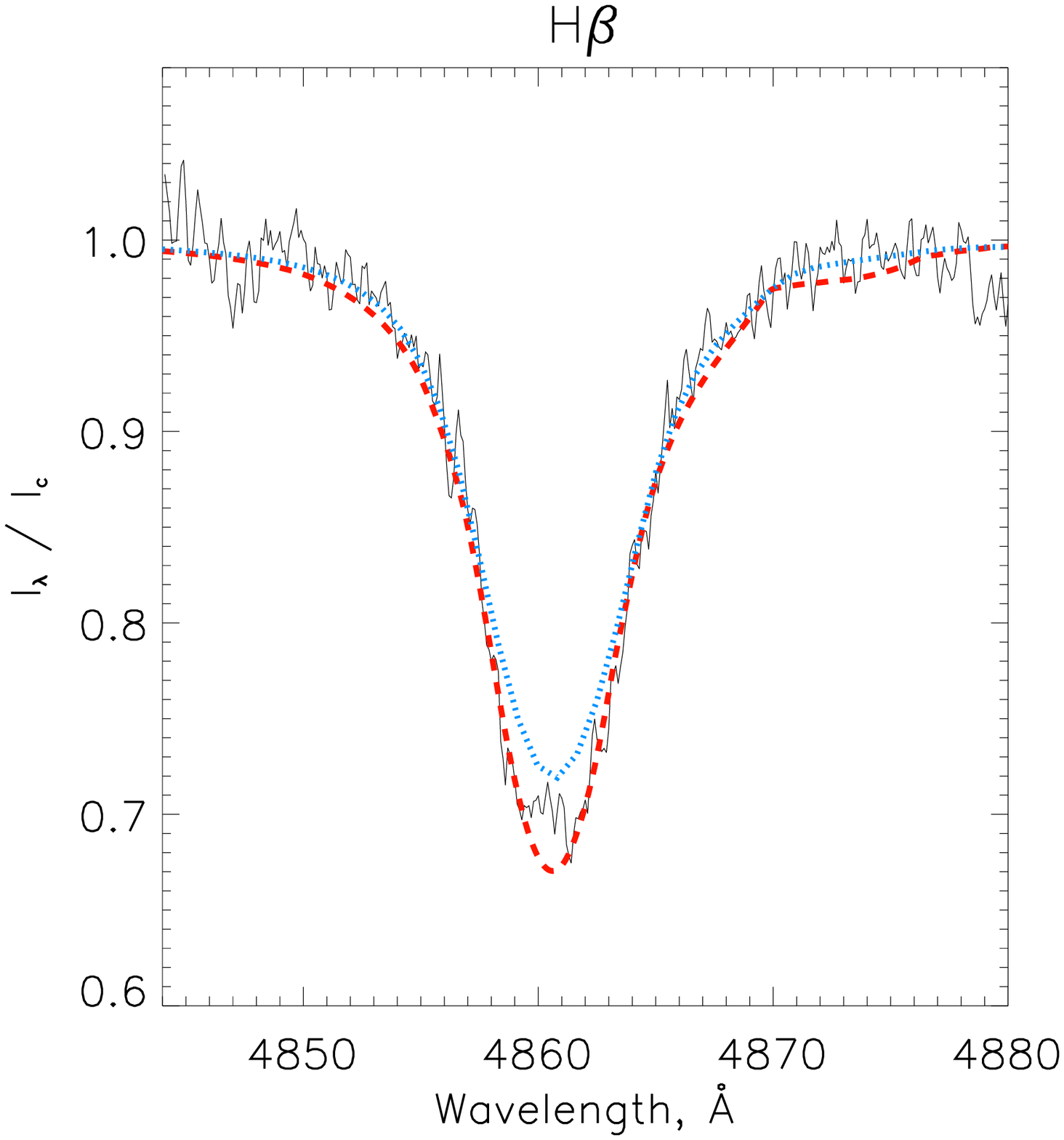}
\includegraphics[scale=0.23,viewport=25 0 554 560,clip]{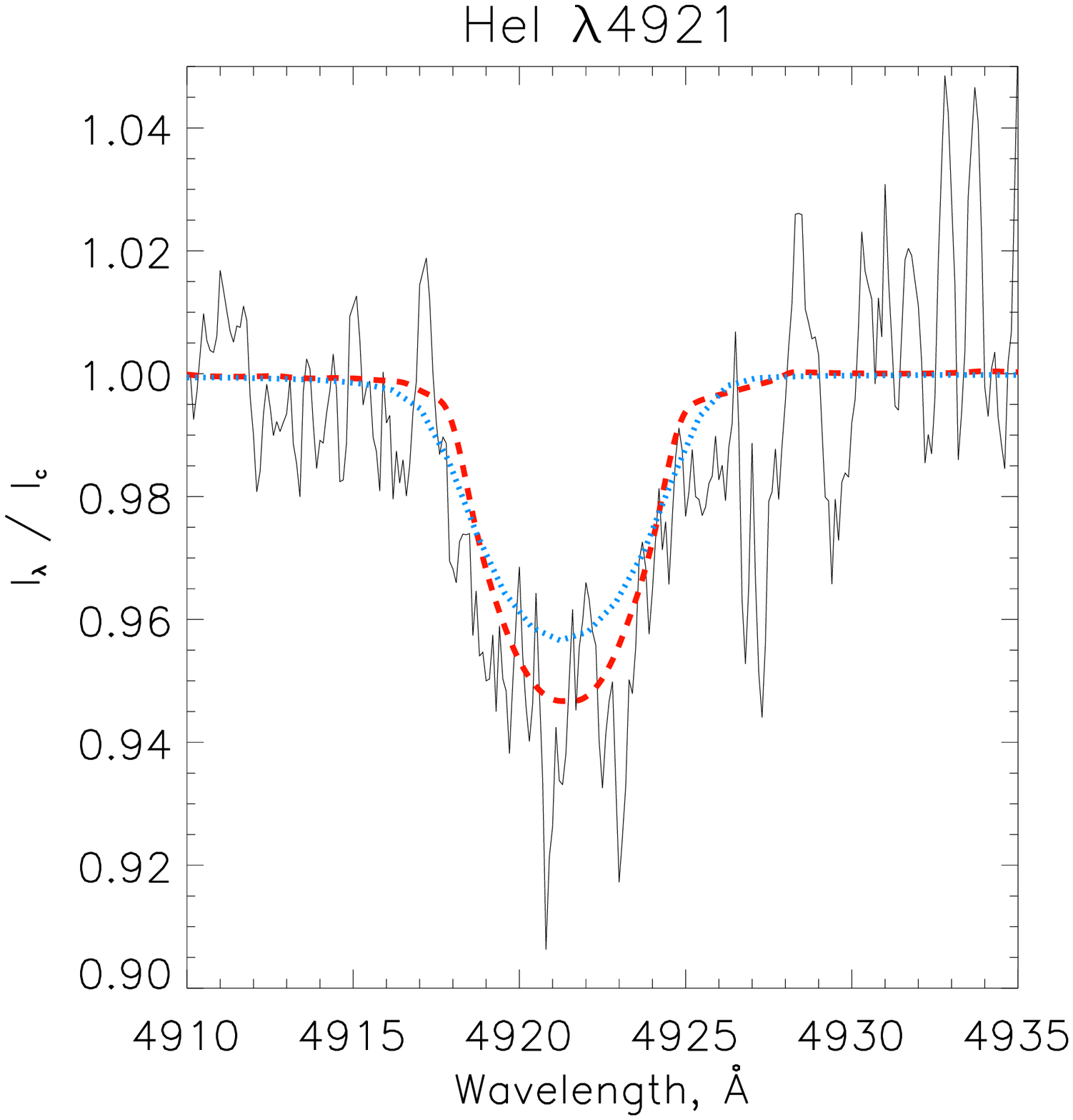}
\includegraphics[scale=0.23,viewport=25 0 554 560,clip]{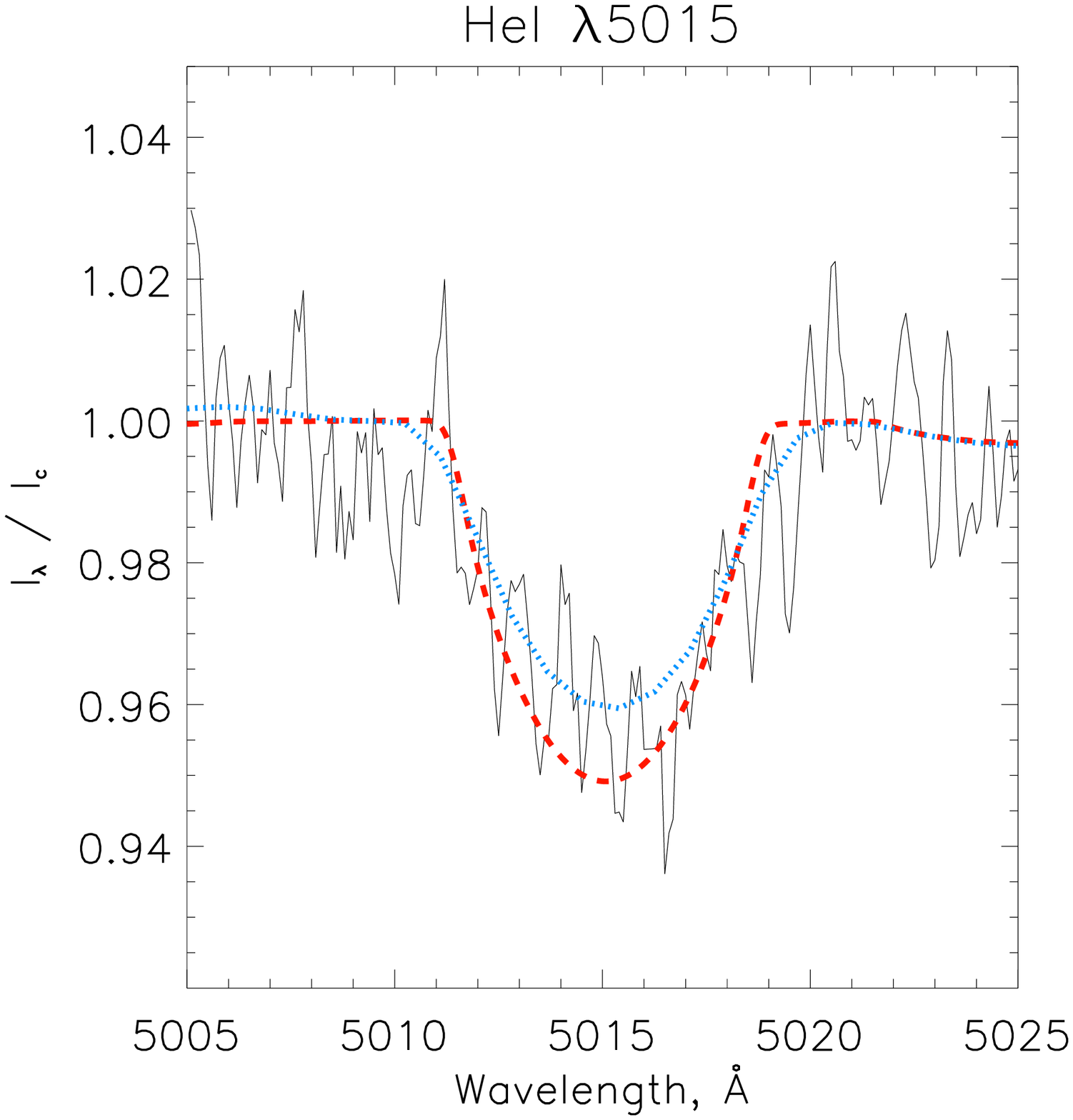}
\includegraphics[scale=0.23,viewport=25 0 554 560,clip]{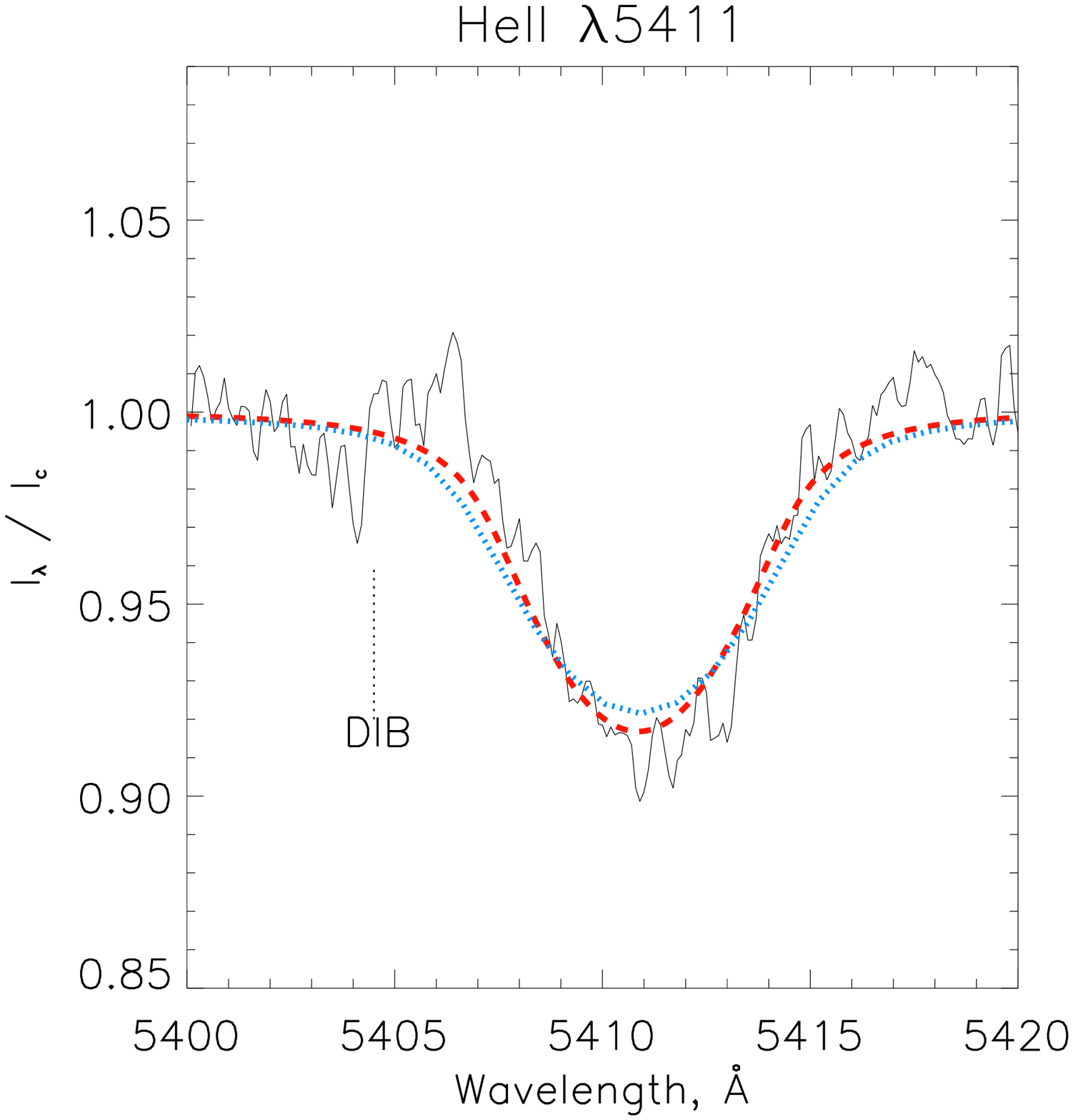}
\includegraphics[scale=0.23,viewport=25 0 554 560,clip]{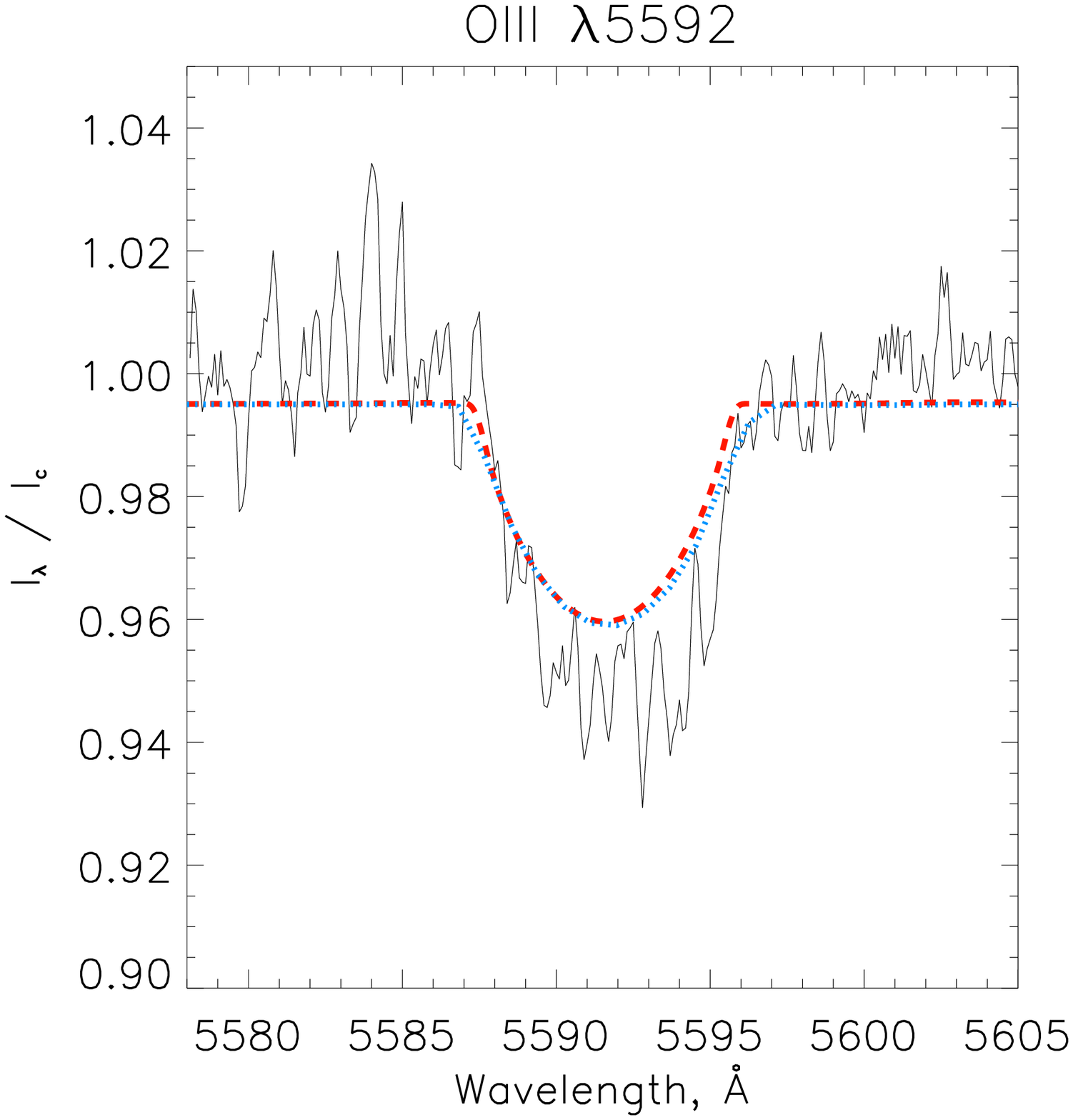}
\includegraphics[scale=0.23,viewport=25 0 554 560,clip]{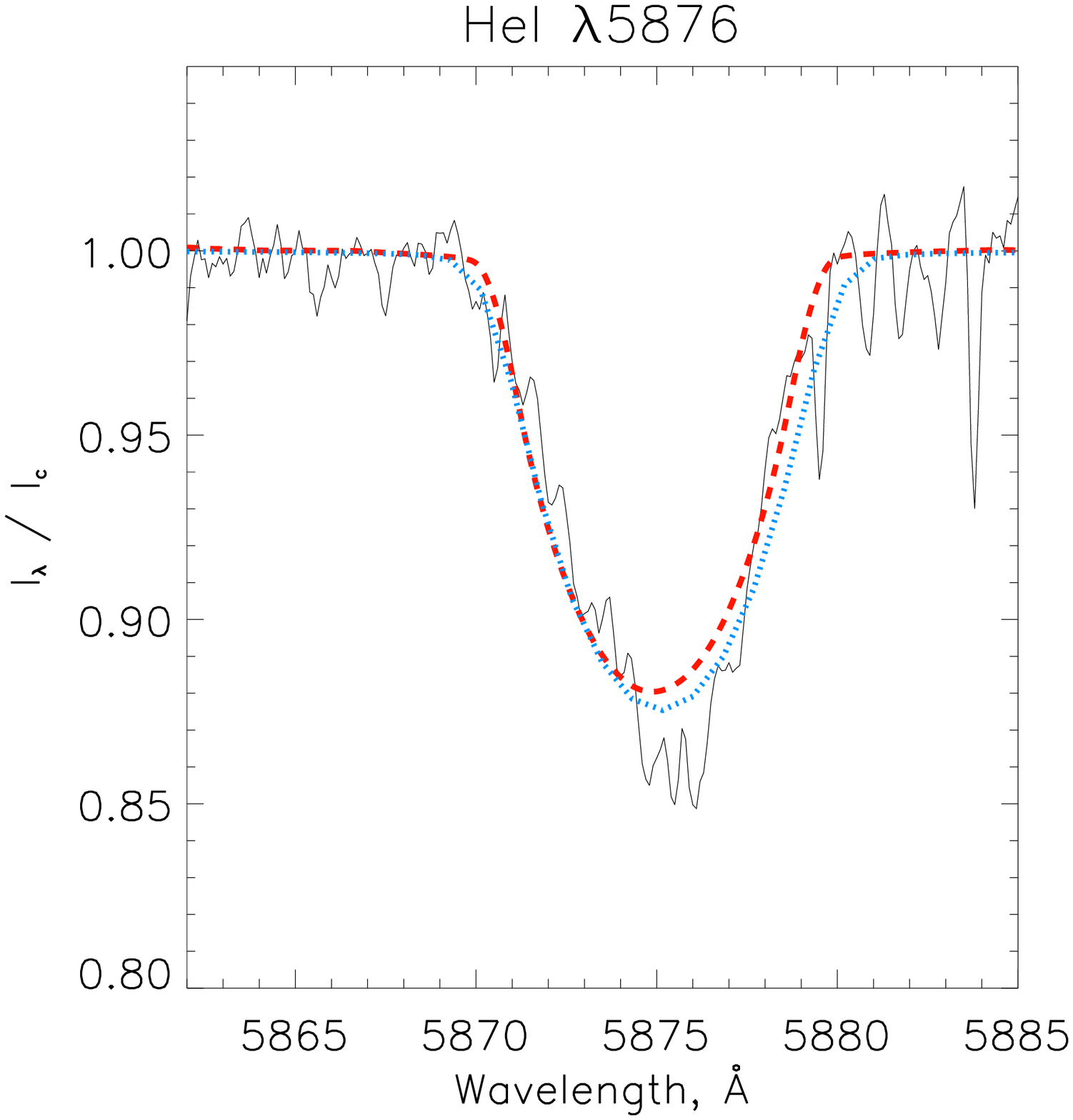}
\includegraphics[scale=0.23,viewport=25 0 554 560,clip]{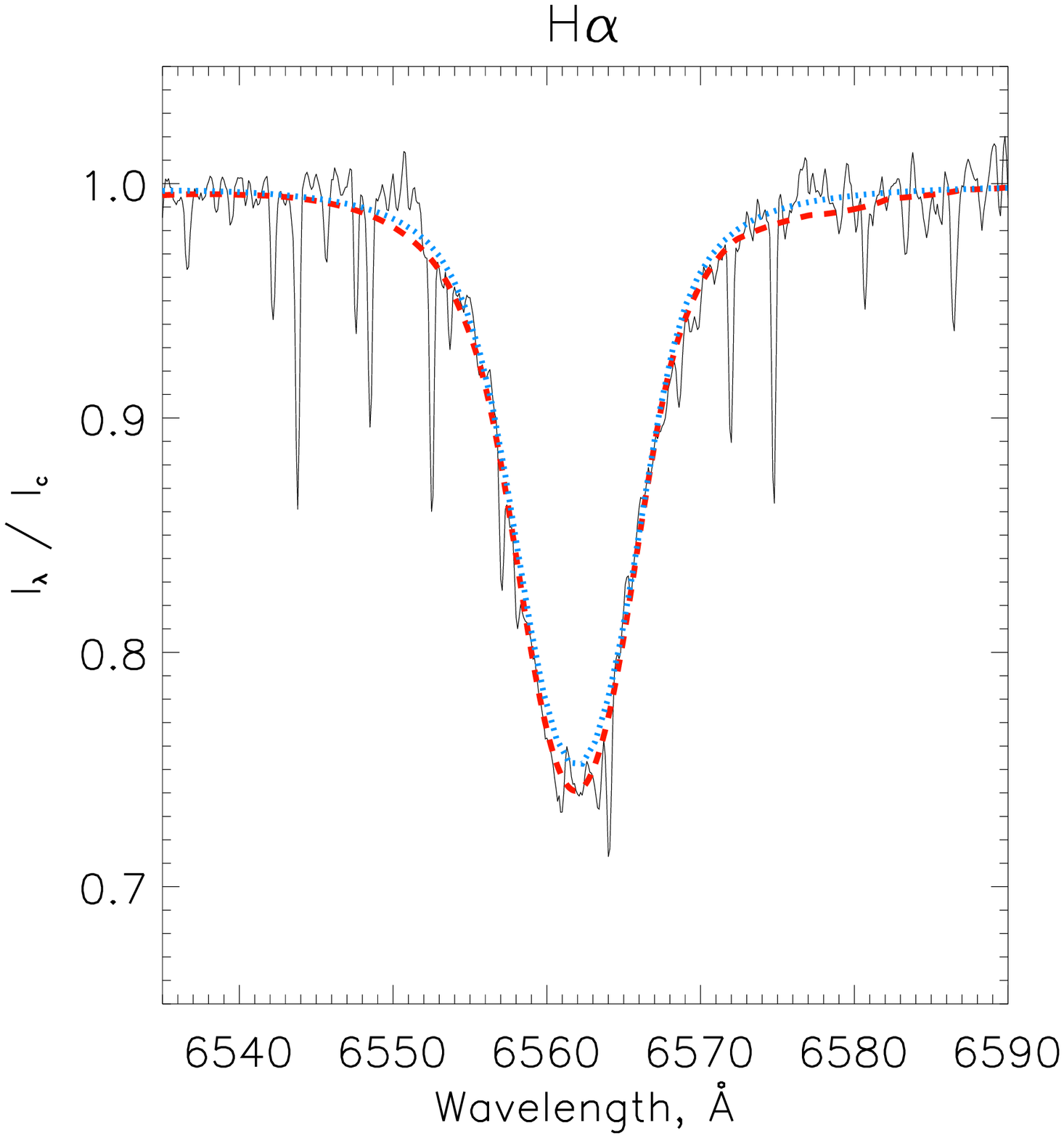}
\caption{Modeling of MT317 (O8 IV). Comparison of the profiles of selected lines with the best model spectra. 
         The solid line shows the observed profile, the blue dotted line -- the TLUSTY-model and the red dashed line -- the CMFGEN-model.}        
\label{fig:spectrum6}
\end{center}
\end{figure*}
\begin{figure*}
\begin{center}
\includegraphics[scale=0.4,viewport=45 0 1320 392,clip]{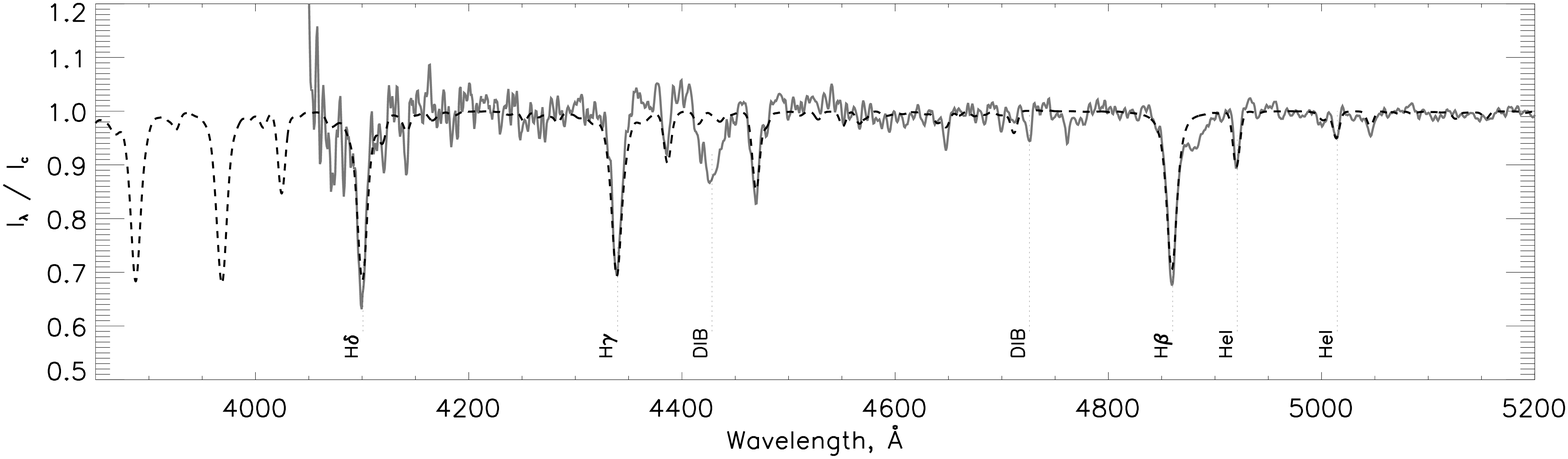}\\
\includegraphics[scale=0.4,viewport=45 0 1320 392,clip]{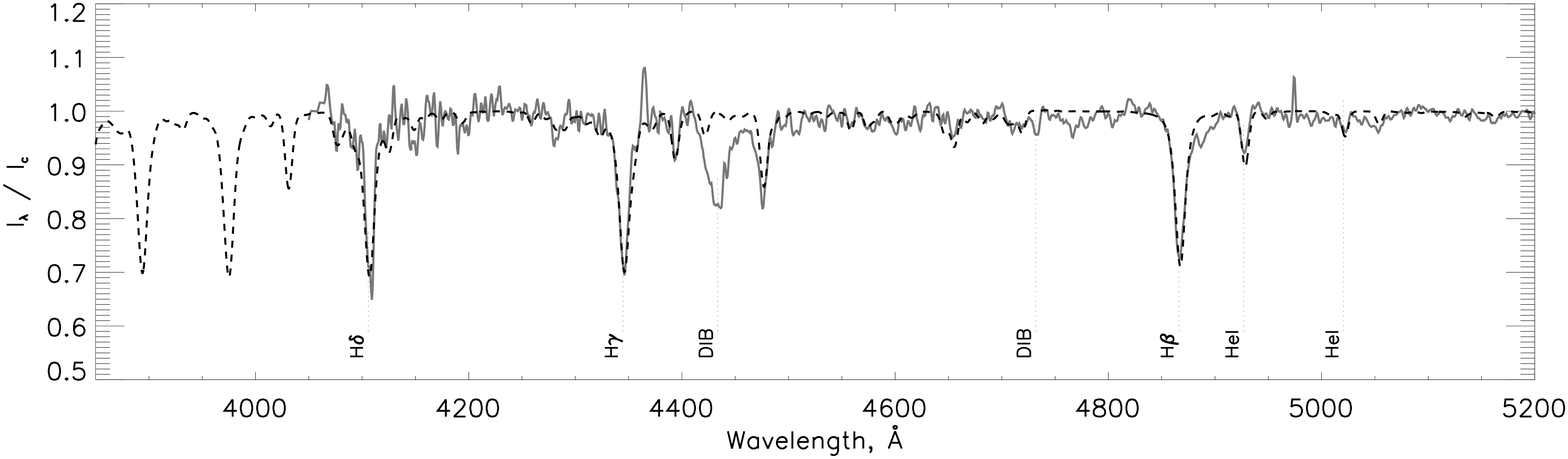}
\caption{Comparison of the observed continuum normalized spectra of MT282 (top) and MT343 (bottom) with the {\sc cmfgen}-model (dashed line). } 
\label{fig:spectrum343}
\end{center}
\end{figure*}

\subsection{Determination of bolometric luminosity and mass loss rate}\label{sec:cmfgen}

       As mentioned above, for determination of wind parameters and bolometric luminosity  we used models computed with {\sc cmfgen} code \citep{Hillier5}. Every model is defined by a hydrostatic stellar radius $R_*$, luminosity $L_*$, mass-loss rate $\dot{M}$, filling factor $f$, wind terminal  velocity $V_\infty$, stellar mass $M_*$ and by the abundances $Z_i$ of contained elementary species. We included  H, He, C, N, O, Si, S, P, Mg, Al and Fe in calculations.
      
       For the photospheric density structure we used the hydrostatic density structure computed with {\sc wind\_hyd} program (one of {\sc cmfgen} utilities). For the wind we used a standard $\beta$-velocity law $(V=V_{\infty}(1-\frac{R_*}{r})^\beta)$ that is connected to the hydrostatic density structure just above the sonic point. We chose $\beta =1.0$ as the default value for our calculation  since it turns out to be typical for O-dwarfs (e.g. \citet{Massa2003,MartinsHD15629}). We did not incorporate the clumping into our models, as the spectra we use lack wind emission lines.
        
       
       The first important step is the determination of a bolometric luminosity. Due to the fact that the stars belong to Cyg~OB2 association, distance for which is known, we were able to  estimate their luminosities. Distance to Cyg~OB2 was estimated in various works using different methods. For example,  spectrophotometric distance measurements span a range that includes 1.5~$\kpc$ \citep{JohnsonMorgan}, 2.1~kpc \citep{distance1966Reddish} 1.7~$\kpc$ \citep{MT91,distance1991Torres-Dodgen} and 1.45~kpc \citep{Hanson2003}. Using double-lined eclipsing binaries Kiminki et al. 2015 found that the distance is 1.33~$\kpc$. In this work, as well as in previous studies \citep{me2013,me2014}, we used an estimation of distance d=1.5~$\kpc$ from \citet{Dambis}, who acquired it from the analysis of line-of-sight velocities and proper motions of OB-associations. Note that the uncertainty of distance of $0.1~\kpc$ results in uncertainty of model stellar magnitude of about 0.14 mag, which is comparable to the errors of apparent stellar magnitudes.
      
       For each sample star we took model  from ``A grid of O star {\sc cmfgen} Models''\footnote{\url{http://kookaburra.phyast.pitt.edu/hillier/web/CMFGEN.htm}} closest to the star in $T_{\rm{eff}}$ and log$\,\textsl{g}$. We changed bolometric luminosities and iteratively refined  the model in a such way that {\it model photometry} matched the observational data ($m_V$ from Table~\ref{tab:nes}). {\it Model photometry} means magnitude in {\it V} filter derived from model spectrum, recalculated for the distance to the Cyg~OB2 association and corrected for the interstellar extinction. The values of the interstellar extinction were taken from \citet{KiminkiAv} and \citet{Chentsov2015}, they are listed in Table~\ref{tab:nes}. On the next step the resulted model was used as the seed model. Then we adjusted the parameters of model including temperature by means of qualitative comparison of computed spectra with observations. 
       
       For measuring $T_{\rm{eff}}$ with {\sc cmfgen}  we compared  intensities of different ion (C{\scriptsize II},{\scriptsize III},{\scriptsize IV}; N{\scriptsize II},{\scriptsize III}, He{\scriptsize I},{\scriptsize II}) lines, i.e we used the traditional ionization-balance method. {\sc cmfgen} measurements of $T_{\rm{eff}}$ are in a good agreement with  {\sc tlusty} estimations.

       For all sample stars there are no estimations of $V_\infty$.  Therefore for each stars we set the value of $V_\infty=2.65~V_{esc}$ \citep{KudritzkiWindsFromHotStars} and measured  an upper limit of mass-loss rate by visual comparison of model spectra with observational data.

\section{Results}\label{sec:results}

      Projected rotational velocities ${\it v\sin i}$ of studied stars have not been estimated previously. The exception is MT259 for which \citet{Herrero1999} estimated ${\it v\sin i}=30$~\kms. High resolution spectra of MT259, MT299 and MT317 allowed us to measure the ${\it v\sin i}$ of these stars. Table~\ref{tab:parameterswind} gives measured values of ${\it v\sin i}$ which were used for determination of other parameters and shows that our estimate is well consistent with one by \citet{Herrero1999}. 
       
      We determined $T_{\mathrm{eff}}$, log$\,\textsl{g}$, $L_*$ and other parameters of the sample stars using the methods described in Sections~\ref{sec:tlusty} and~\ref{sec:cmfgen}. Spectral lines  used to estimate $T_{\mathrm{eff}}$ and log$\,\textsl{g}$ of MT259, MT299 and MT317 with {\sc tlusty}  are listed in Table~\ref{tab:parameters}. On the other hand, for the determination of parameters  with {\sc cmfgen} we compared the overall shape of the model spectra with observational one, not just the separate lines. Figures~\ref{fig:spectrum21},~\ref{fig:spectrum16} and~\ref{fig:spectrum6} show the comparison of several lines from the spectra of MT259, MT299 and MT317  with model ones. They also demonstrate that  both {\sc cmfgen} and {\sc tlusty} successfully describe observed lines. We do not list metal lines used to estimate ${\it v\sin i}$ in Table~\ref{tab:parameters} but show some of these lines in the figures.

      Due to the low  spectral resolution of our spectra for MT282 and MT343 we were unable to estimate their $v\sin i$. Moreover in our spectra of MT282 and MT343 there are no lines of ionized helium, and the silicon lines are not well resolved. Thus it is problematic to constrain the effective temperature simultaneously along with other parameters using the automatic method. To avoid such problems we varied only $T_{\mathrm{eff}}$ and log$\,\textsl{g}$. We also considered not only spectral lines but spectral regions which contains lines useful to constrain $T_{\mathrm{eff}}$. In this study we selected the region $\lambda\lambda$4547--4591 containing Si{\scriptsize III} lines and the region $\lambda\lambda$4665--4731 around He{\scriptsize II} $\lambda$4686 line (see Table~\ref{tab:parameters}). The values of  $T_{\mathrm{eff}}$ and log$\,\textsl{g}$ determined with  {\sc tlusty} became starting parameters for {\sc cmfgen} calculations, results of which are presented in Figure~\ref{fig:spectrum343}. For calculation of their $V_{\infty}$ we supposed that log$\,\textsl{g}=3.9$ and $M_*$ are 20.6~$\Msun$ and 13~$\Msun$ for MT282 and MT343 accordingly.

\begin{table*}
\begin{center}
\caption{Physical parameters and wind properties of the studied Cyg~OB2 stars. $R_{2/3}$ is the radius where the Rosseland optical depth is equal to 2/3, $T_{eff}$ is the effective temperature at $R_{2/3}$.  The last column gives values of upper limits for mass-loss rate. }   
\label{tab:parameterswind}
\begin{tabular}{lll lcc ccc cl}
\hline
Star     &Spec.      &${\rm log L*/\Lsun}$&$T_{\rm{eff}}$& $R_{2/3}$&$M_*$     &log$\,\textsl{g}$&${\it v\sin i}$ &$v_{\rm{turb}}$ &$V_{\infty}$& $\dot{M}\cdot10^{-7}$            \\ 
         &class$^*$  &                   & [kK]         & [$\Rsun$]&[$\Msun$] &                 &[$\kms$]  & [$\kms$]       &[$\kms$]    & [$\rm M_{\odot}\mbox{yr}^{-1}$]\\
         &           &                   &              &          &          &                 &          &                &            &                  \\
\hline 
 MT259   & B0~V      &   $4.4\pm0.04$    &$31.2 \pm 0.7$&  5.3     &$10.3\pm4$& $3.97\pm 0.16$  & $25\pm5$ &$9\pm 1$        & 2230       &  0.3             \\ 
 MT282   &B1~{\bf IV}&   $4.4\pm0.04$    &$25\pm 3$     &  8.42    & $<$26    & $3.9\pm 0.3$    &          &                & 2560       &          \\ 
 MT299   & O7.5~V    &   $4.93\pm0.04$   &$33.3\pm 1.1$ &  8.75    &$17.7\pm7$& $3.78\pm 0.21$  &$200\pm20$&$12\pm 3$       & 2280       &  $1.5$           \\ 
 MT317   &O8~{\bf IV}&   $5.08\pm0.04$   &$32.8\pm 1.0$ & 10.6     &$23\pm10$ & $3.72\pm 0.21$  &$210\pm20$&$12\pm 3$       & 2370       &  $3.5$           \\ 
 MT343   & B1~V      &   $4.3\pm0.04$    & $26\pm 3$    &  6.7     & $<$16.4  & $3.9\pm 0.3$    &          &                & 2280       &        \\
\\
MT259$^+$&           &                   &34.5          &          &          &3.9              & 30       &                &            &                  \\
 \hline
 \multicolumn{11}{l}{$^*$ Spectral types according to our estimations of luminosities.} \\
 \multicolumn{11}{l}{$^{+}$ Data are taken from \citet{Herrero1999}.} \\
\end{tabular}
\end{center}
\end{table*}
\begin{table}
\begin{center}
\caption{Spectral lines and regions used  to estimate $T_{\mathrm{eff}}$ and log$\,\textsl{g}$ of considered Cyg~OB2 stars with {\sc tlusty} atmosphere models.  }
\label{tab:parameters}
\begin{tabular}{l|l}
\hline
Star  & Spectral lines and regions      \\ 
      & used in the analysis            \\
\hline 
      &                                 \\
MT259 & He{\scriptsize I} $\lambda\lambda$5876, 6678 He{\scriptsize II} $\lambda$5411        \\
      & wings of H$\alpha$                                                                   \\
MT299 & He{\scriptsize I} $\lambda\lambda$4921, 5015, 5876, He{\scriptsize II} $\lambda$5411 \\
      & wings of H$\beta$                                                                    \\
MT317 & He{\scriptsize I} $\lambda\lambda$4713, 4921, 5015, 5876                             \\ 
      & He{\scriptsize II} $\lambda\lambda$4542, 4686, 5411                                  \\
      & wings of H$\alpha$ and H$\beta$                                                      \\
MT282 & He{\scriptsize I} $\lambda\lambda$4143, 4388, 4471, 4921, 5015                       \\
      & H$\beta$, H$\gamma$, H$\delta$, C{\scriptsize III} $\lambda$4650                     \\
      & $\lambda\lambda$4547--4591, $\lambda\lambda$4665--4697                               \\
MT343 & He{\scriptsize I} $\lambda\lambda$4143, 4388, 4471, 4921, 5015                       \\
      & H$\beta$, H$\gamma$, H$\delta$, C{\scriptsize II} $\lambda$4267,C{\scriptsize III} $\lambda$4650 \\
      & $\lambda\lambda$4547--4592, $\lambda\lambda$4665--4731                               \\
\end{tabular}
\end{center}
\end{table}

      The derived parameters  are presented in Table~\ref{tab:parameterswind}. 
      For comparison Table~\ref{tab:parameterswind} also gives the parameters of MT259 estimated by \citet{Herrero1999}. The value of $T_{\mathrm{eff}}$ obtained in this study (31.2~kK) is lower than the one given by \citet{Herrero1999} (34.5~kK), and we will discuss it below. The other parameters of MT259 do agree with results of \citet{Herrero1999}.

\begin{figure}
\includegraphics[scale=0.65]{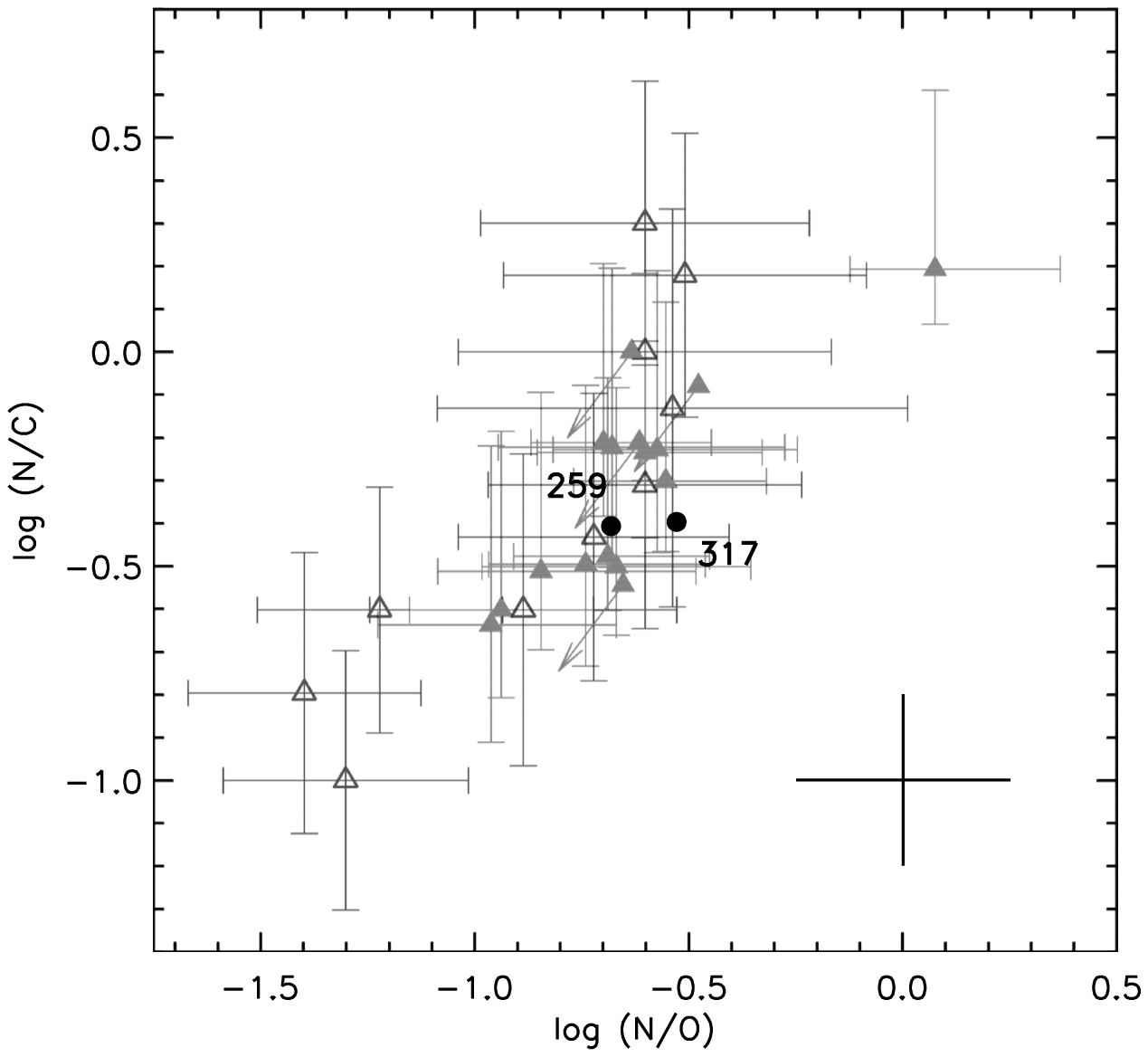}
\caption{$log (N/C)$ as a function of $log (N/O)$ for dwarfs stars. Circles are the studied stars;  filled triangles are Galactic stars from
\citet{MartinsMiMes}; open triangles are SMC dwarfs from \citet{BouretSMC}. The black cross in right lower corner displays typical errors of estimated abundances.}
\label{fig:cno}
\end{figure}
 \begin{table}
 \caption{The abundances of chemical elements estimated with CMFGEN atmosphere models.}
 \label{tab:frac}
 \begin{tabular}{lcccc}
\hline
 Star         &     He/H      & C/H         & N/H        &   O/H        \\
              &               & [$10^{-4}$] &[$10^{-4}$] &  [$10^{-4}$] \\
\hline              
  MT259       &$0.09\pm0.01$  & $2\pm1$     &$0.8\pm0.1$ &   $3.8\pm1$  \\ 
  MT317       &$0.09\pm0.01$  & $5\pm2$     &  $2\pm1$   &   $7\pm2$    \\ 
\end{tabular} 
\end{table}

      All optical helium lines falling into the observed spectral range were  used to constrain the relative He to H abundance. As a result, all Cyg~OB2 stars studied in this work demonstrate nearly solar helium  abundance\footnote{Solar He abundance is $He/H=0.085\pm0.002$ according to  \citet{solarabundance}.} He$\simeq26\%$ ($He/H=0.09\pm0.01$). 
  
      We determined the CNO abundances for MT259 and MT317 using C{\scriptsize II}~$\lambda\lambda 6578, 6583$, C{\scriptsize III}~$\lambda 5826$, C{\scriptsize IV}~$\lambda\lambda 5801.3, 5812$, N{\scriptsize II}~$\lambda\lambda 5666, 5676, 5680, 5686$, O{\scriptsize III}~$\lambda5592.25$  for the first star, and C{\scriptsize IV}~$\lambda\lambda 5801.3, 5812$, N{\scriptsize III}~$\lambda\lambda 4511, 4515, 4634.0, 4640.6 $,  O{\scriptsize III}~$\lambda5592.25$ for the second. Calculated abundances of the elements are given in Table~\ref{tab:frac}. 
      Figure~\ref{fig:cno} shows the locations of MT259 and MT317 on $log (N/C)/ log (N/O)$ diagram, taken from \citet{MartinsMiMes} (the figure~10 in the original article). Although,  MT259 and MT317 are members of the association and lie farther from us than stars considered in the  study of \citet{MartinsMiMes} they do not show any anomalies of chemical composition.

      For MT282, MT343 and MT299, however, we were not able to derive the chemical composition from our spectra due to limited spectral range of our data not including all the necessary diagnostic lines together. Due to that, for the estimation of physical parameters we assumed that chemical composition of MT282 and MT343 is the same as of MT259, while CNO-abundance in MT299 is equal to one of  MT317.
 
\section{Discussion}\label{sec:discuss}
\subsection{Temperatures and Luminosities}\label{sec:luminosity} 

      As it was mentioned in Section~\ref{sec:results} the value of $T_{\mathrm{eff}}$ obtained in this study for MT259 is lower than the one given by \citet{Herrero1999}, although it is close to estimate of \citet{Humphreys} and calibration of \citet{Crowther1997} for B0~V spectral type. This may be related to the difference of stellar atmosphere models used by us and used by \citet{Herrero1999} -- our models, for example, do account for line blanketing. As it was shown by \citet{Repolust} $T_{\mathrm{eff}}$ values obtained with line-blanketed models are lower than values obtained with unblanketed stellar atmosphere models.

      \citet{MartinsBC} presented new calibrations of stellar parameters of O-stars as a function of spectral type based on atmosphere models computed with the {\sc cmfgen} code. 
      They derived two types of effective temperature scales: an {\it observational} one from the results of modeling of individual O-stars and a {\it theoretical} one based on the grid of models. We compared our O-stars MT299 (O7.5~V) and MT317 (O8~IV) with these scales. Values of $T_{\mathrm{eff}}$ estimated by us for MT299 and MT317 are slightly lower than the values that could be obtained from spectral classification of these stars and {\it observational} $T_{\mathrm{eff}}$ scale of \citet{MartinsBC}. If we use {\it theoretical} $T_{\mathrm{eff}}$ scale of \citet{MartinsBC} then our $T_{\mathrm{eff}}$ estimates are in better agreement with spectral class.

      The modeling we performed  allows us to refine the luminosity classes of studied stars. There is uncertainty in luminosity class of MT259. \citet{KiminkiAv} estimated that MT259 is a supergiant (B0~Ib) while \citet{Chentsov} claims that this star is a dwarf (B0~V). Our estimates of log$\,\textsl{g}$ and $L_*$ for this star confirm the classification of \citet{Chentsov}. Luminosity of MT317 is a bit higher than the tabulated value of luminosity for O8~V class, according to \citet{MartinsBC}. Probably MT317 should be classified as O8~IV. \citet{Chentsov2015} claimed that the absorption lines in the spectrum of MT282 are not so deep as ones in the spectrum of MT343 (classified as B1~V dwarf), and probably the stars differ in luminosity class. Our modelling also demonstrated that MT282 is brighter than MT343, and argued in favor of MT282 being the star of B1~IV type. Table~\ref{tab:parameterswind} and~\ref{tab:ages} give the refined luminosity classes.

\subsection{Ages and Masses}\label{sec:ages} 
              
      Estimations of masses and ages are important for reconstruction of star formation history in the Cyg~OB2 region, they are also necessary to define the initial mass function (IMF). 
      Accurate determination of mass and age requires knowing precise location of the star on Hertzsprung-Russell (H-R) diagram, which is often being done based on the tabulated values for a given spectral class in Morgan-Keenan (M-K) system. This leads to uncertainties due to inevitable difference between tabulated data and intrinsic luminosities and temperatures of stars. According to  \citet{Wright2015} for O-type stars these uncertainties are typically small, $\sim$0.02~dex in ${\rm log T_{eff}}$ and $\sim$0.1~dex in ${\rm log L*/\Lsun}$. However for the B-type stars the errors rise significantly to $\sim$0.07~dex in ${\rm log T_{eff}}$ and $\sim$0.2~dex in ${\rm log L*/\Lsun}$ due to classification uncertainties and the larger difference in $T_{\rm{eff}}$ between spectral subclasses. Our  uncertainties obtained as result of modeling are significantly smaller:  $<$0.014~dex in ${\rm log T_{eff}}$ and $<$0.04~dex in ${\rm log L*/\Lsun}$ for O-type stars; $<$0.02~dex in ${\rm log T_{eff}}$ and $<$0.04~dex in ${\rm log L*/\Lsun}$ for the B-type stars. Table~\ref{tab:ages}  clearly demonstrates that the uncertainties in ${\rm log L*/\Lsun}$ and ${\rm log T_{eff}}$ obtained with modeling are smaller than the ones obtained  with M-K classification \citep{Wright2015}.

       \begin{table*}
       \begin{center}
        \caption{Derived masses, luminosities and ages of stars considered in this study, $*$ -- spectroscopic, ${evol}$ -- evolutionary. 
        Spectral classes are according to our estimations of luminosities.  $\dot{M}$ are upper limits for mass-loss rate. }
        \label{tab:ages}
\begin{tabular}{llllll}
\hline
\multicolumn{1}{c}{\bf Star}    &\multicolumn{1}{c}{\bf MT259}&\multicolumn{1}{c}{\bf MT282}  &\multicolumn{1}{c}{\bf MT299}&\multicolumn{1}{c}{\bf MT317}  &\multicolumn{1}{c}{\bf MT343}\\
\multicolumn{1}{c}{Spec. class} &\multicolumn{1}{c}{B0~V}     &\multicolumn{1}{c}{B1~{\bf IV}}&\multicolumn{1}{c}{O7.5~V}   &\multicolumn{1}{c}{O8~{\bf IV}}& \multicolumn{1}{c}{B1~V}    \\                   
\hline                                                       
\multicolumn{6}{c}{{\bf Measured using CMFGEN}}\\
${\rm log L*/\Lsun}$                                     & $4.4\pm0.04$       &$4.4\pm0.04$  &$4.93\pm0.04$      &$5.08\pm0.04$    & $4.3 \pm0.04$\\
${\rm log T_{eff}}$                                      & $4.5\pm0.01$       &$4.4\pm0.05$  & $4.522\pm0.013$   &$4.516\pm0.013$  & $4.42\pm0.05$\\ 
${\rm M_V}$                                              & $-3.27$            &$-3.8$        &$-4.49$            &$-4.89$          & $-3.42$\\
${\rm BC}$                                               & $-2.9$             &$-2.37$       &$-3$               &$-2.98$          & $-2.5$\\
                                                         &                    &              &                   &                 &              \\  
$M_*$,[$\rm \Msun$]                                      & $10\pm4$           &  $<$26       &$18\pm7$           &$23\pm10$        &  $<$16.4     \\ 
$\dot{M}$,[$\rm M_{\odot}\mbox{yr}^{-1}$]                &$3\cdot10^{-8}$     &              &$1.5\cdot10^{-7}$  &$3.5\cdot10^{-7}$&              \\
$\dot{M}_{Vink}$,[$\rm M_{\odot}\mbox{yr}^{-1}$]         &$8.4\cdot10^{-9}$   &              &$1.44\cdot10^{-7}$ &$2.1\cdot10^{-7}$&              \\
\hline    
\multicolumn{6}{c}{{\bf Measured using  ($\rm \log{g},\,T_{\rm eff}$)} } \\
$M_{evol}$,[$\rm \Msun$]                                 &$18^{+2}_{-1}$      & $11\pm1$      &$24^{+4}_{-2}$     &$24^{+4}_{-2}$   & $12\pm1$   \\
$log L_{evol}/L_{\odot}$                                 &$4.63\pm 0.05$      &$4.05\pm0.05$  &$5.03\pm0.07$      &$5.06\pm0.07$    & $4.18\pm0.05$\\
 Age,[Myr]                                               &$5.9 \pm0.9$        &$14 \pm2$      &$5.6\pm0.4$        &$5.9 \pm0.4$     & $12.1\pm1.5$ \\ 
\hline
\multicolumn{6}{c}{{\bf Measured using ($\rm L_*,\,T_{\rm eff}$) }} \\
$M_{evol}$,[$\rm \Msun$]                                 &$16\pm1$            &$12.9\pm0.5$  &$22\pm1$           & $24\pm1$        & $12.7\pm0.5$ \\ 
 Age,[Myr]                                               &$1-4.5$             &$14.1\pm1.3$  &$5.1\pm0.7$        &$5.8\pm0.4$      & $12\pm2$     \\ 
                                                         &                    &              &                   &                 &              \\ 
\hline
\multicolumn{6}{c}{{\bf Measured by  \citet{Wright2015} using M-K classification  }} \\                                                        
   ${\rm log L_{evol}/\Lsun}$                            & $4.2^{+0.14}_{-0.17}$ &       & $4.92^{+0.05}_{-0.06}$ & $5.04^{+0.04}_{-0.07} $ &  $4^{+0.13}_{-0.2}$            \\
   ${\rm log T_{eff~evol}}$                              & $4.44^{+0.06}_{-0.08}$&       & $4.54\pm0.02$          & $4.54\pm0.02$           &  $4.39^{+0.08}_{-0.06}$            \\  
   $M_{evol}$,[$\rm \Msun$]                              & $12.9^{+1.3}_{-3.1}$  &       & $23.4^{+1.1}_{-3.2}$   & $24.8^{+1.7}_{-3.4}$    &  $10.6^{+1.2}_{-2.1}$            \\
    Age,[Myr]                                            & $7.63^{+2.37}_{-4.46}$&       & $3.65^{+1.07}_{-2.35}$ & $4.28^{+0.75}_{-2.4}$   &  $>$10            \\                                                        
\end{tabular}
\end{center}
 \end{table*}
\begin{figure*}
\begin{center}
\includegraphics[viewport=39 0 475 330,clip]{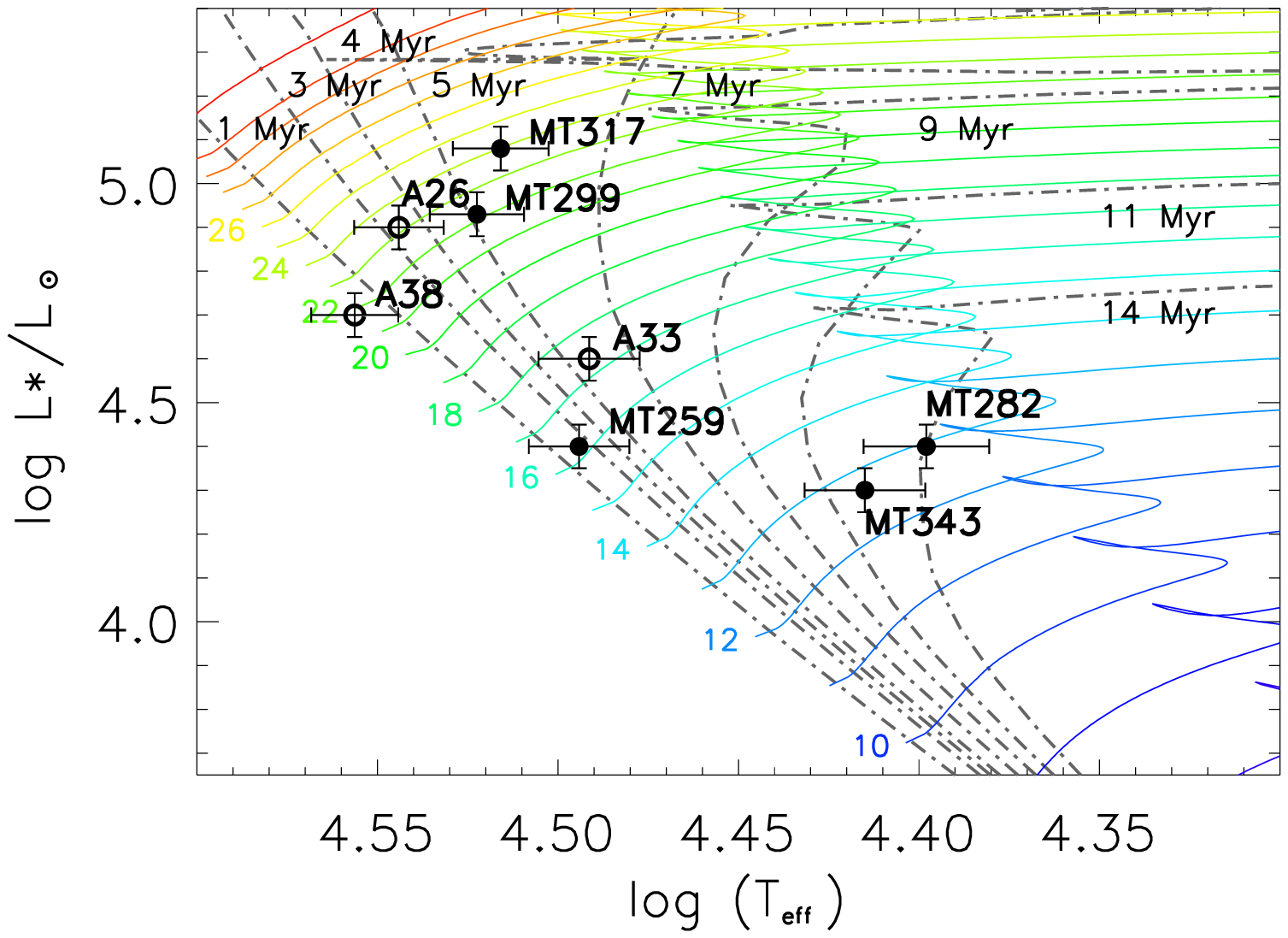}\\ 
\caption{HR diagram for the stars of the present study (filled circles) and the OB-dwarfs A26 (O9.5~V), A33 (B0.2~V) and A38 (O8~V) from \citet{Negueruela} (open circles).  
         Evolutionary tracks (solid lines) and isochrones (dashed-dotted lines) are  from \citet{Ekstrom}.}
\label{fig:gtlum}
\end{center}
\end{figure*}
\begin{figure*}
\begin{center}
\includegraphics[viewport=39 0 475 330,clip]{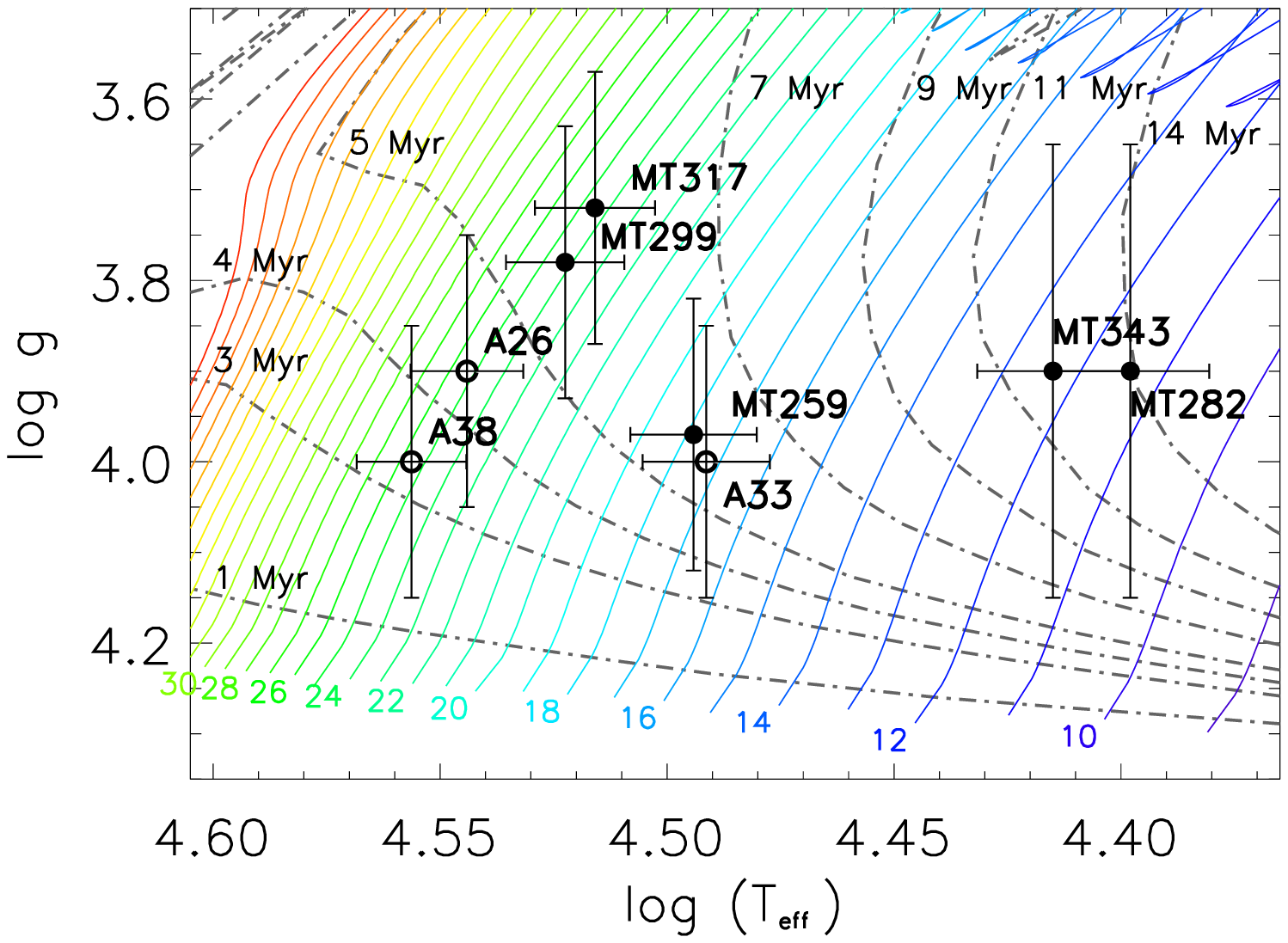}\\
\caption{$\log{g},\,T_{\rm eff}$ diagram for the stars of the present study (filled circles) and the OB-dwarfs A26 (O9.5~V), A33 (B0.2~V) and A38 (O8~V) from \citet{Negueruela} (open circles).  
         Evolutionary tracks (solid lines) and isochrones (dashed-dotted lines) are  from \citet{Ekstrom}.} 
\label{fig:gtmass}
\end{center}
\end{figure*}

            Figures~\ref{fig:gtlum} and~\ref{fig:gtmass} show the location of studied stars in the H-R diagram and $\log{g}~-~T_{\rm eff}$ diagram as well as evolutionary tracks and isochrones from the Geneva database \citep{Ekstrom}, constructed using the online calculator\footnote{http://obswww.unige.ch/Recherche/evol/-Database-} taking into account the effects of rotation. The rotation rate is $V_{ini}/V_{crit}=0.4$.

            Table~\ref{tab:ages} gives the comparison of the values of mass, luminosity  and age for each star, estimated using the diagrams and from the spectral modelling. The index ``${evol}$'' means that the quantity is measured using the H-R or $\log{g}~-~T_{\rm eff}$ diagram and evolutionary tracks.           
            Also Table~\ref{tab:ages} gives the same parameters obtained from the H-R diagram according to M-K classification \citep{Wright2015}. As we can see from the table, the H-R and $\log{g}~-~T_{\rm eff}$ diagrams provide similar values  of masses and close values of ages. The luminosities estimated with {\sc cmfgen} $L_{*}$ and with $\log{g}~-~T_{\rm eff}$ diagram $L_{evol}$ are relatively close. It once again shows that $\log{g}~-~T_{\rm eff}$ together with the current models of stellar evolution is very useful instrument for determination of  masses and ages of stars with unknown distances. The $\log{g}~-~T_{\rm eff}$ diagram combined with isochrones provides ages, and together with evolution tracks -- gives stellar mass, and therefore stellar luminosity $L_{evol}$.
                
            From the Table~\ref{tab:ages} we also can see, that $M_*$ is equal to $M_{evol}$ within the errors.  It is also seen that $M_*$ tends to be lower than $M_{evol}$.  This tendency can be due to the so-called {\it mass discrepancy} -- systematic overestimate of evolutionary masses $M_{evol}$ compared to spectroscopically derived masses $M_*$ (see e.g. \citet{Herrero92}). \citet{Markova2015} suggested that the reason for this effect is the neglection of turbulent pressure in {\sc fastwind} and {\sc cmfgen} atmospheric models. Moreover, in our case the tendency is less pronounced for the stars with higher masses. This can be consistent with trends founded by \citet{Markova2015}. But it is hard to make any certain conclusions taking into account that the uncertainties in $M_*$ are significantly higher than ones in  $M_{evol}$. The uncertainties in $M_*$ result from uncertainties in $\log{g}$ (about 0.2~dex), which are linked with narrow spectra range and, therefore, small number of spectral lines, in particular, hydrogen lines used in our analysis. The increase of analyzed spectral range is the way to improve the accuracy.

            \citet{MT91,Hanson2003}  have noted that the star formation in Cyg~OB2 is non-coeval. \citet{Drew} found clustering of A-type stars at distance of 20~arcmin south of the center  of Cyg~OB2 using data from the INT/WFC Photometric H$\alpha$ Survey (IPHAS).  One of the interpretations is that the cluster of  A-type stars and the already known OB star concentration are parts of the same association. And probably there were two main episodes of star formation, or there is a substantial age spread \citep{Drew}. \citet{Wright} analyzed the properties of stars in two fields in Cyg~OB2 using deep Chandra X-ray point sources catalog and found that these two fields are different by age of star generation.  \citet{Comeron12} concluded that members of the association display a mixture of ages ranging from less than 3~Myr to over 10~Myr, indicating a long sustained rate of star formation.  Also \citet{Comeron12} claimed that Cyg~OB2  extends beyond the area occupied by the youngest and hottest  stars and  that the southern part  of the association is clearly older. 
            \citet{Wright2015} presented list of massive stars in Cyg~OB2  which is most complete as of now. The age distribution of stars more massive than 20~$\Msun$ and  down to $\sim$B0.5~V spectral type showed that star formation process started at least $\sim$6-8~Myr ago. Between 1 and 7~Myr ago star formation occurred more or less continuously with probable excess between 4-5 Myr ago \citep{Wright2015}. Cyg~OB2  occupies a much larger region than typical young compact star clusters or associations. Based on spatial size of Cyg~OB2 and observed age spread \citet{Wright2015} suggested that Cyg OB2 was not born as a single star cluster but as a ``distribution'' of smaller groups or clusters of stars with a range of stellar ages.

            According to the H-R diagram the ages of MT299 and MT317 stars lie in the range between 5-6~Myr (see Table~\ref{tab:ages}), while MT259 is younger than 5~Myr, and  MT282 and MT343 belong to older population of Cyg~OB2. In $\log{g}~-~T_{\rm eff}$ diagram MT259 lies in the same area as MT299 and MT317 stars (5-6 Myr), while MT282 and MT343 stars are still in $>10$~Myr region. The ages estimated by \citet{Wright2015} are consistent with our estimates within the 2 standard errors. Thus comparing our results with results of previous studies we can conclude that the studied stars belong to different subclusters of Cyg~OB2: MT259 (B0~V), MT299 (O7.5~V) and MT317 (O8~V) are 4-6~Myr, MT282 (B1~IV) and MT343 (B1~V) are older cluster with 12-14~Myr age. 
             
            In addition to the studied stars, in Figures~\ref{fig:gtlum} and~\ref{fig:gtmass}  we marked the A26 (O9.5~V), A33 (B0.2~V) and A38 (O8~V) OB-dwarfs from \citet{Negueruela}. On the one hand this shows that the number of studied dwarfs belonging to Cyg~OB2 is quite small. Before our work, among the dwarf stars of Cyg~OB2 association, only these three stars, as well as  MT259 (B0~V) and MT29 (O7~V) studied by \citet{Herrero1999}, were modelled. On the other hand, the locations of A26  and A38 together with MT299 and  MT317 stars demonstrates the continuity of  star formation process in the association. 
            
\begin{figure}
\begin{center}
\includegraphics[scale=0.4,viewport=45 20 530 530,clip]{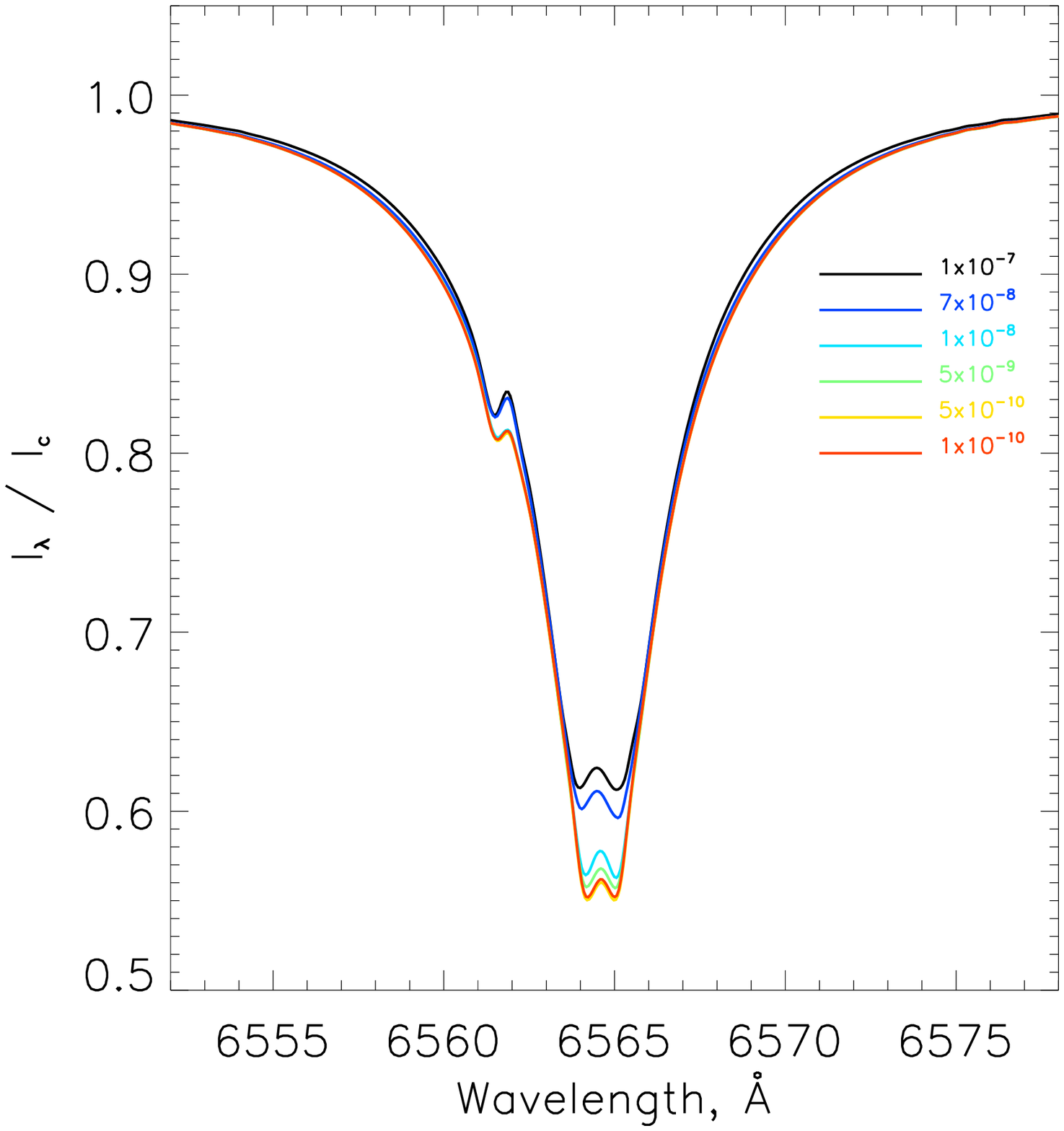}
\includegraphics[scale=0.4,viewport=45 20 530 530,clip]{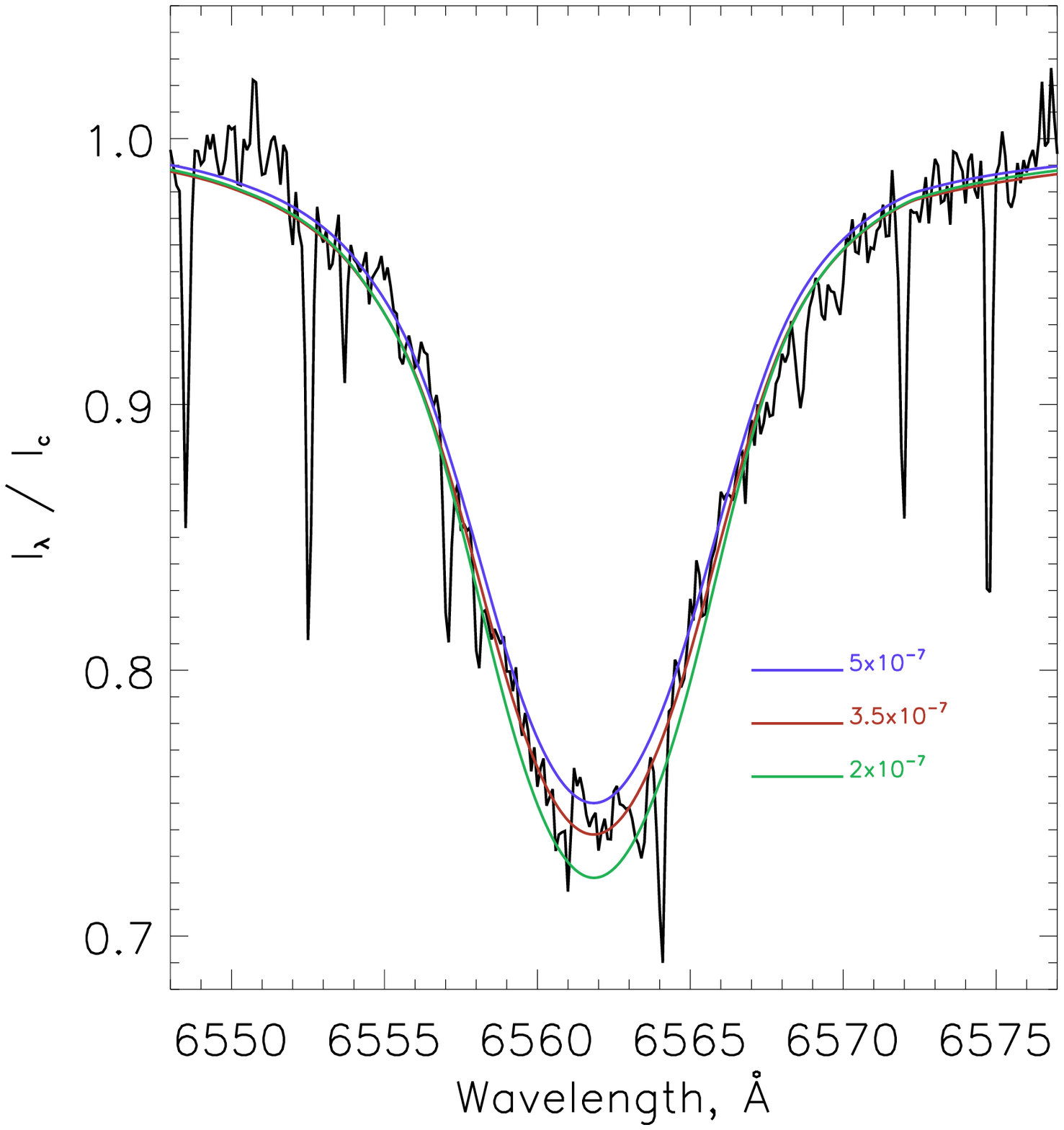}
\caption{Upper panel: changes of profile of  $H_{\alpha}$ line with the decrease of mass loss rate. Model luminosity is $1.2\cdot10^5~\Lsun$, terminal velocity is $V_{\infty}=1000~\kms$. Lower panel: comparison of $H_{\alpha}$ line profile from  the spectrum of MT317 (O8 IV)  with models computed for different mass loss rates. Mass loss rate is in units of $\rm M_{\odot}\mbox{yr}^{-1}$. } 
\label{fig:masslossrate}
\end{center}
\end{figure}
 
\subsection{Mass loss rate}\label{sec:mdot}
      
            O-dwarfs show spectral lines with P-Cyg profiles only in the ultraviolet (UV) range.  Therefore UV spectra are necessary for an accurate assessment of their mass loss rates. \citet{Marcolino09} demonstrated that when $\dot{M}$ falls below $10^{-7}\rm M_{\odot}\mbox{yr}^{-1}$ the profile of $H_{\alpha}$ line becomes insensitive to its further reduction.  Thus our estimates of $\dot{M}$ obtained based on optical spectra are just upper limits.  As an example, the upper panel of Figure~\ref{fig:masslossrate}  demonstrates  changes of $H_{\alpha}$ line profile in model spectra with the decrease of mass loss rate. The lower panel also shows the example of fitting of $H_{\alpha}$  line in the spectum of MT317 (O8~IV) with the model spectra calculated for different values of mass loss rate. 
  
            Tables~\ref{tab:ages}  shows $\dot{M}$ predicted by the theory \citep{Vink2000,Vink2001},  marked as $\dot{M}_{Vink}$. The measured mass loss rates are consistent with $\dot{M}_{Vink}$ predicted.

\section{Conclusions}\label{sec:conclusions}

            We investigated five stars -- MT259 (B0~V),  MT282 (B1~IV), MT299 (O7.5~V),  MT317 (O8~V) and MT343 (B1~V) -- belonging to Cyg~OB2 association. Using {\sc tlusty} and {\sc cmfgen} codes we estimated their physical parameters, including luminosity, mass loss rate, projected rotation velocity and chemical composition of the atmospheres. According to our modelling we reclassify MT282 and MT317 from dwarfs to subgiants owing to their higher luminosities. H-R and $\log{g}~-~T_{\rm eff}$ diagrams show that the ages of  MT259, MT299 and MT317 are between 4-6 Myr like the ones of most part of Cyg~OB2 stars. On the other hand, MT282 and MT343 belong to the older population of the association, their ages $>10$~Myr.

            In the article we examined in detail the method of automatic comparison of observed spectra with {\sc tlusty}-model. We demonstrated that the method works well and it can be used for a simple estimations of stellar parameters of OB-stars  based on both high and low resolution spectra. 
            
            We  hope that this study will be useful for calibration of the {\it spectral class -- luminosity -- effective temperature} ratio, as the numbers of OB-stars with known distance are still small.

\section*{Acknowledgements}
The observations at the 6-meter BTA telescope were carried out with the financial support of the Ministry of Education and Science of the Russian Federation (agreement No. 14.619.21.0004, project ID RFMEFI61914X0004). We used spectral data retrieved from the ELODIE archive at Observatoire de Haute-Provence (OHP) and database of stellar evolution group at the Geneva Observatory. The study was supported by the Russian Foundation for Basic Research (projects no. 14-02-31247,14-02-00291). Olga Maryeva thanks the grant of Dynasty Foundation. AS was supported by The Ministry of Education and Science of the Russian Federation within the framework of the research activities (project no. 3.1781.2014/K). Also we would like to thank  the anonymous referee for valuable comments.

\begin{appendix}
\section{Estimation of $\chi^2_{\rm t}$}\label{sec:chi}  


       The choice of $\chi^2_{\rm t}$ given by Eq.~\ref{eq:chit} in Section~\ref{sec:tlusty} is based on the analysis of spectra of O-stars ALS~8476 (O9~V) and BD+35$^\mathtt{o}$1201 (O9.5~V). Spectra of ALS~8476 and BD+35$^\mathtt{o}$1201 were obtained on the Russian 6-m telescope with the high-resolution Nasmyth Echelle Spectrograph (NES) \citep{NES1}, spectral resolution is R=60000. Continuum normalization of these spectra had been performed manually with {\sc dech} program \citep{DECH}. The signal-to-noise ratio (S/N) of these spectra in the vicinity of He{\scriptsize I} $\lambda$5876 line is 80 and 110 for  the ALS~8476 and BD+35$^\mathtt{o}$1201 spectra, respectively. But (S/N) is significantly lower ($\sim 30$) in the blue region of spectra resulting in relatively high parameter errors of ALS~8476 and BD+35$^\mathtt{o}$1201.
       
       The projected rotational velocities ${\it v\sin i}$ of ALS~8476 and BD+35$^\mathtt{o}$1201 were estimated with Full Width at Half Maximum (FWHM) method (see e.g. Herrero 1992). The metal lines were used in the case of  ALS~8476 while in the case of BD+35$^\mathtt{o}$1201 we used He lines. The obtained values are ${\it v\sin i}=18\pm4$ and $217\pm16$~\kms for ALS~8476 and BD+35$^\mathtt{o}$1201, respectively.

       The effective temperature and surface gravity of ALS~8476 and BD+35$^\mathtt{o}$1201 were initially estimated by us using the method based on the $T_{\rm{eff}}$-log$\,\textsl{g}$ diagnostic diagram (see e.g. \citet{Herrero92}) with hydrogen and helium lines. The estimated values are $T_{\mathrm{eff}}=32000\pm1500$~K and log$\,\textsl{g}=4.1\pm0.2$~dex for  ALS~8476 and $T_{\mathrm{eff}}=32000\pm1500$~K and log$\,\textsl{g}=3.9\pm0.2$~dex for  BD+35$^\mathtt{o}$1201. Then the parameters of these two stars were refined using the method for high resolution spectra described in Section~\ref{sec:tlusty} with different $\chi^2_{\rm t}$ values. The final $\chi^2_{\rm t}$ was chosen to provide the values of parameter errors comparable to the ones obtained with the method based on the $T_{\rm{eff}}$-log$\,\textsl{g}$ diagnostic diagram. With this chosen value of $\chi^2_{\rm t}$ the refined parameters are $T_{\mathrm{eff}}=32000\pm1000$~K and log$\,\textsl{g}=4.2\pm0.2$~dex for  ALS~8476 and $T_{\mathrm{eff}}=32500\pm1100$~K and log$\,\textsl{g}=4.1\pm0.2$~dex for  BD+35$^\mathtt{o}$1201. This estimates are consistent with those obtained with $T_{\rm{eff}}$-log$\,\textsl{g}$ diagnostic diagram.
       
\section{Testing the method of effective temperature and surface gravity determination}\label{sec:test}

         
       To test the method of automatic spectral analysis described in the Section~\ref{sec:tlusty} we used spectra of the following stars: HD~15629 (O5~V((f))), HD~34078 (O9.5~V). Moreover to test the applicability of this method for low resolution spectra of B-stars we used the spectra of HD~36591 (B1~V) and HD~42597 (B1~V). The parameters of HD~15629, HD~34078 and HD~36591 are known from the literature  \citep{MartinsHD15629,Lefever2010} while the spectrum of HD~42597 had not previously been modeled.

       The spectra of HD~15629, HD~42597 and HD~36591 were acquired from  Elodie archive of Observatoire de Haute-Provence  (\url{http://atlas.obs-hp.fr/elodie/intro.html}) \citep{Elodie}. The spectrum of HD~34078 obtained on the 1.2-m telescope of the Kourovka Astronomical Observatory of the Ural Federal University with the fiber-fed high resolution echelle spectrograph (UFES) \citep{PanchukUFES}. Continuum normalization of all spectra had been performed manually with {\sc dech} program \citep{DECH}. The spectral resolutions of Elodie and UFES spectra are 42000 and 30000, correspondingly. Signal-to-noise ratio (S/N) of Elodie spectra in the vicinity of He{\scriptsize I} $\lambda$5876 line is higher than 100, while for the UFES spectrum it is 75.

\begin{table*}
\begin{center}
\caption{Testing of the spectral analysis method. The comparison of estimated parameters of  HD~15629 (O5~V((f))), HD~34078 (O9.5~V), HD~36591 (B1V), HD~42597 (B1V) with the literature data.  }  
\label{tab:test}
\begin{tabular}{l|l|l|l}
 \multicolumn{1}{c|}{Star}&$T_{\rm{eff}}$, [kK]& log$\,\textsl{g}$, [dex]&   \multicolumn{1}{c}{Source}   \\
 \hline
 \hline
                   &                    &                         &                                \\
 {HD~15629 }       & {$40.8\pm 1.1$}    &  {$3.81\pm 0.12$}       &He{\scriptsize II} $\lambda\lambda$4200, 4542, He{\scriptsize I} $\lambda\lambda$4471, 4921, 5876, wings of H$\delta$, H$\gamma$, H$\beta$, H$\alpha$\\
 \cline{2-4}
                   &                    &                         &                                \\   
  {O5V ((f))}      & {$41.1\pm 1.1$}    &  {$3.84\pm 0.12$}       &He{\scriptsize II} $\lambda$4542, He{\scriptsize I} $\lambda\lambda$4471, 4921, wings of H$\beta$ \\
  \cline{2-4}
                   &                    &                         &                                                                      \\
                   & {$41  \pm 2.0$}    &  {$3.75\pm 0.1$}        &  \citet{MartinsHD15629}   \\
\hline
\hline                    
 {HD~34078 }       &                    &                         &He{\scriptsize II} $\lambda\lambda$4200, 4542, 4686, 5411,  He{\scriptsize I} $\lambda\lambda$4143, 4388, 4471, 4713, 4921, 5015\\            
 {O9.5V}           &  {$33.0\pm 0.7$}   &  {$4.12\pm 0.18$}       & He{\scriptsize I} $\lambda\lambda$ 5047, 5876, 6678, 7065, wings of H$\delta$, H$\gamma$,  H$\beta$, H$\alpha$  \\ 
 \cline{2-4}
                   &                    &                         &                               \\
                   &  {$33.2\pm 0.7$}   &  {$4.15\pm 0.18$}       & He{\scriptsize II} $\lambda$5411, He{\scriptsize I} $\lambda\lambda$5876, 6678, wings of H$\alpha$ \\
 \cline{2-4}
                   &                    &                         &                         \\
                   &  {$33.0\pm 2.0$}   &  {$4.05\pm 0.1 $}       &\citet{MartinsHD15629}   \\
\hline
\hline
HD~42597           &                    &                         & High Resolution spectrum. Si{\scriptsize II} $\lambda\lambda$4128, 4130, 5056, 6347 \\
 B1V               &  $22.8\pm1.2$      &   $3.98\pm0.19$         &  Si{\scriptsize III} $\lambda\lambda$4552, 4567, 4574, 4819, 4829, 5740, wings of H$\delta$, H$\gamma$,  H$\beta$, H$\alpha$ \\         
\cline{2-4}
                   &                    &                         &                         \\
                   &  $23.7\pm2.8$      &  $4.0\pm0.3$            & Low Resolution  spectrum.  Region $\lambda\lambda$4547--4591, region $\lambda\lambda$4665--4731  \\
\hline
\hline 
HD~36591           &                    &                         & High Resolution spectrum. Si{\scriptsize II}  $\lambda\lambda$4128, 4130, 5056, 6347   \\
B1V                &  $26.1\pm0.8$      &   $4.15\pm0.19$         &  Si{\scriptsize III} $\lambda\lambda$4552, 4567, 4574, 4819, 4829, 5740, wings of H$\delta$, H$\gamma$,  H$\beta$, H$\alpha$ \\  
\cline{2-4}
                   &                    &                         &                         \\                   
                   &  $26.3\pm 2.3$     &  $4.2 \pm 0.3$          & Low Resolution spectrum. Region $\lambda\lambda$4547--4591, region $\lambda\lambda$4665--4731 \\
\cline{2-4} 
                 &                      &                             &                        \\
                 &  {$26\pm1$  }        &    {$3.9\pm0.1$ }           &{\citet{Lefever2010}}   \\
                 &  {$27\pm1$  }        &    {$4.0\pm0.2$  }          &{\citet{Morel2008}}     \\
                 &  {$27\pm0.3$ }       &    {$4.12\pm0.05$ }         &{\citet{Przybilla2008}} \\
\end{tabular}
\end{center}
\end{table*}

\begin{figure*}
\begin{center}
\includegraphics[scale=0.2,viewport=20 0 560 560,clip]{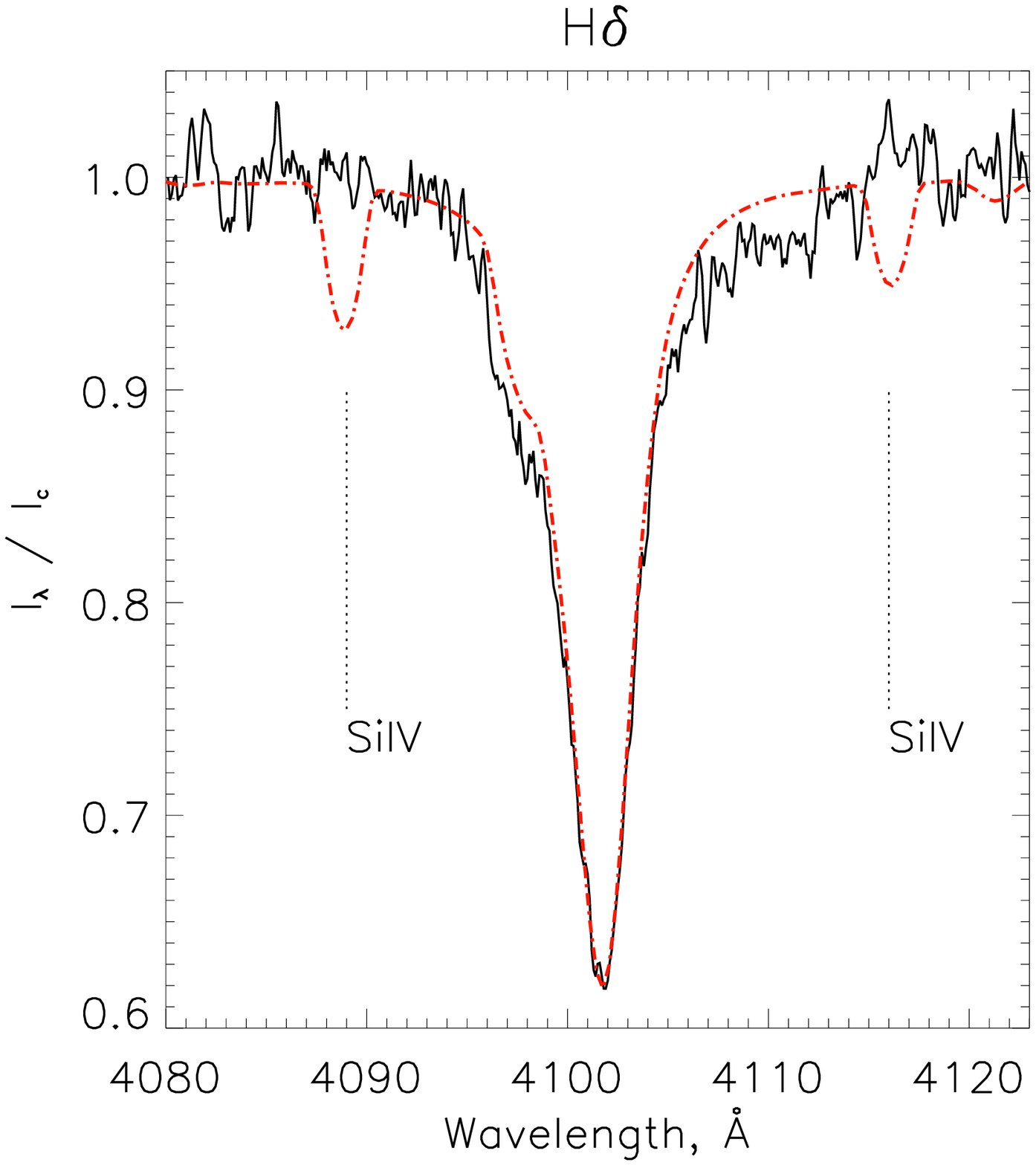}
\includegraphics[scale=0.2,viewport=20 0 560 560,clip]{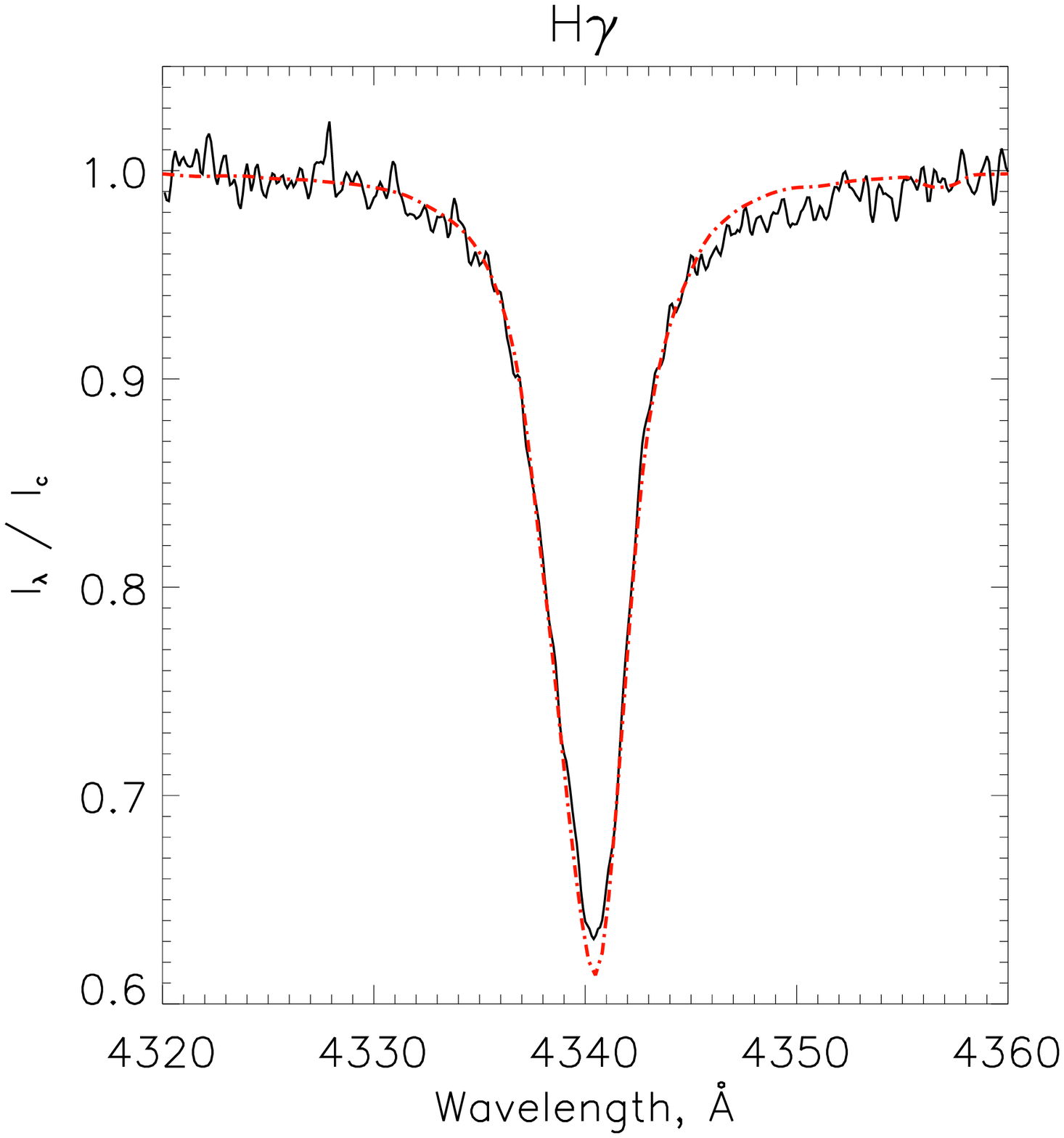}
\includegraphics[scale=0.20,viewport=0 0 560 560,clip]{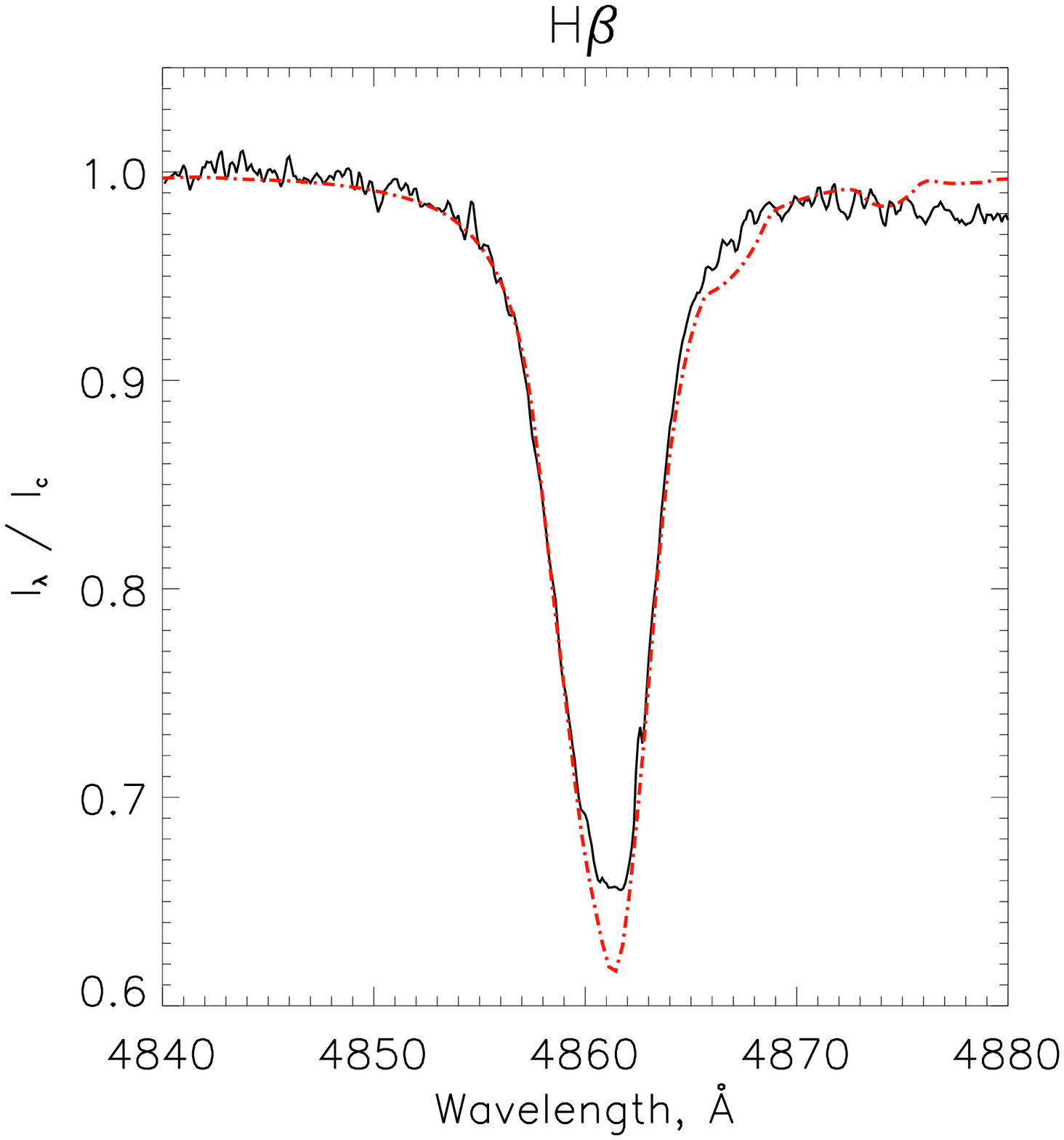}
\includegraphics[scale=0.20,viewport=20 0 560 560,clip]{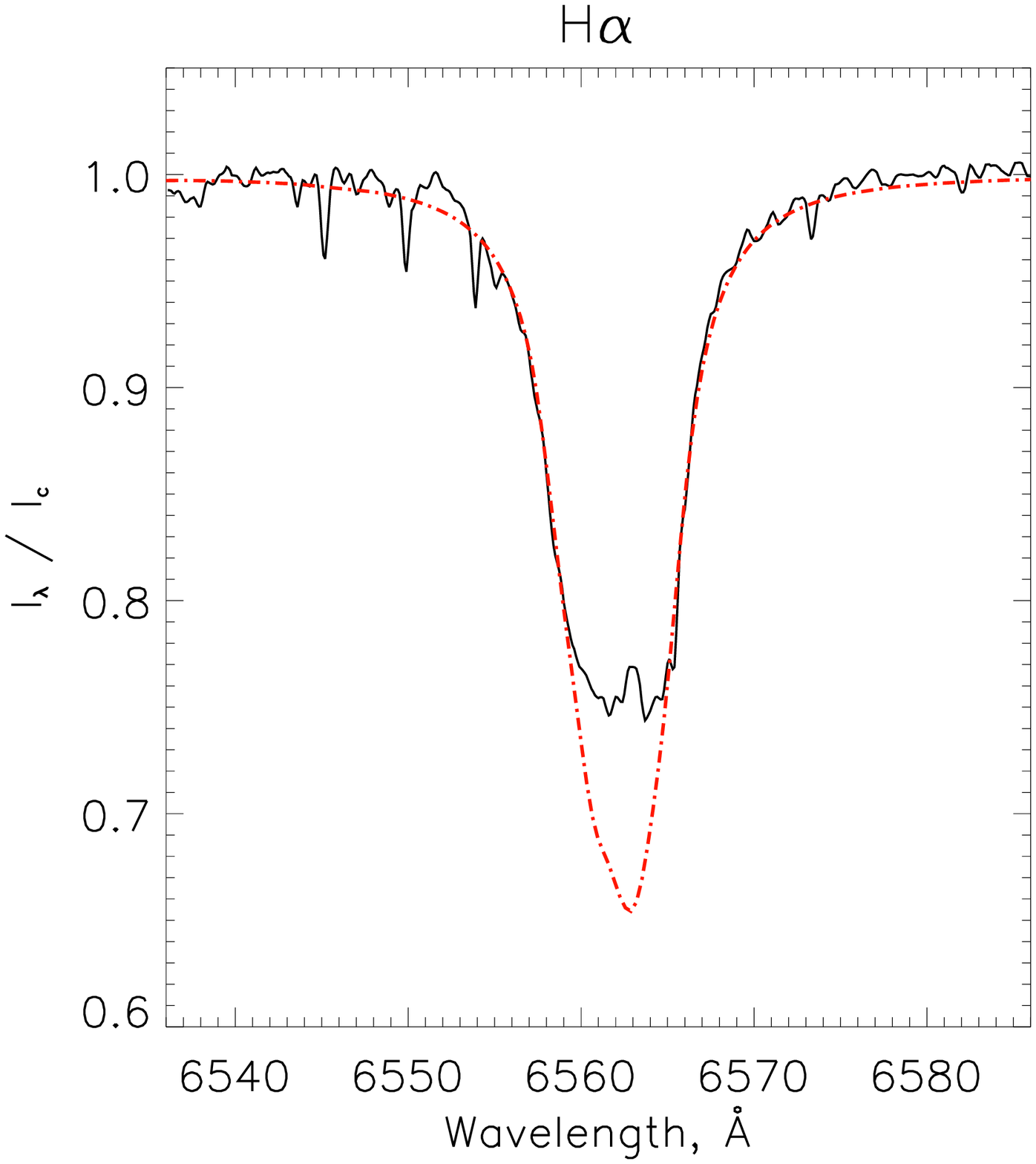}
\includegraphics[scale=0.20,viewport=20 0 580 560,clip]{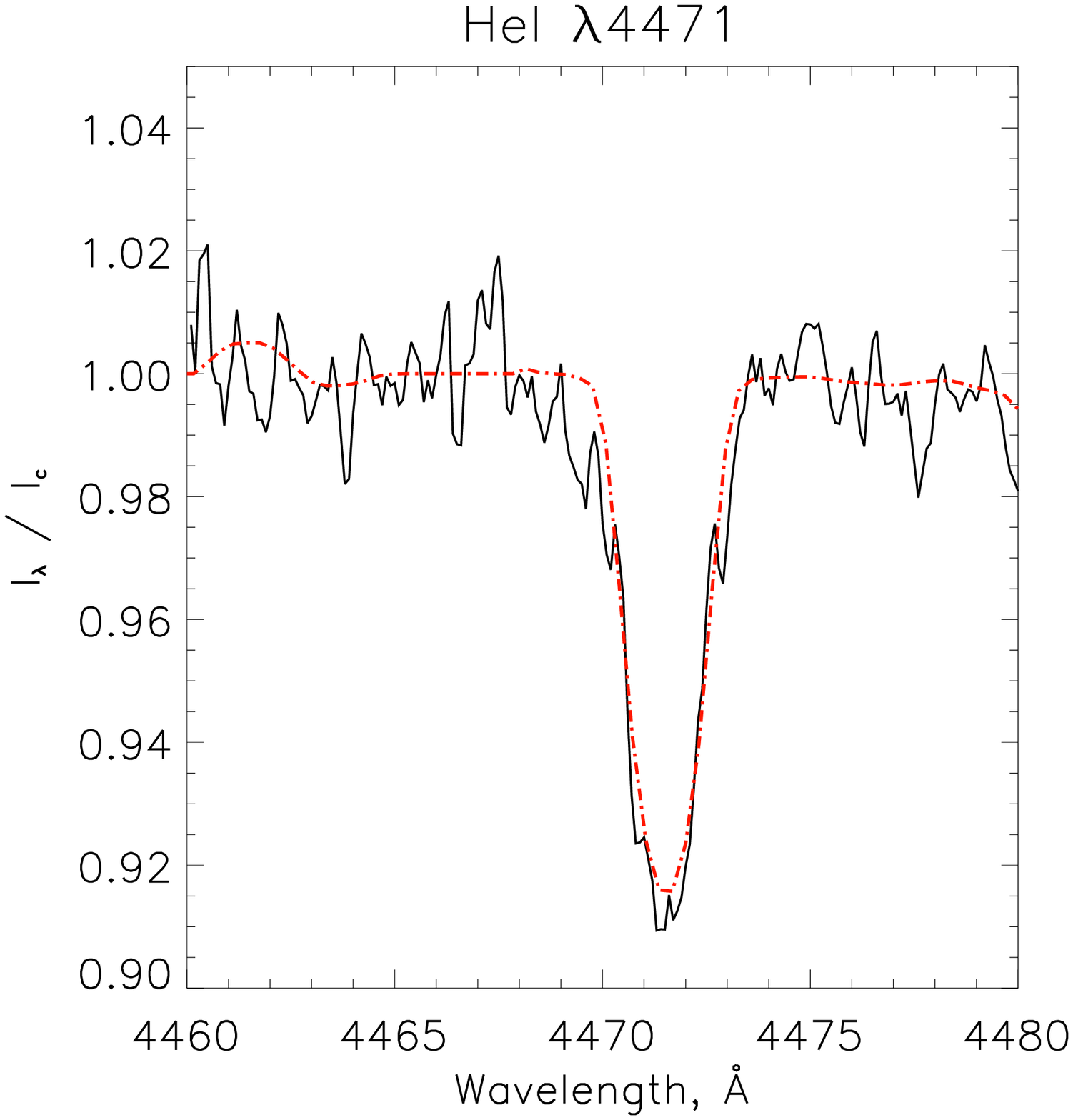}
\includegraphics[scale=0.20,viewport=20 0 560 560,clip]{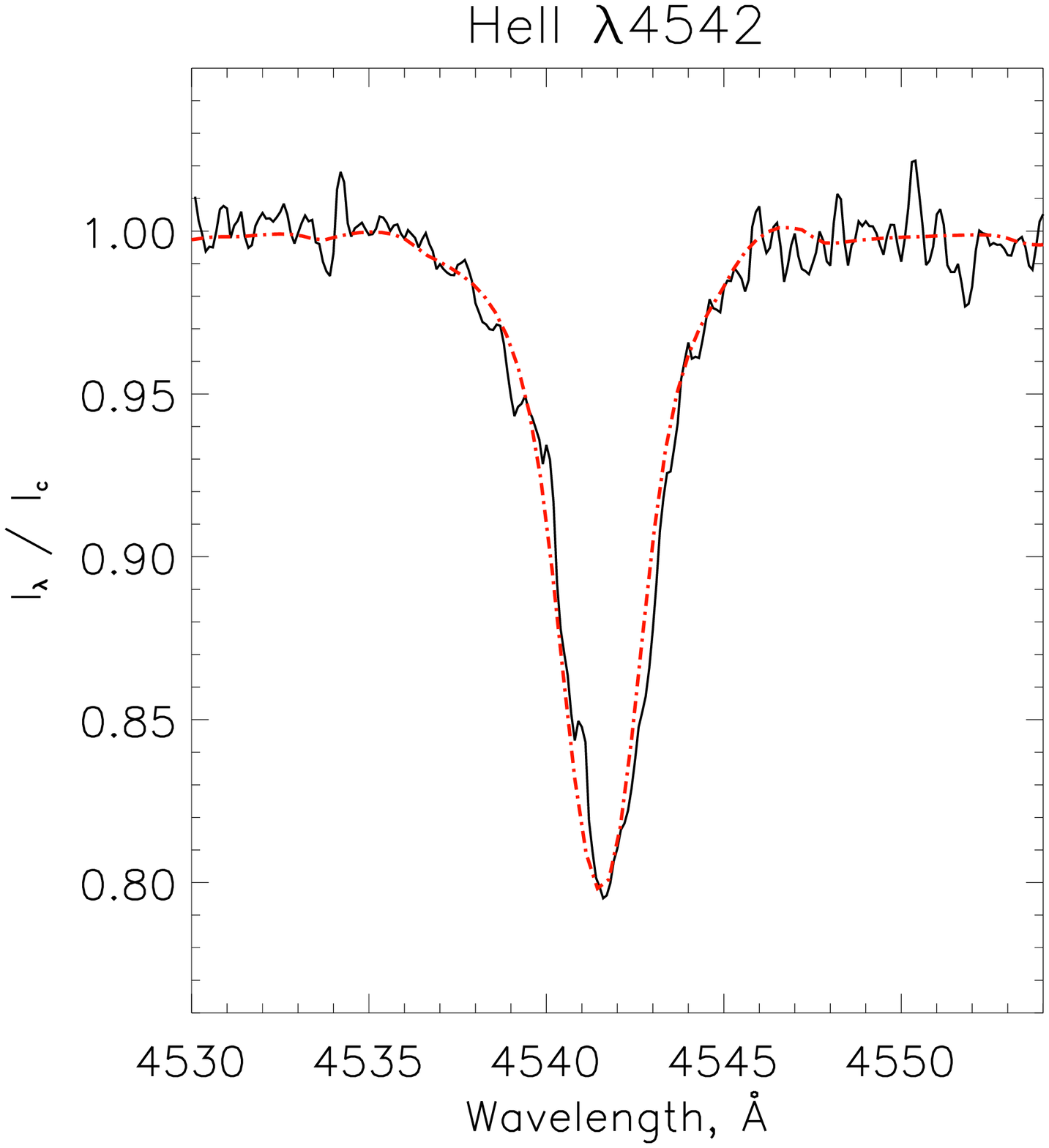}
\includegraphics[scale=0.20,viewport=20 0 580 560,clip]{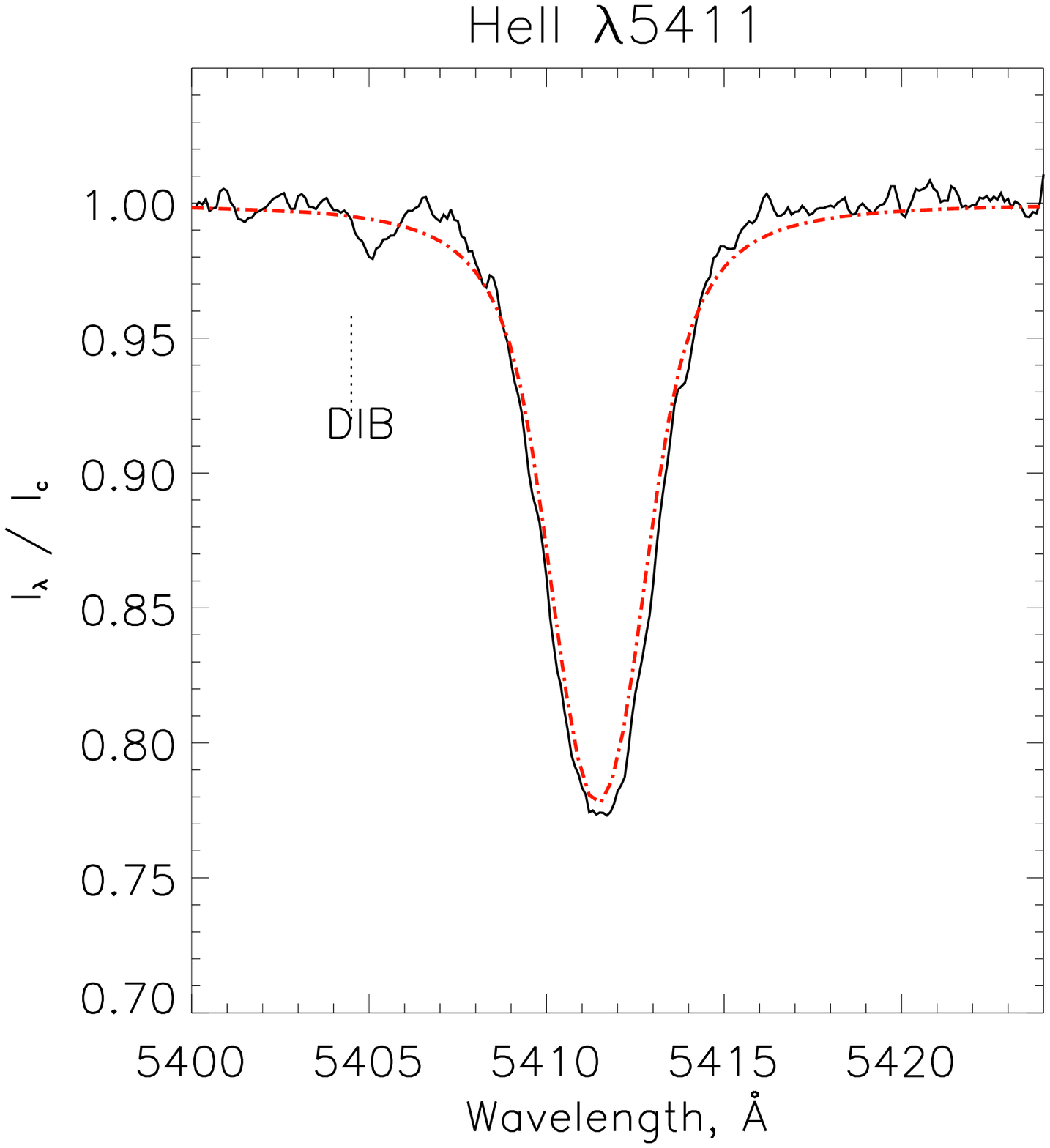}
\includegraphics[scale=0.20,viewport=20 0 560 560,clip]{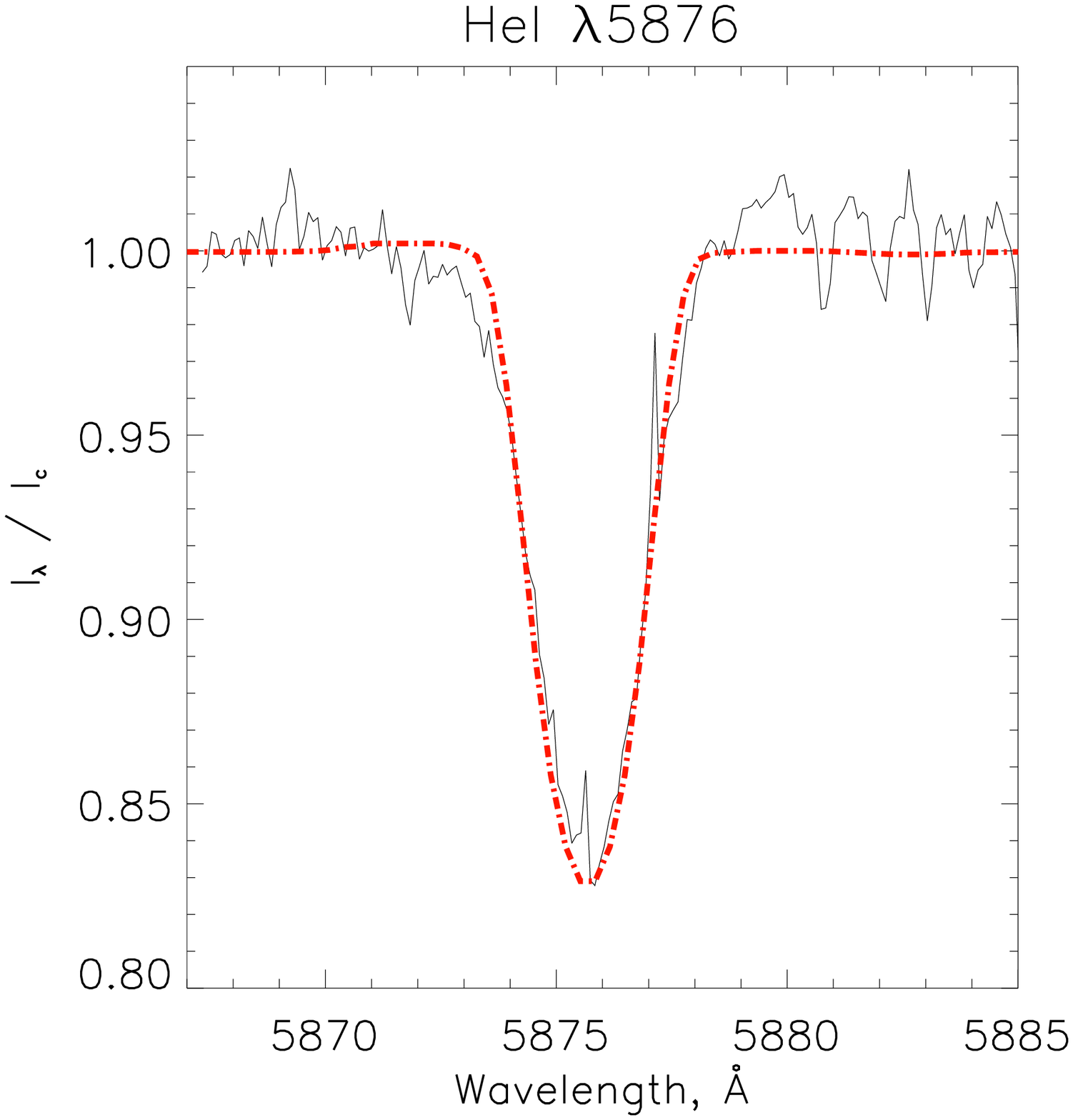}
\caption{Testing of the spectral analysis method. Modeling of HD~15629 (05 V ((f))). Comparison of the profiles of selected lines with the best model spectra. The solid line shows the observed profile, and the dashed-dotted line -- the  TLUSTY-model. The spectrum was obtained with Elodie echelle spectrograph.  }
\label{fig:spectrum15629}
\end{center}
\end{figure*}
\begin{figure*}
\begin{center}
\includegraphics[scale=0.2,viewport=20 0 560 560,clip]{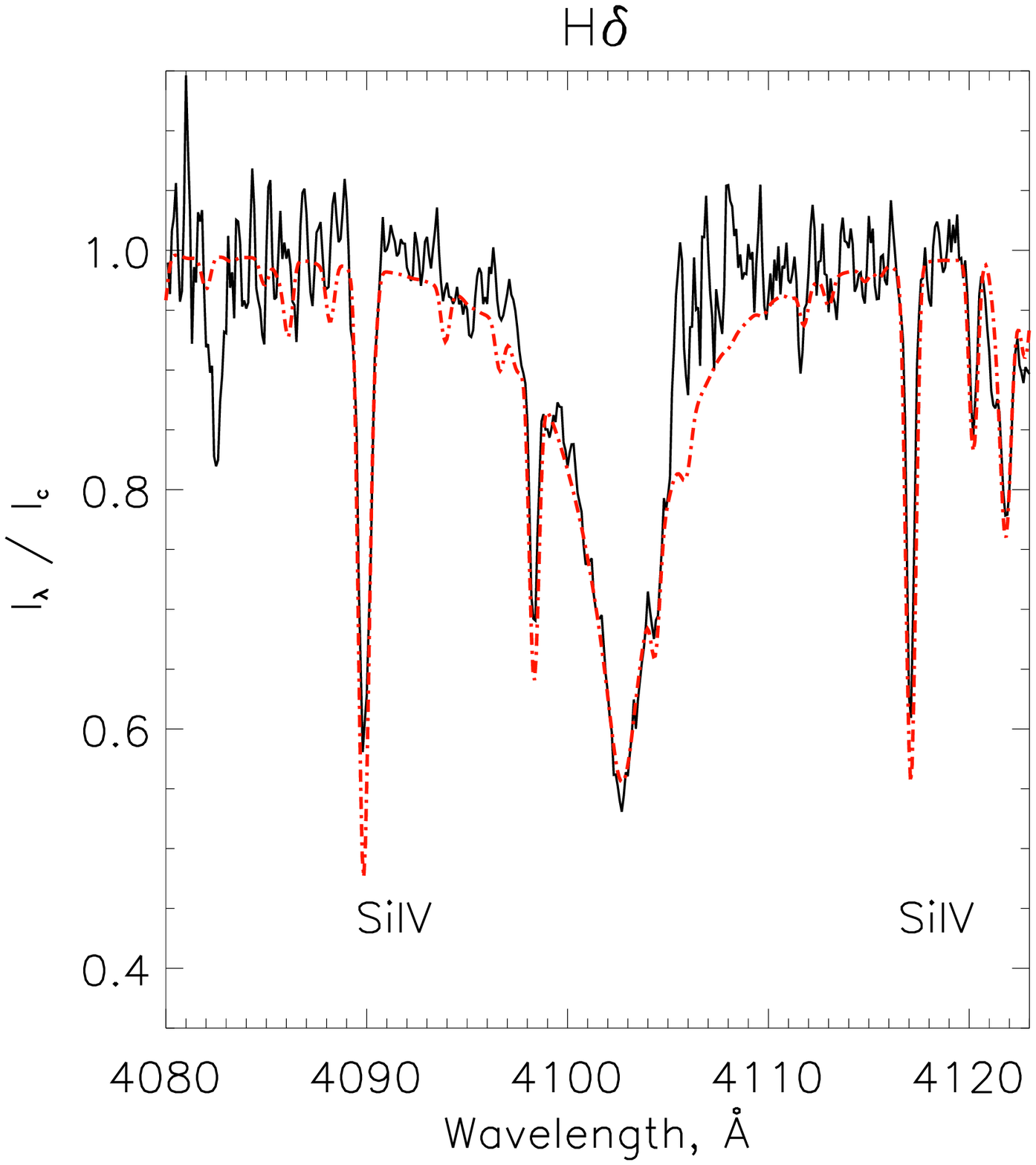}
\includegraphics[scale=0.2,viewport=20 0 560 560,clip]{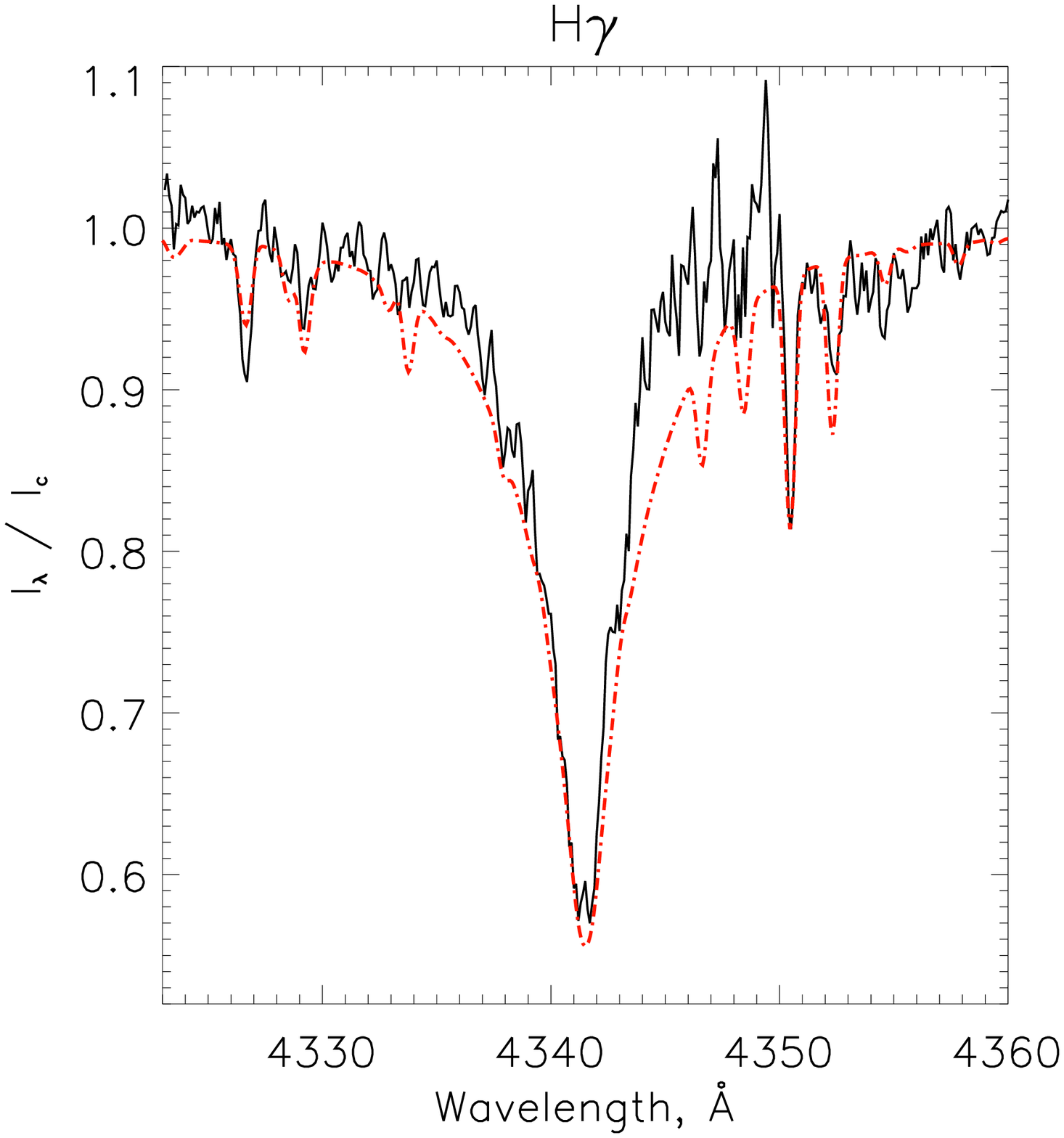}
\includegraphics[scale=0.20,viewport=0 0 560 560,clip]{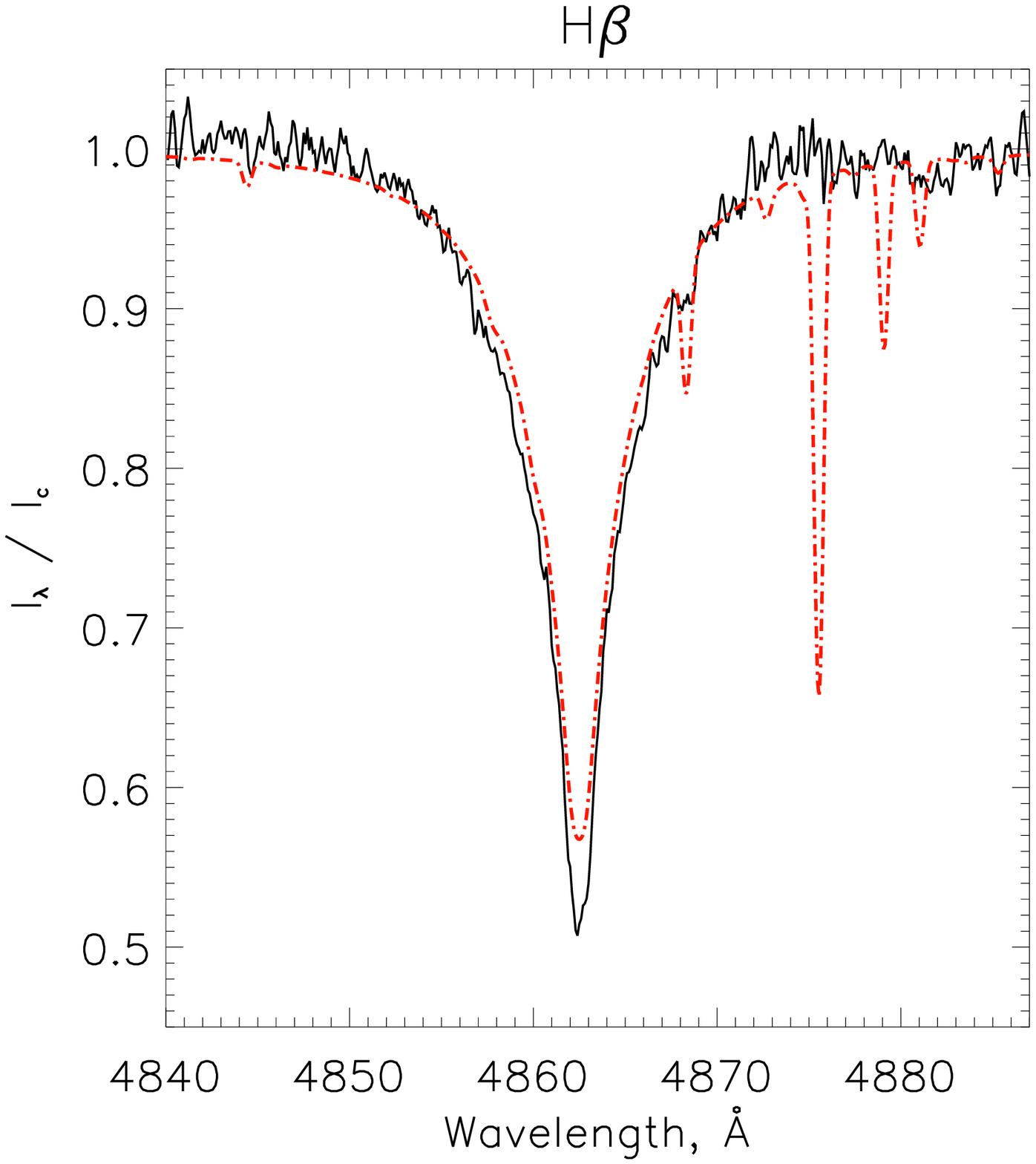}
\includegraphics[scale=0.20,viewport=20 0 560 560,clip]{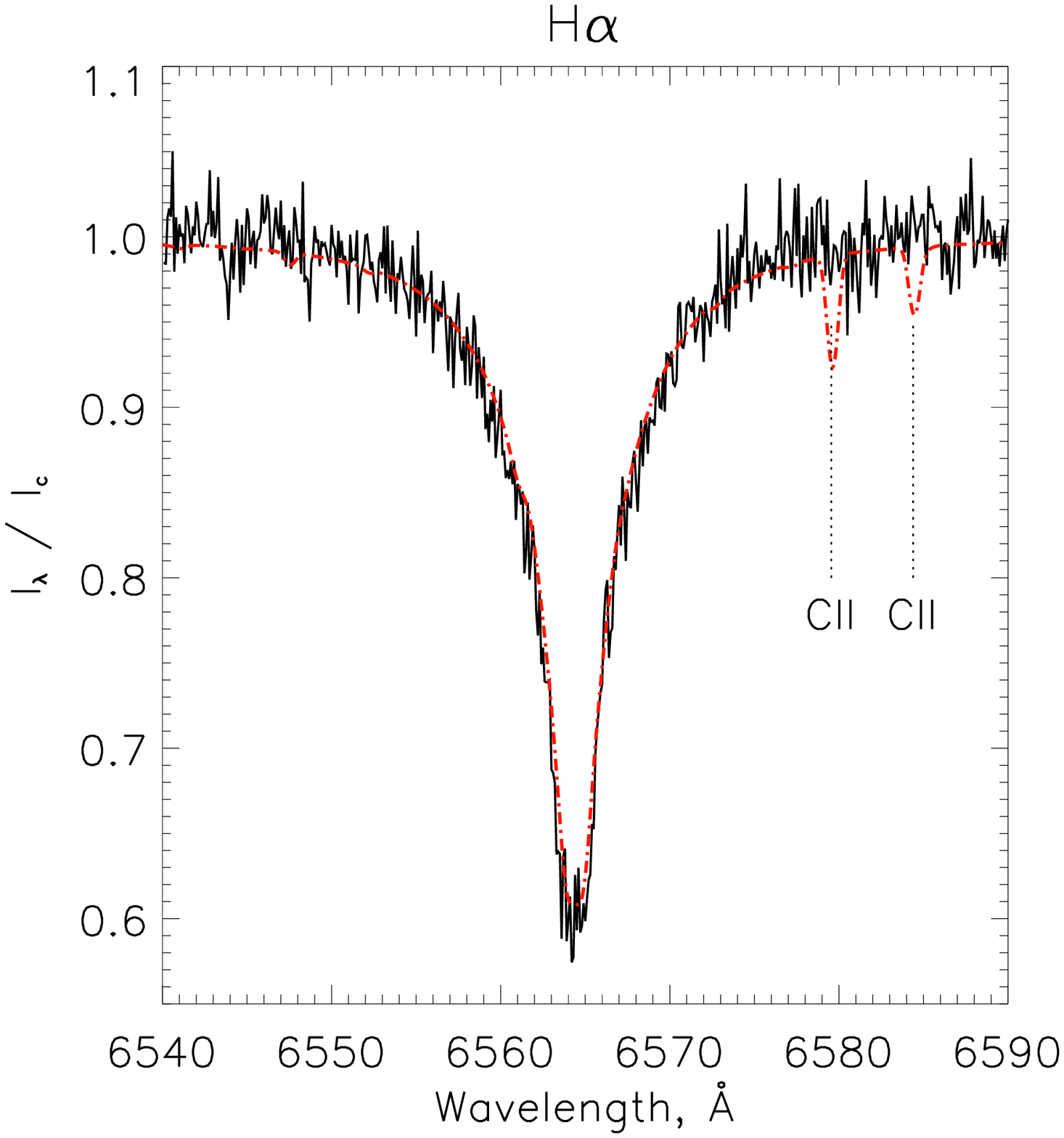}
\includegraphics[scale=0.20,viewport=20 0 580 560,clip]{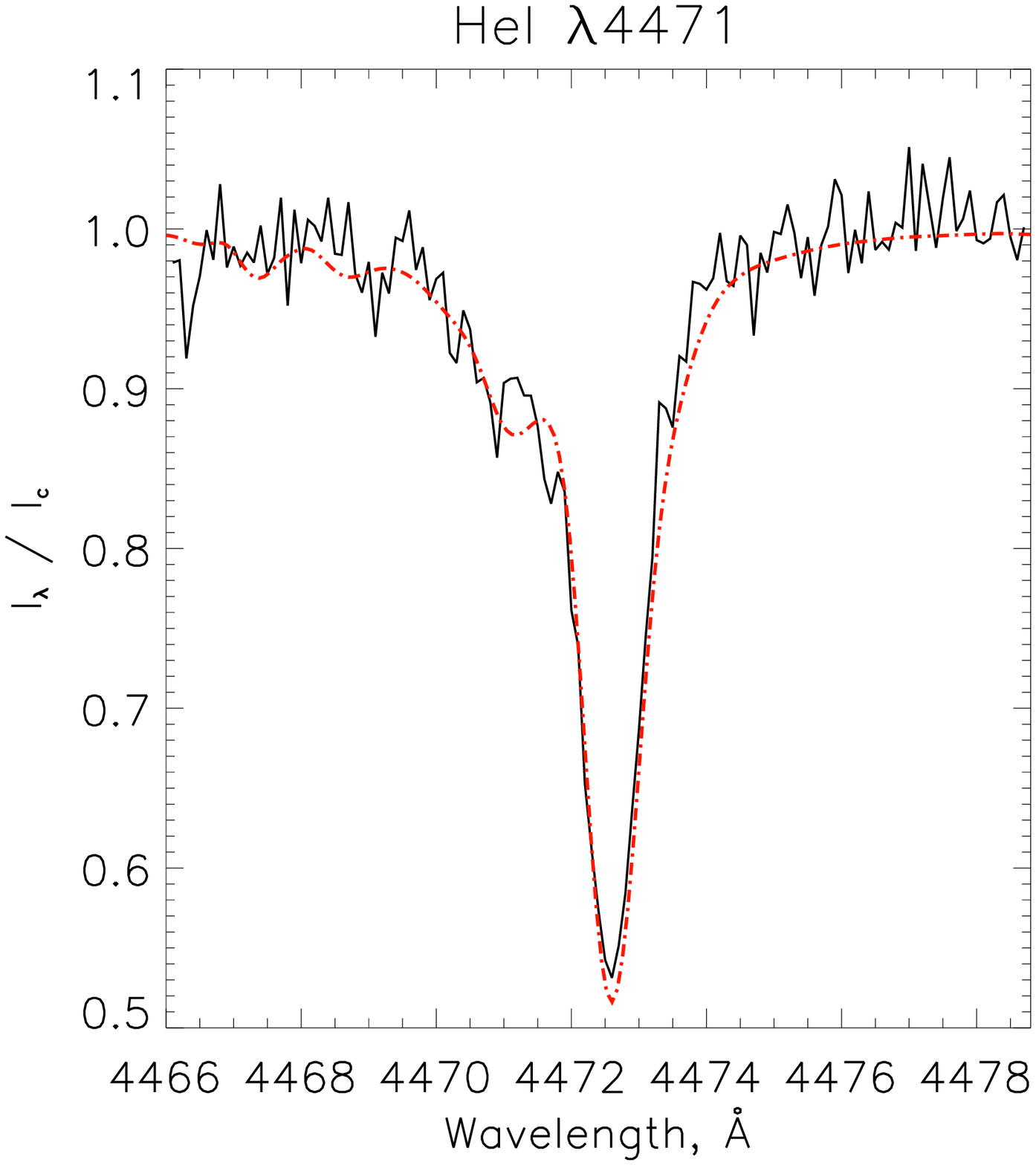}
\includegraphics[scale=0.20,viewport=20 0 560 560,clip]{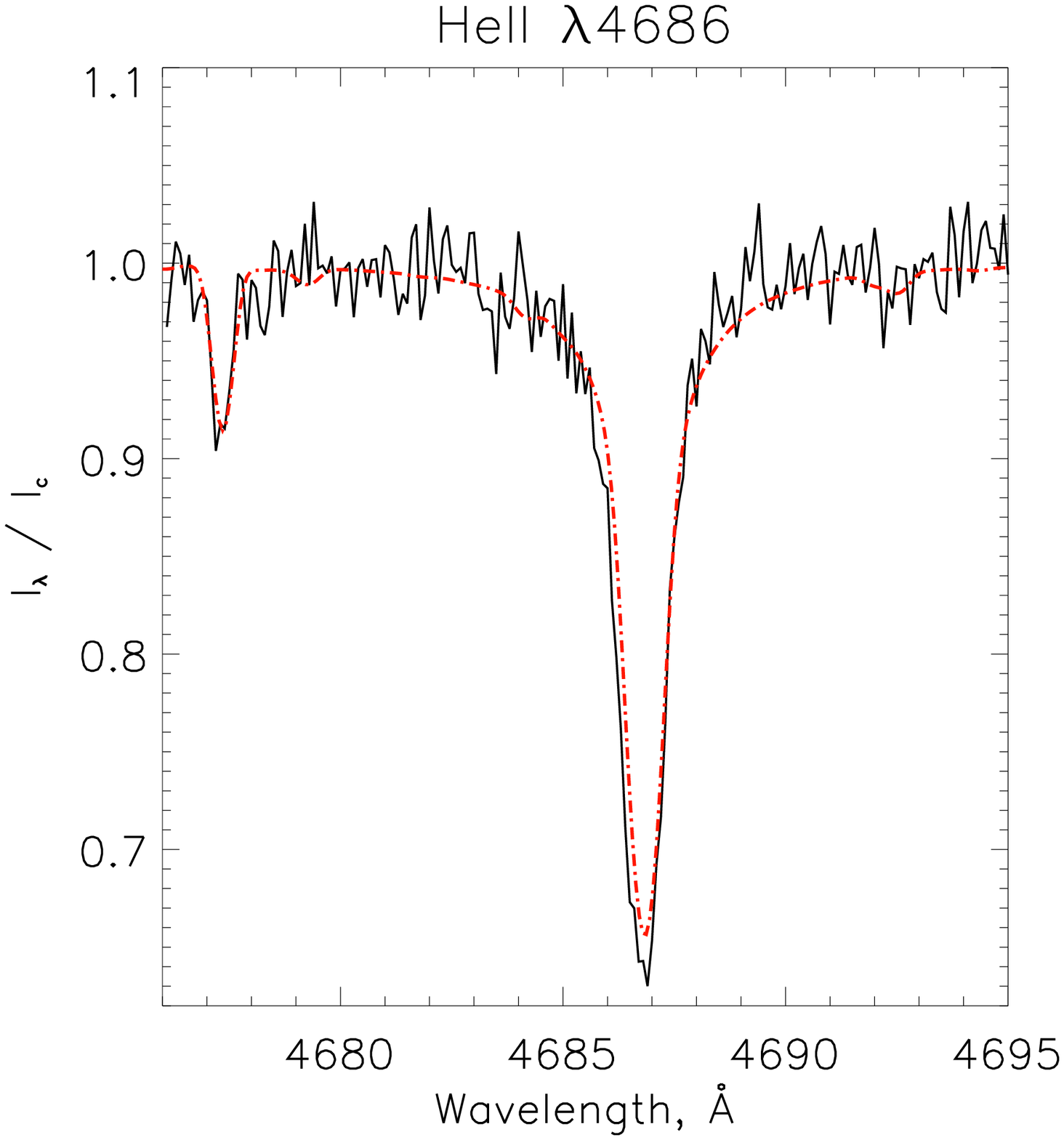}
\includegraphics[scale=0.20,viewport=20 0 580 560,clip]{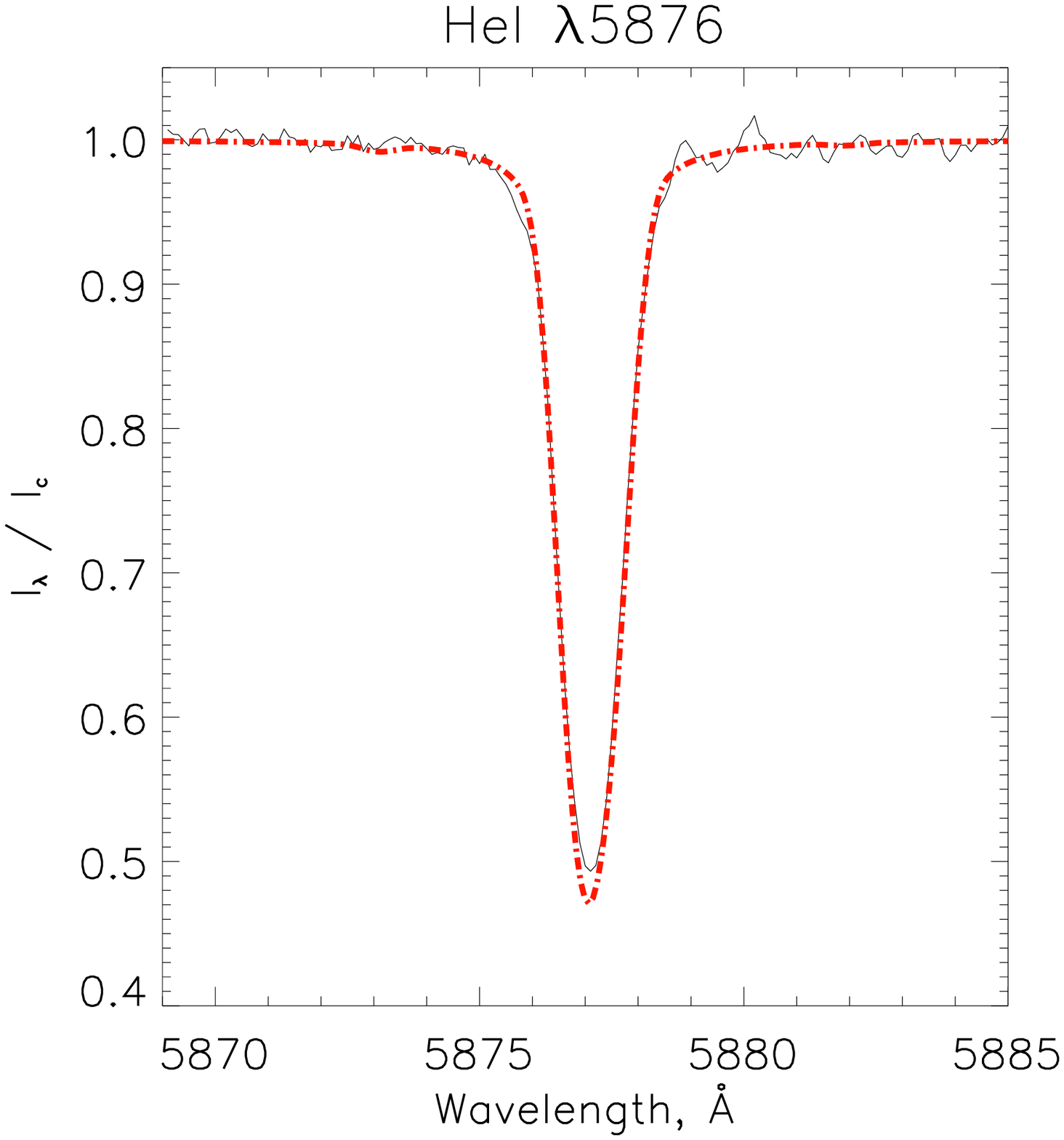}
\includegraphics[scale=0.20,viewport=20 0 560 560,clip]{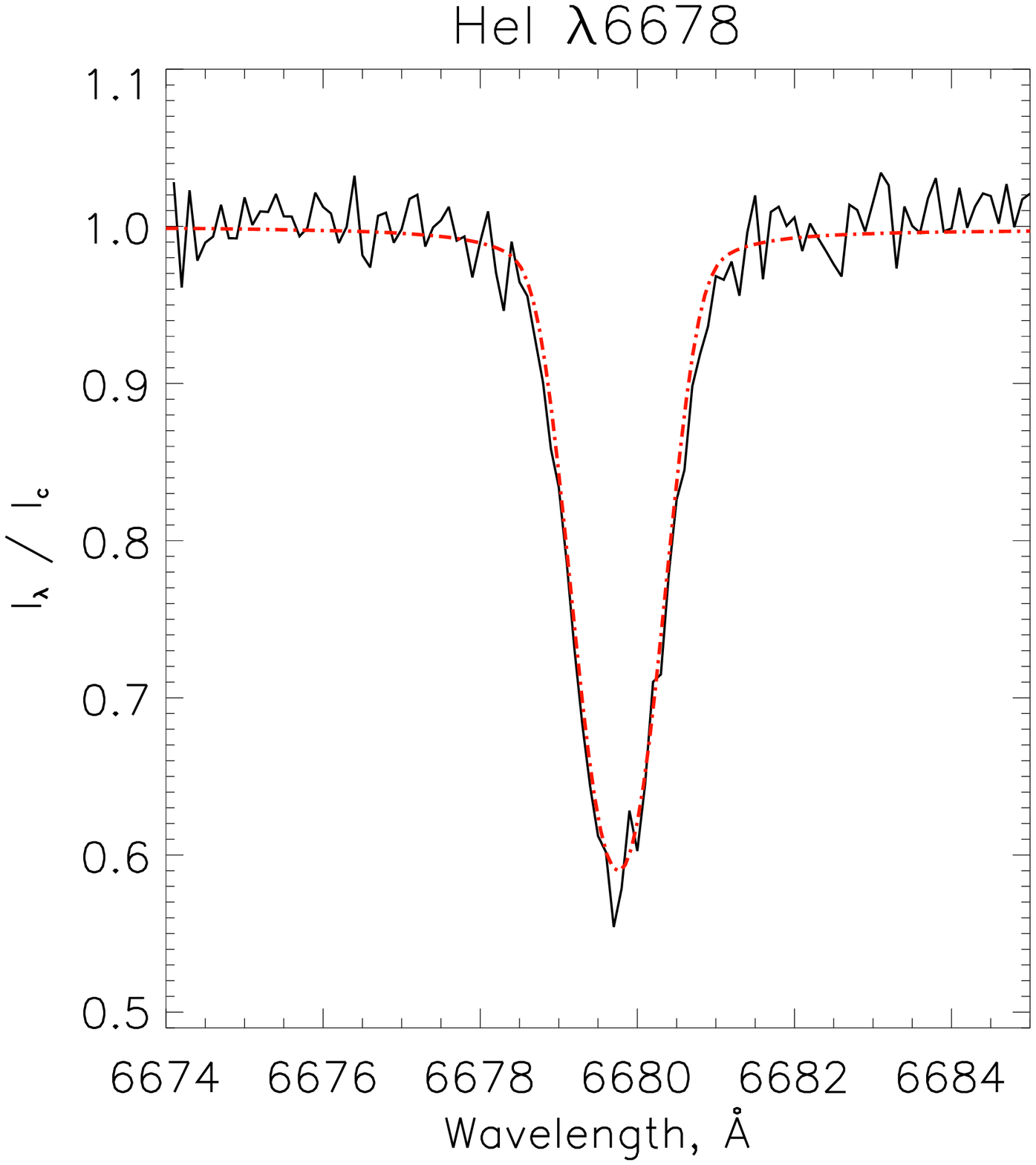}
\caption{Testing of the spectral analysis method. Modeling of HD~34078 (O9.5 V). Comparison of the profiles of selected lines with the best model spectra. The solid line shows the observed profile, and the dashed-dotted line -- the TLUSTY-model. The spectrum was obtained with UFES spectrograph. }
\label{fig:spectrum34078}
\end{center}
\end{figure*}

       ${\it v\sin i}$ values for HD~15629, HD~36591 and  HD~42597  were adopted to be the same as in \citet{MartinsHD15629,SimonDiaz2014} and \citet{Braganca12}, and equal to 90, 9 and 87~$\kms$ correspondingly. There are several different estimates of ${\it v\sin i}$ for HD~34078 -- \citet{Herrero92} give ${\it v\sin i}=40 ~\kms$, while \citet{Gies} obtained ${\it v\sin i}=27~\kms$, \citet{MartinsHD15629} adopted ${\it v\sin i}=40~\kms$. We obtained ${\it v\sin i}=21\pm3 ~\kms$  from the Fourier transform of the profiles of metal lines (see e.g. \citet{SimonDiaz2007}) and used this value in the analysis. 

       We estimated the parameters of HD~15629 and HD~34078 with method given in Section~\ref{sec:tlusty} using 9 and 18 spectral lines, correspondingly.  We also estimated their parameters with the smaller number of lines which is equal to the number of lines used to estimate parameters of MT259. From Table~\ref{tab:test} it is clear that parameters estimated with different number of spectral lines are in good agreement with each other and with the parameters obtained by \citet{MartinsHD15629} who used  {\sc tlusty} and {\sc cmfgen} codes. Moreover we estimated ${\rm He/H}=0.08\pm0.02$ for HD~15629 that is in agreement with ${\rm He/H}=0.08$ obtained by \citet{Repolust}. Figures~\ref{fig:spectrum15629}~and~\ref{fig:spectrum34078} show the results of fitting of HD~15629 and HD~34078 observed spectra with the best model.

       To test the method for low resolution spectra we initially estimated $T_{\mathrm{eff}}$ and log$\,\textsl{g}$ of HD~42597 and HD~36591 with the procedure described in Section~\ref{sec:tlusty} and synthetic spectra computed using stellar atmosphere models from the {\sc bstar2006} \citep{LanzHubeny2007}. The synthetic spectra were computed with different values of Teff, logg, $v_{\rm{turb}}$ and He/H. The final values of $T_{\mathrm{eff}}$ and log$\,\textsl{g}$ are given in Table~\ref{tab:test}. The parameters of HD~36591 are well consistent with the estimates of other authors summarized in Table~\ref{tab:test}. Then we estimated $T_{\mathrm{eff}}$ and log$\,\textsl{g}$ of HD~42597 and HD~36591 using low resolution spectra of these stars. These low resolution spectra were obtained from high resolution spectra by convolving with the instrumental profile with the width corresponding to SCORPIO resolution. We also had added Poisson noise in these low resolution spectra to make their S/N comparable to S/N of our spectra of MT282 and MT343 stars. Estimates of  $T_{\mathrm{eff}}$ and log$\,\textsl{g}$ obtained with the same lines and spectral regions as were used for MT343 are presented in Table~\ref{tab:test}. These estimates are well consistent with those obtained for high resolution spectra. Thus the method described in Section~\ref{sec:tlusty} allows to obtain reliable estimates of $T_{\mathrm{eff}}$ and log$\,\textsl{g}$ in the case of low spectral resolution. But the parameter errors obtained from low resolution spectra are relatively high that is mostly due to lower S/N of these spectra. These errors are consistent with those that can be estimated comparing the observed and synthetic spectra by eye. Thus, for the analysis of low resolution spectra we used the same value of $\chi^2_{\rm t}$ that was found for high resolution spectra.
       
\end{appendix}  

\bibliographystyle{pasa-mnras}
\bibliography{Maryevabib}
       
\end{document}